\documentclass[aps,prc,twocolumn,superscriptaddress,nofootinbib,tightenlines,preprintnumbers,showkeys,floatfix,10pt]{revtex4-2}

\usepackage{graphicx}
\usepackage{setspace}
\usepackage{amsmath}
\newcommand{\trace}{\operatorname{tr}}

\usepackage{url}
\usepackage{tikz}
\usepackage{forest}
\usepackage[super]{nth}
\usepackage{gensymb}
\usepackage{multirow}

\usepackage{mathtools}

\usepackage{isotope}
\usepackage{braket}

\newcommand{\smel}[3]{\braket{#1 | #2 | #3}}

\usepackage{xcolor}

\begin{document}

\title{\textit{Ab initio} Gamow density matrix renormalization group for broad nuclear many-body resonances}

\author{A. Sehovic}
\email{asehovic@fsu.edu}
\affiliation{Department of Physics, Florida State University,
Tallahassee, Florida 32306, USA}

\author{K. Fossez}
\email{kfossez@fsu.edu}
\affiliation{Department of Physics, Florida State University,
Tallahassee, Florida 32306, USA}
\affiliation{Physics Division, Argonne National Laboratory, Lemont,
Illinois 60439, USA}

\author{H. Hergert}
\email{hergert@frib.msu.edu}
\affiliation{Facility for Rare Isotope Beams, East Lansing, Michigan 48824, USA}

\begin{abstract}
\begin{description}
    \item[Background] The reach of \textit{ab initio} theory has greatly increased in recent decades. However, predicting the location of the drip lines remains challenging due to uncertainties in nuclear forces and difficulties in describing nuclei that behave as open quantum systems.
    \item[Purpose] In this work, we extend the \textit{ab initio} Gamow Density Matrix Renormalization Group (G-DMRG) approach to the regime of broad many-body resonances to pave the way for systematic tests of nuclear forces in light exotic nuclei.
    \item[Methods] To stabilize calculations, we introduce a new truncation scheme in the reference space, and propose an orbital ordering based on entanglement considerations. 
    We then show how continuum couplings increase entanglement in the many-body problem, and propose a new truncation scheme to stabilize the renormalization and accelerate calculations in extreme conditions. 
    Finally, we demonstrate that natural orbitals can be used to efficiently describe broad resonances by introducing a new ordering scheme and by redefining the reference space based on occupations.
    \item[Results] Leveraging our findings, we propose a recipe to converge \textit{ab initio} G-DMRG calculations and apply it in low-lying states of \isotope[5,6]{He} and \isotope[4]{H}, demonstrating control of the renormalization and the emergence of convergence patterns. 
    We also obtain the first direct \textit{ab initio} calculation of the $J^\pi = {1/2}^+$ ground state of \isotope[5]{H}.
    \item[Conclusions] We demonstrate that entanglement due to continuum couplings can be controlled in extreme conditions and successfully extend the G-DMRG approach in the regime of broad many-body resonances.
\end{description}
\end{abstract}

\maketitle

\section{Introduction}
\label{sec_intro}

In low-energy nuclear physics, predicting the limits of nuclear stability with respect to the number of protons ($Z$) and neutrons ($N$), or drip lines, remains an ambitious challenge~\cite{baumann12_1556,thoennessen13_1843,crawford24_3100}. 
Away from stability, changes in the nuclear mean field lead to shell evolution~\cite{sorlin08_2379,otsuka20_2383} and deformation in the so-called islands of inversion~\cite{brown22_2667}. 
Even though there remains considerable work to fully understand their emergence from nuclear forces and many-body effects, the basic mechanisms behind these phenomena are relatively well understood. 
At the drip lines, however, new intriguing phenomena appear such as halo structures~\cite{jensen04_233,tanihata13_549} and exotic decay modes~\cite{zhou22_2560,pfutzner23_2859}, whose description requires a new interdisciplinary paradigm in which weakly bound and unbound nuclei are treated as open quantum systems (OQSs)~\cite{dobaczewski07_17,michel09_2,rotter15_2464}, \textit{i.e.} systems coupled to an environment of scattering states and decay channels~\cite{michel21_b260}. 
In practice, this paradigm translates into the theoretical unification of nuclear structure and reactions~\cite{navratil16_1956}, in which couplings between bound and scattering states must be accounted for when solving the quantum many-body problem.

The experimental exploration of the drip lines is currently a high priority in nuclear science~\cite{johnson20_2389,crawford24_3100,brown25_3117}, making it critical to extend our predictive power away from stability. 
A promising avenue is to rely on \textit{ab initio} theory~\cite{hebeler15_1872,hergert20_2412}. 
In low-energy nuclear physics, \textit{ab initio} methods start from nucleons as degrees of freedom, take nuclear forces adjusted on few-body data alone as input~\cite{epelbaum09_866,machleidt11_414,epelbaum12_856,hammer20_2651}, and converge to an exact solution of the quantum many-body problem in some limit. 
In the past few decades, bound-state \textit{ab initio} calculations have reached medium-mass~\cite{hergert20_2412} and even closed-shell heavy nuclei~\cite{hu22_2632,bonaiti25}.
However, recent results show that stronger constraints on nuclear forces might be needed to better reproduce observables in exotic nuclei, notably to locate the drip lines where other approaches lose their predictive power~\cite{hergert20_2412,stroberg21_2483,hu25}.

In this work, we extend \textit{ab initio} theory in light exotic nuclei with extreme $N/Z$ imbalances where, from a computational perspective, quasi-exact results can realistically be obtained. 
Among the 45 known light nuclei, defined here as nuclei with $A = N + Z < 12$, five have halo ground states (\isotope[6,8]{He}, \isotope[8]{B}, \isotope[11]{Li}, \isotope[11]{Be}), 21 are unbound, and most have low-lying unbound excited states.
Thus, the light sector is a rich testing ground where emergent phenomena and resonant structures can reveal unique information about nuclear forces. 
Of course, as is discussed below, nuclear forces have already been tested to some degree in this region, but more systematic explorations that also include challenging states are needed.

At present, there are essentially two classes of \textit{ab initio} approaches to the description of many-body resonances in light nuclei. 
The first is based on the resonating group method (RGM)~\cite{wheeler37_2315,fliessbach82_638}, where the structure of bound clusters is calculated with, for instance, the no-core shell model (NCSM)~\cite{barrett13_688}, and then the continua between the clusters (their relative motions) are handled \textit{via} coupled-channel equations, like in the NCSM-RGM and its extension, the NCSM with continuum (NCSMC)~\cite{navratil16_1956}. By assuming specific partitions,
these approaches can describe systems like $\isotope[6]{He} $ and $\isotope[9]{He}$ as $\isotope[4]{He}+n+n$~\cite{baroni13_753} and $\isotope[8]{He}+n$~\cite{vorabbi18_1977}, respectively, and reach sufficient precision to actually test nuclear forces. However, these approaches are difficult to extend beyond two- and three-body decay or reactions.

In contrast to RGM-based methods, the second class of approaches formally solves the problem of scaling to an arbitrary number of partitions of the system by generalizing the many-body problem into the complex-energy plane using the quasistationary formalism~\cite{baz69_b3}.
In this formalism, resonant phenomena are approximated as Gamow (or Siegert) states~\cite{gamow28_500,siegert39_132}, which are solutions of the Schr\"odinger equation with outgoing boundary condition, \textit{i.e.} complex-energy eigenstates of a non-Hermitian many-body Hamiltonian~\cite{moiseyev11_b72}. 
We note that Gamow states, and non-Hermitian quantum mechanics, are rigorously defined in the rigged Hilbert space (RHS)~\cite{gelfand61_b1,gelfand68_b2,maurin68_b4}.
Methods in the quasistationary formalism are typically based either on the uniform complex-scaling (CS) method~\cite{moiseyev11_b72}, or on the Berggren basis (BB) expansion~\cite{berggren68_32,berggren93_481}.
In principle, the two methods provide equivalent results~\cite{masui14_876}, but in many-body applications the flexibility of the BB can be exploited to design efficient truncation schemes. 
This situation could change with new extrapolation techniques for Gamow states leveraging bound and narrow resonance information~\cite{yapa23_2926,yapa25_3250}.

In the CS method, the wave function is rotated from the real axis into the complex (momentum) plane in order to access resonant states associated with poles of the scattering matrix in the \nth{2} Riemann sheet. 
It was used successfully to describe systems that were partitioned into up to five subsystems~\cite{myo20_2674}, including \textit{ab initio} Faddeev-Yakubovsky calculations~\cite{lazauskas18_2032,lazauskas19_2363}, and efforts are underway to match these achievements using complex-scaled finite-volume discrete variable representation calculations~\cite{yu24_3045}.

The BB expansion takes a more direct approach and expands the wave function explicitly in terms of bound states, resonances, and scattering states. 
It has found many applications, and here we only mention the Gamow shell model (GSM)~\cite{michel09_2,michel21_b260}, the Gamow density matrix renormalization group (G-DMRG)~\cite{rotureau06_15,rotureau09_140}, and their respective \textit{ab initio} (no-core) formulations~\cite{papadimitriou13_441,shin16_1860}.

All the approaches mentioned above can describe narrow many-body resonances as well as broad few-body resonances, but options are limited when it comes to broad many-body resonances. 
Here, we define a broad many-body resonance as \emph{an unstable quantum state, whose lifetime is barely larger than the interaction timescale of its constituents, and whose dynamics cannot be reduced to the effective interaction of a few substructures.} 
Examples of such resonances are the ground states of \isotope[8]{C}~\cite{brown14_2616} and \isotope[9]{N}~\cite{charity23_2960} which are sequential four- and five-proton emitters, respectively, or the ground states of \isotope[4-7]{H}. 
Other examples of unbound many-body states of interest, although narrower are the five-neutron $J^\pi = {1/2}^+$ ground state of \isotope[9]{He}~\cite{fossez18_2171,votaw20_2353}, or the four-neutron ground state of \isotope[28]{O}~\cite{kondo23_2923}.
It is the purpose of this work to extend \textit{ab initio} G-DMRG calculations toward the regime of such broad many-body resonances.

The G-DMRG method was proposed in Ref.~\cite{michel04_93} and implemented in Refs.~\cite{rotureau06_15,rotureau09_140} for effective shell model Hamiltonians. 
A few years later, the G-DMRG was extended to solve the \textit{ab initio} problem. 
It was first applied to the single-particle neutron resonance in \isotope[5]{He}~\cite{papadimitriou13_441}, and later to the low-lying resonances in \isotope[6]{Li}~\cite{shin16_1860}. However, it failed to provide converged results due to computational limitations. 
The most recent \textit{ab initio} application was the description of the so-called tetraneutron or four-neutron system~\cite{fossez17_1916}, where the strength of the interaction was artificially increased to overbind the system.
In parallel, the method was used with effective shell model Hamiltonians to study halo structures and many-body resonances in neutron-rich Mg~\cite{fossez16_1793}, O~\cite{fossez17_1927,jones17_1973}, He~\cite{fossez18_2171}, Li~\cite{mao20_2375}, and F isotopes~\cite{fossez22_2540}. 
At present, the G-DMRG method is one of the most powerful many-body approaches for studying multi-nucleon resonances, but past applications have shown that significant developments are needed to handle broad many-body resonances.

In Sec.\ref{sec_formalism}, we introduce the formalism, followed by a discussion of theoretical and computational developments in Sec.\ref{sec_developments}. 
In Sec.\ref{sec_results}, we demonstrate how these new techniques can be combined to efficiently describe resonances in selected few- and many-body systems. 
Finally, we present our conclusions in Sec.~\ref{sec_conclusion}.
\section{Formalism}
\label{sec_formalism}

\subsection{Broad many-body resonances and quasistationary formalism}
\label{ssec_resonance}

In quantum mechanics, a resonance is a time-dependent phenomenon where a system forms at an energy $E_r$, with a decay width $\Gamma$, and subsequently decays. 
Here, we are concerned with atomic nuclei consisting of protons and neutrons. 
The interaction between nucleons is mediated by the strong force, whose characteristic timescale is of the order of $\Delta t \sim \hbar/(m c^2) \approx {10}^{-24} \, \text{s}$, where $m \approx 939.0 \, \text{MeV}$ is the mass of a nucleon. Naturally, this timescale affects the interpretation of dynamical processes. 
For example, a fast reaction process may provide insufficient time for the formation of a compound nucleus if nucleons in the projectile can only interact with a few nucleons on the surface of the target and immediately separate. 
For a nucleus to exhibit a resonance, all the nucleons must interact with each other before separating, meaning they must form a proper many-body system. 
Intuitively, the larger the system, the more time will be necessary for this to happen. This constraint constitutes the limit of the system's existence.
Quantifying this limit in nuclei is difficult~\cite{fossez16_1335}, but if we assume that i) the nucleus is spherical, ii) nucleons move at the Fermi velocity of $v_F = 0.27c$, and iii) all the nucleons will interact with each other if at least one has enough time to move across the nucleus, we arrive at a minimal existence time of about $10^{-23}$-$10^{-22} \, \text{s}$ for a medium-sized nucleus of mass $A = 10$-40. 
Experimentally, the threshold for existence is usually set at $10^{-22} \, \text{s}$~\cite{thoennessen04_1165}, but it can probably be somewhat lower in light nuclei.

Assuming that we have a proper resonance phenomenon, we can treat the formation of the system and its subsequent decay as a two-step process. 
If the system has enough time to ``forget'' how it was formed, its state can be approximately described as stationary before the decay. 
This is the essence of the quasistationary formalism. 
In this picture, the system is in a ``state'' with a well-defined energy $E_r > 0$, coupled to a continuum of scattering states, rendering it unstable. 
The stronger these couplings, the larger the energy spread (or width) $\Gamma$ and the shorter the half-life $T_{1/2}$. 
For exponential decay, the half-life is directly related to the width since $T_{1/2} = \hbar \ln(2) / \Gamma$. 
Using this expression and the time-energy uncertainty $\Delta E\Delta t \approx \hbar$, we see that the width can be understood as the uncertainty on the energy $\Delta E \sim \Gamma$, due to the time-dependent nature of the phenomenon.

We note that this result can be used to derive a limit on the width of a resonance. 
Assuming the experimental limit of $T_{1/2} \approx 10^{-22} \, \text{s}$, we get a maximal width of $\Gamma \approx 4.5 \, \text{MeV}$. 
However, we must take this result with a grain of salt. 
If the width is so large that $\Gamma/2 > E_r$, according to the statistical interpretation of the quasi-stationary formalism, there is a non-zero probability to observe an unbound system with a negative energy $E_r - \Gamma/2 < 0$. 
This obviously cannot be the case, and thus it is conventional to consider broad resonances to be those states that satisfy $\Gamma/2 \sim E_r$. 
This sets the limit of validity for the quasistationary formalism.

For a state with well-defined energy, the time-dependent Schr\"odinger equation yields a separable solution of the form $\psi(\vec{r}) \exp(-iEt/\hbar)$. 
To account for decay, Gamow and Siegert's idea was to solve the stationary radial part with outgoing boundary condition. 
At positive energy, this approach yields solutions with complex energies,
\begin{equation}
    \tilde{E} = E_r - i\frac{\Gamma}{2}
    \label{eq_E_G}\,,
\end{equation}
where the real part gives the energy position $E_r$ of the resonance, and the imaginary part is directly proportional to its width $\Gamma$.
These solutions are known as Gamow or Siegert states. 
Fundamentally, a Gamow state approximates a resonance phenomenon by a stationary radial wave function that grows exponentially (particles are leaving the system), combined with a time-dependent wave function that decreases exponentially, as can be seen by inserting Eq.~\eqref{eq_E_G} in the time-dependent part of the wave function:
\begin{equation}
    e^{-\frac{i}{\hbar} \tilde{E} t} = e^{-\frac{i}{\hbar} E_r t} e^{-\frac{\Gamma}{2\hbar} t}
    \label{eq_Et}
\end{equation}
This compensation between the radial and time-dependent parts of the wave function ensures the conservation of the flux of particles.

\begin{figure}[h!]
    \centering
	\includegraphics[width=.9\linewidth]{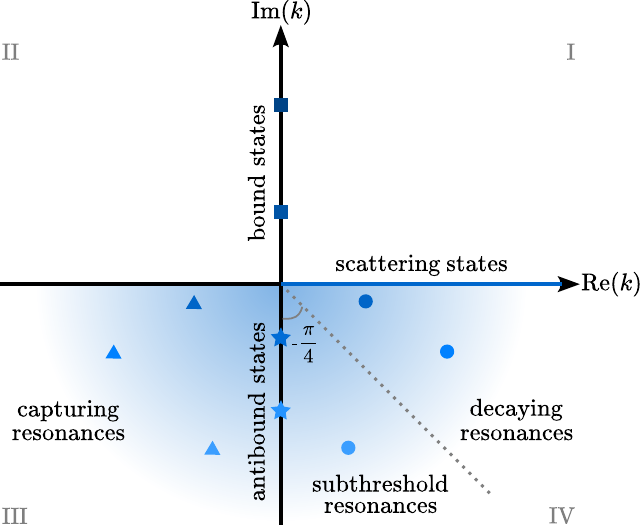}
  \caption{Illustration of the different types of poles of the single-particle $S$ matrix in the complex momentum plane. The continuum of scattering states is represented by a thick blue line on the positive side of the real axis.}
	\label{fig_poles}
\end{figure}

At the single-particle level, the complex energy $\tilde{E}$ of a Gamow state is directly related to the linear momentum $\tilde{k}$ by the usual dispersion relation ${ \tilde{E} = (\hbar^2 \tilde{k}^2)/(2m) }$. 
Furthermore, complex momenta of Gamow states are associated with poles of the single-particle scattering matrix ($S$-matrix) as illustrated in Fig.~\ref{fig_poles}. 
We note in passing that~\cite{papadimitriou13_441,michel21_b260}:
\begin{equation}
    \tilde{k}^2 = \frac{2m\tilde{E}}{\hbar^2} = \frac{2m E_r}{\hbar^2} \left(1 - i\frac{\Gamma}{2 E_r}\right)
    \label{eq_kcx}
\end{equation}
which again illustrates that broad resonances are defined by $\Gamma/2 \sim E_r$. 
Note that $E_r$ must be replaced with the appropriate separation energy in the many-body case.

Before we continue, we must emphasize an important point related to the interpretation of complex expectation values: There is a difference between the so-called pole energy obtained in the quasi-stationary formalism, and the measured energy position and width~\cite{myo23_2689}. 
While this difference is negligible for narrow resonances, it can be noticeable for broad resonances~\cite{klaiman10_3288}. 
In fact, it can be argued that it is the modulus of complex expectation values (and their phase) that are observables~\cite{moiseyev11_b72}. 
For convenience, in this work we rely on Berggren's interpretation of complex expectation values, which remains approximately valid even in the strong continuum coupling regime, but we recognize that one must be careful when discussing broad resonances.

Poles of the ${ S }$-matrix are symmetric with respect to the transformation ${ k \to -k^* }$, which is equivalent to the time-reversal operation in this context. 
By inspecting the asymptotic behavior of their radial wave functions $e^{i\tilde{k}r}$, we can distinguish four types of poles. 
Assuming ${ \kappa, \gamma > 0 }$, we have i) bound states defined by ${ k = i\gamma }$ (squares); ii) antibound or virtual states defined by ${ k = -i\gamma }$ (stars); iii) decaying resonances defined by ${ k = \kappa - i\gamma }$ (circles); and capturing resonances defined by ${ k = -\kappa - i\gamma }$ (triangles). 
Except for bound states, which can be normalized using the norm in the Hilbert space, all other poles are in the \emph{unphysical Riemann sheet} of the complex-momentum plane, represented by a color gradient in Fig.~\ref{fig_poles} where the shaded part represents the region of physical interest.

To help the reader gain intuition about the relation between energy positions, widths, and complex momenta, we show the evolution of $E_r$ and $\Gamma$ with the position of $k$ in the complex plane in App.~\ref{app_cx_energy}. 
In addition, writing $\tilde{k} = |\tilde{k}| e^{i\theta}$, we derive the conditions on $\theta$ that delineate the region in the $(E_r,\Gamma)$ plane where $0 \leq E_r \leq \Gamma/2$, and show that it is defined by $-\pi/8 \geq \theta \geq -\pi/4$. 
There are relevant physical states in this region that satisfy $\Gamma/2 \sim E_r$, like the $J^\pi = 2^-$ state in \isotope[4]{H} described in this work.

States in the \nth{4} quadrant with poles below the ${ -\pi/4 }$ line are usually called subthreshold resonances, antibound resonances, or virtual resonances. 
While a quantum system cannot be prepared in such states, subthreshold poles can affect observables like low-energy scattering cross sections. 
For example, the large scattering length of the neutron-neutron interaction is a result of the existence of an n-n antibound state.

The situation is more complicated in many-body systems.  
Directly solving the radial Schr\"odinger equation by imposing outgoing boundary conditions is impractical beyond three-body decay. 
It is equally difficult to find many-body resonances by searching for the corresponding many-body poles of the S-matrix.

The quasistationary formalism offers an efficient approach to the problem of multi-particle decay by approximating resonant phenomena using Gamow states, resulting in a stationary non-Hermitian problem. 
The key idea is to express the many-body problem in a basis that can represent any many-body outgoing asymptotic\footnote{This is a subtle point discussed in the last paragraph of Sec.~\ref{ssec_berggren}}, and then to solve the problem using linear algebra for complex-symmetric matrices.

\subsection{Berggren basis}
\label{ssec_berggren}

To access many-body resonances with the G-DMRG method, we rely on the Berggren basis~\cite{berggren68_32,berggren93_481}. 
It is an extension of the Newton basis~\cite{newton82_b6} in the complex-momentum plane using Cauchy's integral theorem, and thus has the particularity of explicitly including single-particle Gamow states. 

To build the Berggren basis for a given partial wave $(l,j)$, the first step is to choose a potential guided by physical arguments, and to solve the one-body Schr\"odinger equation with outgoing boundary conditions to find its bound and low-lying resonant states, as shown in Fig.~\ref{fig_poles}. 
The goal is usually to find a potential whose eigenstates are as close as possible to the natural orbitals of the system of interest, defined as eigenstates of the one-body density matrix~\cite{brillouin33,lowdin55_2499,lowdin56_2498}.

\begin{figure}[b]
  \centering
	\includegraphics[width=.9\linewidth]{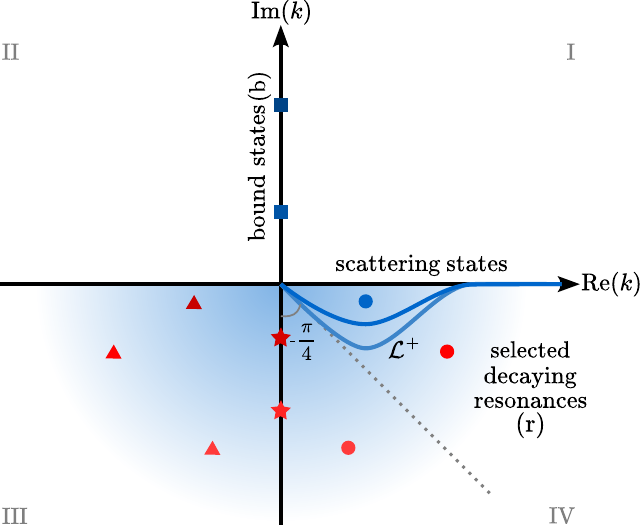}
	\caption{Illustration of the Berggren basis in the complex momentum plane. Poles of the $S$ matrix that are excluded are marked in red, while those included are marked in blue. Two valid deformations of the continuum of scattering states $\mathcal{L}^+$ are indicated by blue contours in the \nth{4} quadrant.}
	\label{fig_Berggren}
\end{figure}

Then, as shown in Fig.~\ref{fig_Berggren}, the Berggren basis is obtained by combining three classes of states: i) all the bound states (b), ii) selected resonant states (r) that are expected to play an important role for the problem at hand, and iii) the complex-energy scattering states along a contour $\mathcal{L}^+$ in the complex momentum plane that encloses the selected resonant states (and only those) before going back to the real axis. 
The completeness relation can be written as 
\begin{equation}
    \sum_{n \in (b,r)} \ket{u(k_n)}\bra{u(k_n)} + \int_{\mathcal{L}^+} dk \, \ket{u(k)}\bra{u(k)} = \hat{1}
    \label{eq_BB}
\end{equation}
where $n$ runs over the bound and resonant states. 
Cauchy's integral theorem ensures that the shape of the contour is irrelevant as long as it encloses the same resonant states. 
In practice, the Berggren basis can be used like any other discrete basis once the integral along the contour has been discretized using a quadrature method.

We note that Berggren proposed to interpret the magnitude of the imaginary part of a complex quantity associated with an observable as the uncertainty on the real part of that quantity, \textit{i.e.} the observable itself, due to treating a time-dependent phenomenon as stationary~\cite{berggren68_32,gyarmati72_775,berggren96_23}. 
This point is discussed in Sec.~\ref{sec_id} when we look at the eigenvalues of complex-symmetric densities.

Finally, it must be noted that even though the Berggren basis can be a complete basis at the single particle level, there is no guarantee that the resulting many-body basis will be complete for the many-body resonance of interest. 
This is a subtle issue which has to do with the poles of the many-body $S$-matrix and the ability of the many-body basis to reconstruct the correct many-body asymptotic.
In principle, one can always extend the contour defining the Berggren basis further into the \nth{4} quadrant of the complex-momentum plane and use a finer discretization. 
However, the issue manifests itself by making calculations unstable, and thus it can be difficult to find a good starting basis, especially for broad resonances.

\subsection{Hamiltonian}
\label{ssec_H}

In this work, our starting point are \textit{ab initio} nuclear Hamiltonians in second quantization:

\begin{equation}
    H = \sum_{i,j} \smel{i}{\hat{T}}{j} \hat{a}_i^\dagger \hat{a}_{j} + \frac{1}{2} \sum_{k,l,m,n} \smel{k,l}{\hat{V}^\text{NN}}{m,n} \hat{a}_k^\dagger \hat{a}_l^\dagger \hat{a}_{n} \hat{a}_{m}
    \label{eq_H}
\end{equation}
where ${ \hat{T} }$ is the kinetic energy operator and ${ \hat{V}^\text{NN} }$ is the nucleon-nucleon (NN) interaction. 
In this work, we employ the two-body chiral interaction $\text{NNLO}_\text{opt}$~\cite{ekstrom13_865} with renormalization cutoff $\lambda = 1.7 \, \text{fm}^{-1}$ in $\text{V}_{\text{low-}k}$~\cite{bogner03_413}. 
While this interaction does not include explicit three-body forces and does not benefit from more recent developements of chiral EFT, it is sufficient for our purpose. 
At present, we focus on the method and will explore more interactions in the future.

To obtain the translationally invariant intrinsic Hamiltonian, we remove the kinetic term due to the center-of-mass motion. 
This leaves the two-body interaction unchanged and the intrinsic kinetic energy is given by:

\begin{align}
    \hat{T}_\text{int} 
    &= \left(1 - \frac{1}{A}\right) \sum_{i=1}^A \frac{\vec{p}_i^2}{2m} - \frac{1}{A} \sum_{i < j = 1}^A \frac{\vec{p}_i.\vec{p}_j}{2m} \\
    &= \left(1 - \frac{1}{A}\right) \hat{T} + \hat{T}^{(2)}
    \label{eq_Tint}
\end{align}
where ${ \hat{T}^{(2)} }$ is the two-body recoil term.

In practice, we calculate the one- and two-body matrix elements in the laboratory (or center-of-mass) frame, so that they are expressed in terms of orbitals from a generating potential which will ideally resemble the mean field of the system. 
While this is trivial for one-body operators, interaction matrix elements require a series of basis changes. 
They are originally given in a relative-momentum basis made of spherical Bessel functions. 
The first step is to express them in the harmonic oscillator (HO) basis in relative coordinates. 
Obviously, we have to assume that the range of the potential is finite so that the matrix elements remain accurate. 
Then, using the Brody-Moshinsky transformation~\cite{moshinsky59_1559}, we can obtain the matrix elements in a HO basis in the laboratory frame. 
Finally, a last basis change is made to express the matrix elements in the Berggren basis. 
It has been shown that the latter can be done accurately due to the finite range of the potential~\cite{hagen06_14,papadimitriou13_441}.

\subsection{Gamow density matrix renormalization group}
\label{sec_dmrg}

Formally, the nuclear, condensed matter, and quantum chemistry many-body problems are all similar. 
For instance, like electrons occupying atomic sites in materials, or atomic and molecular orbitals in molecules, nucleons can be seen as occupying nuclear orbitals. 
Furthermore, nuclei are finite systems, with a number of nucleons ranging from a few to a couple of hundreds, which is similar to the quantum chemistry problem. 
However, there are also a few key differences: 
Nuclei consist of two types of fermions, namely protons and neutrons, and they are self-bound. 
In addition, nuclear forces generate strong correlations between nucleons forming bound structures, making the nuclear problem highly 
non-perturbative.
For example, adding just one nucleon to a nucleus or providing some excitation energy can often result in a dramatic change of structure.

The DMRG method~\cite{peschel99_b245,schollwock05_479}, originally introduced in condensed matter physics~\cite{white92_488,white93_491} and now widely used in quantum chemistry as well~\cite{wilson08_b371,marti10_3192,chan11_3194,wouters14_3248,baiardi20_3193}, efficiently solves the many-body problem by expressing wave functions in a factorized form that reflects the entanglement within the system. Such states are commonly known as a tensor network states~\cite{orus14_2015,okunishi22_3196}. 
The DMRG performs particularly well when the entanglement between particles is relatively low and only a limited number of many-body basis states is required to approximate the wave function.

In nuclear physics, the DMRG method was first applied to solve the $M$-scheme (valence) shell model problem~\cite{dukelsky99_2004,dukelsky01_2003,pittel01_2008,dukelsky02_1568} for bound states. 
However, the method exhibited convergence issues because the usual DMRG truncation slowly breaks spherical symmetry~\cite{pittel03_2007,papenbrock05_837}. 
This led to a reformulation of the DMRG method for the $J$-scheme shell model~\cite{pittel03_2007,dukelsky04_446,pittel06_274,pittel08_2005,thakur08_2006,pittel09_2012} which, by construction, always preserves the spherical symmetry. 
Despite this achievement, the convergence of the DMRG method in the harmonic oscillator basis, or even the Hartree-Fock basis, remained slow. 
New evidence suggests that entanglement in bound nuclei scales with the size of the system, \textit{i.e.} obeys a volume law~\cite{pazy23_3230,gu23_3201}, which could explain prior observations.

There have been recent attempts to overcome this issue by using ordering techniques based on entanglement measures~\cite{legeza03_3199,rissler06_3191,barcza11_3247,legeza15_2055} or to reduce the problem of strong correlations by applying the DMRG method with \textit{ab initio} Hamiltonians renormalized by the in-medium similarity renormalization group (IMSRG) method~\cite{tichai23_2940,tichai24_3190}. 
In any case, more studies of entanglement in nuclear systems~\cite{robin21_2471,hengstenberg23_3227,perez23_3249,kruppa21_3198,kruppa22_3200,kovacs25_3197} are needed. 
For instance, recent results show that a proton-neutron factorization of the nuclear wave function could provide an efficient ansatz in certain cases~\cite{johnson23_3195,gorton24_3228}. 
This was indeed observed in Refs.~\cite{perez23_3249,tichai24_3190}.

The G-DMRG approach~\cite{rotureau06_15,rotureau09_140} generalizes the DMRG method to the complex-energy plane using the Berggren basis for the description of open quantum systems. In this approach, the system corresponds to discrete states (bound states and resonances) and the environment to scattering states. We note that formulations of DMRG for OQSs exist in other contexts~\cite{weimer21_3210}. 

When the system-environment entanglement is relatively low, like in narrow resonances, the G-DMRG method is efficient and can reach model spaces beyond standard configuration-interaction methods, in both phenomenological~\cite{fossez17_1927,fossez18_2171,fossez22_2540} and \textit{ab initio}~\cite{papadimitriou13_441,shin16_1860,fossez17_1916} studies.

\begin{figure}[b!]
  \centering
  \includegraphics[width=.9\linewidth]{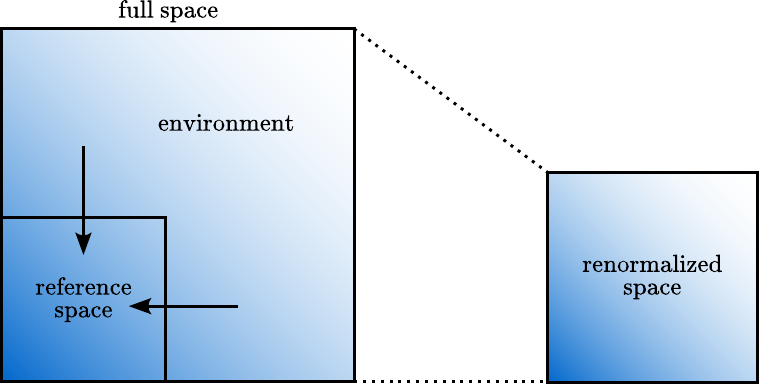}
  \caption{Illustration of the Wilsonian renormalization of the Hamiltonian in the original formulation of DMRG. The Hamiltonian is originally built in the reference space, and then the effect of the environment is ``absorbed'' by the renormalization. The Hamiltonian obtained, expressed in a renormalized space, approximates the full Hamiltonian.}
  \label{fig_DMRG}
\end{figure}

In its original formulation~\cite{white92_488,white93_491}, 
the DMRG method starts from a small reference space providing a reasonable approximation of the state of interest, and then iteratively builds and optimizes the many-body wave function. 
This is done by constructing a reduced density matrix every time a new, previously excluded orbital is added in the active space, and by selecting only the relevant couplings based on occupations. 
In short, the DMRG method selects the most probable configurations for the targeted state. 
The Hamiltonian matrix built using these selected configurations is thus renormalized in the Wilsonian sense~\cite{marti10_3192}, as illustrated in Fig.~\ref{fig_DMRG}.

In contrast, the modern formulation of DMRG usually represents the wave function using matrix product states~\cite{rommer97_3246} (MPS), a particular type of tensor network state~\cite{orus14_2015,okunishi22_3196}, and optimizes it iteratively. 
By explicitly leveraging the structure of entanglement in the wave function, the MPS version of the DMRG method is quite flexible and able to handle large model spaces without dealing with the numerical renormalization of the Hamiltonian~\cite{peschel99_b245,schollwock05_479}. 
The G-DMRG method presented in this work is still based on the Wilsonian renormalization, but
we will explore an MPS formulation of the G-DMRG in the future. 
Below, we present the method in detail and then discuss the consequences of having a non-Hermitian Hamiltonian.

The first step in the DMRG algorithm is the selection of a number of orbitals based on physical considerations, and to build a fixed many-body reference space $\mathcal{A}$ in which a reasonable approximation of the targeted state can be obtained. 
We also refer to the space $\mathcal{A}$ as the \emph{system}. 
The choice of the reference space is critical, as discussed in Sec.~\ref{sec_reference_state}.

The space $\mathcal{A}$ contains all the $n$-body states with 
$n=1, ..., A$ that can be built using the selected orbitals. 
These many-body states are constructed as Slater determinants (SDs) with well-defined total angular momentum and parity ${ j_n^{\pi_n} }$, as defined in Eq.~\eqref{eq_SD} for $n=A$, and organized into families ${ f_\mathcal{A} = \{ n, j_A^{\pi_A} \} }$. 
\begin{equation}
    \Psi^{f_\mathcal{A}}(1, 2, ..., A) = \frac{1}{\sqrt{A!}}
    \begin{vmatrix}
        \psi_{i_1}(1) & \psi_{i_2}(1) & ... & \psi_{i_A}(1)\\
        \psi_{i_1}(2) & \psi_{i_2}(2) & ... & \psi_{i_A}(2)\\
        \vdots & \vdots & \ddots & \vdots \\
        \psi_{i_1}(A) & \psi_{i_2}(A) & ... & \psi_{i_A}(A)
    \end{vmatrix}
    \label{eq_SD}
\end{equation}
The numbers $1, 2, ..., A$ refer to the particles, and the indices $i_n$ to the orbitals occupied in the SD. 
Symbolically, we write the states in $\mathcal{A}$ as:
\begin{equation}
    \ket{\text{SD}^{f_\mathcal{A}}_a}
    \label{eq_SDa}
\end{equation}
where $a$ is the index of the SD state. 

We then build the Hamiltonian matrix in the reference space and diagonalize it to obtain the reference state $\ket{\Psi^{A, J^\pi}}_{0}$.
This state plays a critical role in G-DMRG as highlighted in Sec.~\ref{sec_id}.

The rest of the single-particle orbitals is what we often call the \emph{environment} or \emph{medium}. 
Strictly speaking, the environment is all the many-body states not included in the reference space. 

At the first iteration ($i=1$), an orbital from the environment is selected, and all the possible $n$-body states with 
$n=1, ..., A$ 
that can be built are added in a space $\mathcal{B}$, initially empty. 
As for the states in $\mathcal{A}$, we denote the states in $\mathcal{B}$ as:
\begin{equation}
    \ket{\text{SD}^{f_\mathcal{B}}_b}_{i=1}
    \label{eq_SDb}
\end{equation}

Finally, $A$-body states coupled to the desired total angular momentum and parity $J^\pi$ of the targeted state are built by coupling the states in $\mathcal{A}$ and $\mathcal{B}$ to form the space ${ \mathcal{A} \otimes \mathcal{B} }$.
We denote these states:
\begin{equation}
    \ket{\text{SD}^{A, J^\pi}_{a,b}}_{i=1} = 
    \left\{ 
    \ket{\text{SD}^{f_\mathcal{A}}_a} \otimes \ket{\text{SD}^{f_\mathcal{B}}_b}_{i=1} 
    \right\}^{A, J^\pi}
    \label{eq_SDab}
\end{equation}
The Hamiltonian matrix is expressed in this enlarged space and diagonalized to obtain the new eigenstate:
\begin{equation}
    \ket{\Psi^{A, J^\pi}}_{i=1} = \sum_{a,b} C_{a,b}^{i=1} \ket{\text{SD}^{A, J^\pi}_{a,b}}_{i=1}
    \label{eq_Psi1}
\end{equation}
We leave the discussion of the selection of the eigenstate for later. 

The reference space being fixed, if we want to optimize the representation of the targeted state in Eq.~\eqref{eq_Psi1} we need to build the 
reduced density matrix for the subspace $\mathcal{B}$ by performing a trace over the $\mathcal{A}$ degrees of freedom. 
Here, because we work in the $J$ coupling scheme, the reduced density matrix is calculated for fixed values of ${ j_B^{\pi_B} }$:
\begin{equation}
    \rho_{b,b',i=1}^{\mathcal{B}}(j_B^{\pi_B}) = \sum_{a} C_{a(j_A^{\pi_A}), b(j_B^{\pi_B})}^{i=1} C_{a(j_A^{\pi_A}), b'(j_B^{\pi_B})}^{i=1} 
    \label{eq_rho_red_1}
\end{equation}
The indices $a(j_A^{\pi_A})$ and $b(j_B^{\pi_B})$ indicate values of $a$ and $b$ matching the total spin and parity indicated in parentheses.

We then diagonalize the reduced density matrix to obtain the optimal representation of $\mathcal{B}$ for the targeted state:
\begin{equation}
    \hat{\rho}^{\mathcal{B}}_{i=1} \ket{ \phi_\alpha }_{i=1} = \omega_{\alpha,i=1} \ket{ \phi_\alpha }_{i=1}
    \label{eq_rho_red_1_diago}
\end{equation}
In the DMRG, the eigenvalues $\omega_{\alpha,i=1}$ are real, but in the G-DMRG they are complex. 
This point will be discussed in detail in Sec.~\ref{sec_id}. 
For the moment, the only important point is that the trace of the reduced density matrix always equals one, which implies that the imaginary parts of the eigenvalues cancel~\cite{rotureau09_140}. 

The DMRG truncation consists in keeping only the eigenvectors of the reduced density matrix, \textit{i.e.} the linear combinations of states in $\mathcal{B}$, that contribute significantly to the wave function. 
Traditionally, this was realized by keeping a fixed number of eigenvectors at each iteration, the so-called fixed bond dimension in condensed matter applications. 
Here, we use the so-called dynamical block state selection method, \textit{i.e.} we keep only the eigenvectors whose summed eigenvalues do not exceed a user-defined threshold. 
In the Hermitian case, these truncations are optimal in the sense of minimizing the error on the representation of the targeted state.
Given the probabilistic interpretation of the eigenvalues $\omega_{\alpha,i=1}$, which can be viewed as occupations in $\mathcal{B}$ for the targeted state, we truncate the space $\mathcal{B}$ by keeping only the $N$ eigenstates $\ket{ \phi_\alpha }_{i=1}$ whose eigenvalues satisfy
\begin{equation}
    \left| 1 - \operatorname{Re}\left( \sum_{\alpha=1}^N \omega_{\alpha,i=1} \right) \right| < \varepsilon
    \label{eq_DMRG_epsilon}
\end{equation}
The quantity $\varepsilon$ is the truncation error, and in the limit $\varepsilon \to 0$, we recover the full CI problem. 
The number of retained states, $N$, is equivalent to the bond dimension in condensed matter applications, and can vary from one iteration to the next. 
We note here that we apply the truncation based on the real part of the eigenvalues $\omega_{\alpha,i}$. 

At the second iteration ($i=2$), another orbital from the environment is selected, and all the possible $n$-body states with 
$n=1, ..., A$ 
that can be built using this state, denoted $\ket{\text{SD}^{f_\mathcal{B}}_b}_{i=2}$, are built. 
These states are coupled to the $N$ eigenvectors $\ket{ \phi_\alpha }_{i=1}$ retained at the previous iteration:
\begin{equation}
    \ket{\Phi_{b'}}_{i=2} = \ket{\text{SD}^{f_\mathcal{B}}_b}_{i=2} \otimes \ket{ \phi_\alpha }_{i=1}
    \label{eq_B2}
\end{equation}
and form the new content of $\mathcal{B}$. 

These states can be coupled to the reference space to form the space ${ \mathcal{A} \otimes \mathcal{B} }$, and the cycle is repeated until the states in the environment are exhausted. 
This process is commonly referred to as the \emph{warm-up} phase. 
The DMRG method is obviously order-dependent, meaning that the results depend on the order in which the states of the environment are included. 
This point is discussed in Sec.~\ref{sec_ordering}. 
To address this issue, most DMRG applications perform \emph{sweeps} where the process described above is repeated in reverse order, and then back in the original order, etc., until convergence is achieved. 
In the G-DMRG, as explained in Sec.~\ref{sec_nat}, sweeps can be avoided when using the Berggren basis together with natural orbitals.

By relying on the Berggren basis to include continuum couplings, the G-DMRG method deals with non-Hermitian Hamiltonians. 
The renormalized Hamiltonian matrix obtained at each iteration is complex-symmetric and yields complex-energy eigenvectors. 
As a consequence, the variational principle is not applicable and the targeted state must be found by other means. 
Moreover, because of the complex-energy scattering states entering the Berggren basis, only a small subset of all the many-body eigenvectors are physical states. 
Finding physical states is a common problem in all complex-energy approaches, which is usually referred to as the \emph{identification problem}. 
As discussed in Sec.~\ref{sec_id}, a simple solution was found for narrow resonances in the context of the Gamow Shell Model, but it can fail for broad many-body resonances, and it must be adapted to the G-DMRG method due to its iterative nature. 
The main consequence of the identification problem is to force the reference space $\mathcal{A}$ to be built exclusively using orbitals associated with poles of the single-particle scattering matrix, the so-called \emph{pole space}.

Another important consequence of non-Hermiticity is that the reduced density matrix is complex-symmetric and has complex eigenvalues (see App.~\ref{app_cx_density} for details). 
Given that i) we can interpret the real part as an occupation, and the absolute value of the imaginary part as an uncertainty on the occupation due to the decaying nature of the system, and ii) the trace of the reduced density matrix must equal one at the end of a G-DMRG calculation, we can generalize the DMRG truncation by applying it on the real part of the eigenvalues, \textit{i.e.} on the occupations. 
We note that our approach differs from Ref.~\cite{tu22_3243}. 
A more serious problem is the presence of eigenvalues with negative real parts (occupations), which is possible with complex-symmetric densities. 
The problem of the positivity of the reduced density matrix is a known~\cite{cuevas13_3237,kshetrimayum17_3211,weimer21_3210} and difficult problem~\cite{kliesch14_3236,werner16_3235}. 
This point is discussed in Sec.~\ref{sec_id}.

\section{Theoretical and computational developments}
\label{sec_developments}

In this section, we present theoretical and computational developments for the description of broad many-body resonances with the G-DMRG method. 
We first discuss in Sec.~\ref{sec_reference_state} a new truncation that can be applied during the construction of the reference space, and the importance of the reference state for the convergence. 
We then propose in Sec.~\ref{sec_ordering} an empirical solution to the problem of the optimization of the energy and width with respect to the orbital ordering, and justify it using quantum information arguments. 
In Sec.~\ref{sec_id}, we treat the problem of the identification of physical states in the G-DMRG method. We propose a new stabilization technique in Sec.~\ref{ssec_kappa} to reduce the entanglement due to excessive continuum couplings.
Finally, in Sec.~\ref{sec_nat}, we discuss the use of natural orbitals in the G-DMRG method and how they affect calculations of broad many-body resonances. 

\subsection{Benchmarks}
\label{sec_benchmark}

The G-DMRG code used in this work is a major upgrade to the version used in previously published studies. In addition to the newly developed methods discussed in the following sections, we made changes to the data structures, parallelization, and many of the core routines. 
For that reason, before discussing the extension of G-DMRG in the broad resonance regime, we first benchmark the G-DMRG approach against the No-core GSM (NCGSM), a full configuration interaction (FCI) method that has been used in numerous publications and thoroughly tested, on the well-bound isotope \isotope[3]{He}.

While we do not attempt to compare our results with experimental data, for completeness, we remind the reader that the $J^\pi = {1/2}^+$ ground states of \isotope[3]{H} and \isotope[3]{He} have experimental energies of $-8.482 \, \text{MeV}$ and $-7.718 \, \text{MeV}$, respectively, and the ground state of \isotope[4]{He} has an energy of $-28.30 \, \text{MeV}$.

We first considered a model space ``HO'' built upon the spherical harmonic oscillator basis. 
As a reminder, in the HO basis, orbitals are labelled by their radial and orbital quantum numbers $n$ and $l$, respectively. 
The total number of quanta associated with a given orbital $i$ defined by $(n_i,l_i)$ is $N_i = 2n_i+l_i$. 
In many-body calculations, it is customary to consider model spaces limited by the so-called $N_\text{max}$ truncation~\cite{barrett13_688}. 
This truncation consists in keeping only Slater determinants that satisfy: 

\begin{equation}
	\sum_{i=1}^A N_i \leq N_\text{min}(A) + N_\text{max}
	\label{eq_Nmax}
\end{equation}
where $A$ is the number of particles and $N_\text{min}(A)$ is the minimum number of quanta required to place all $A$ nucleons in the lowest possible orbitals. 
This truncation preserves translational invariance by separating the center-of-mass motion and allows for exact factorization of spurious center-of-mass states.

In the following, we considered a maximum number of quanta $N_\text{max} = 10$ and a maximum angular orbital momentum $l_\text{max} = 2$. 
By definition, results in the HO basis are variational and provide an optimal representation for well-bound states with spherical symmetry. 
The HO length was fixed to $b_\text{HO} = 1.6 \, \text{fm}$ to minimize the energy of \isotope[3]{H} and \isotope[3]{He}.

To demonstrate that using the Berggren basis does not cause any issue, we also considered a second model space ``BB(real)'' in which the $l=2$ partial waves were kept in the HO basis, but the $l \leq 1$ partial waves were replaced by the Berggren basis (see Sec.~\ref{ssec_berggren} for details). 
Each $l \leq 1$ partial wave included one bound orbital and 30 real-energy scattering states with momenta between 0.0 and $4.0 \, \text{fm}^{-1}$. 
The proton and neutron bound orbitals $0s_{1/2}$, $0p_{3/2}$, and $0p_{1/2}$ have energies of about $-20.0 \, \text{MeV}$, $-7.0 \, \text{MeV}$, and $-1.0 \, \text{MeV}$.

The third model space ``BB(cx)'' was identical to the second one, except that the $l=1$ real-energy scattering states were replaced by complex-energy scattering states taken along a contour in the complex-momentum plane, starting at $k=0.0$ and connecting the points $k=0.25-i0.10$, $k=1.0$, and $4.0$ (all in $\text{fm}^{-1}$).
Results for this model space demonstrate that using complex-energy scattering states does not cause issues.

Finally, for the sake of completeness, we considered a fourth model space ``BB(cx-pole)'' in which all partial waves were in the HO basis, except for the proton and neutron $p_{3/2}$ partial waves, where we have one resonant orbital (decaying Gamow state) and the same set of complex-energy scattering orbitals as in the ``BB(cx)'' model space. 
This model space shows that having complex-energy discrete states does not cause issues either.

\begin{figure}[htb]
  \centering
	\includegraphics[width=.9\linewidth]{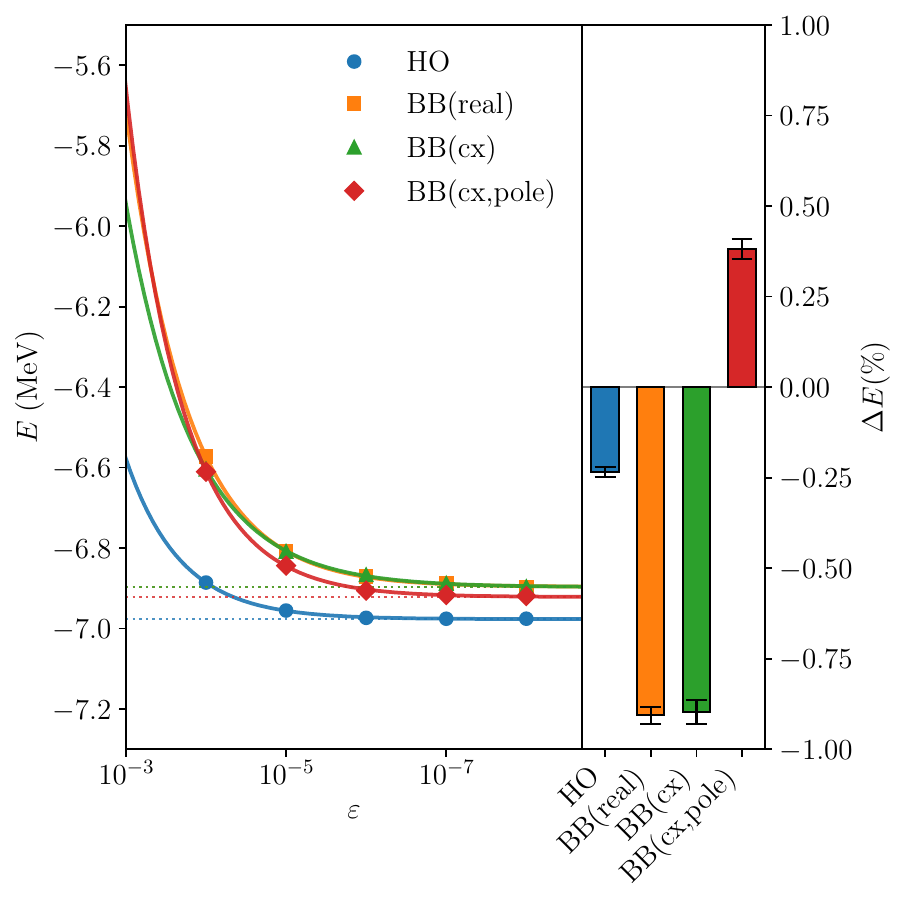}
	\caption{Ground-state energy of \isotope[3]{He} as a function of the DMRG truncation $\epsilon$. The horizontal lines denote the extrapolated G-DMRG energies. The relative energy difference with the exact NCGSM result (in percent) is shown on the right panel.}
	\label{fig_bench_3He}
\end{figure}

In the G-DMRG calculations, the reference space included the proton and neutron $0s_{1/2}$, $0p_{3/2}$, $0p_{1/2}$, $0d_{5/2}$, and $0d_{3/2}$ orbitals. 
As in previous G-DMRG studies, the remaining orbitals were essentially ordered by their energy (See $E$-ordering in Sec.~\ref{sec_ordering}).
Then, for a given system, the ground-state energy was calculated for different values of the DMRG truncation threshold $\epsilon$, and extrapolated in the limit $\epsilon \to 0$ using $E(\epsilon) = E + \alpha e^\beta$, as prescribed in Ref.~\cite{papadimitriou13_441}. 
For illustration, the convergence of the ground-state energy of \isotope[3]{He} with $\epsilon$ is shown in Fig.~\ref{fig_bench_3He} (left panel) in all four considered model spaces.

The relative error $\Delta E$ between the NCGSM and G-DMRG results, shown in the right panel, is calculated as
\begin{equation}
	\Delta E = \frac{E^\text{NCGSM} - E^\text{G-DMRG}}{|E^\text{NCGSM}|}\,,
\end{equation}
and it remains below 1\% in magnitude for all model spaces considered. 
Note that $\Delta E < 0$ indicates that the G-DMRG energy is above the NCGSM one. 
In addition, differences between model spaces are all less than 0.1 MeV in magnitude. 
We also checked that, in the HO basis, the G-DMRG method continued to pass the benchmark when applying different particle-hole (p-h) truncations in ground-state calculations of \isotope[4]{He}.

We did not have sufficient computational resources to obtain exact NCGSM results in resonant systems such as \isotope[4]{H} or \isotope[5]{He}, but we were able to obtain results with a 2p2h truncation. 
These are provided in Sec.~\ref{sec_results} alongside our G-DMRG results, and we found excellent agreement.

\subsection{Construction of the reference state}
\label{sec_reference_state}

In the Wilsonian formulation of the DMRG method used in this work, the starting point is the reference state, obtained by diagonalizing the Hamiltonian in the reference space $\mathcal{A}$. 
A large reference space provides a better approximation of the targeted state, resulting in a reduced number of iterations to converge, but it also comes at the cost of computationally more expensive iterations. 
As discussed in Sec.~\ref{sec_id}, the identification of physical states in G-DMRG calculations makes the choice of the reference space even more critical. 

The size of the reference space is largely controlled by the number of particles that must be accommodated, but also by the nature of the targeted state. 
Indeed, while low-lying states can be approximated well with the lowest-energy orbitals, excited states require the inclusion of higher orbitals, hence increasing the number of possible particle-hole excitations.

Usually, when dealing with many-body resonances, the dimension of the \emph{many-body} reference space does not need to be increased compared to bound-state calculations with the same number of particles. 
However, the \emph{single-particle} basis is usually significantly larger because of the discretization of the continuum of scattering states entering the Berggren basis. 
Broad resonances, which have a rapidly changing outgoing many-body asymptotic, are even more demanding in that regard.
This results in a significant increase in the number of G-DRMG iterations. 
It is therefore imperative to keep the computational cost of each iteration as low as possible.

One way to achieve this is, of course, to improve the single-particle basis so that a minimal reference space can provide an excellent approximation of the targeted state. 
Another is to optimize the order in which orbitals are added from the environment. 
These options are discussed in Secs.~\ref{sec_nat} and \ref{sec_ordering}, respectively.

Here, we are concerned with the convergence of calculations with respect to the size of the reference space, and how to apply truncations in that space to include additional orbitals while keeping the computational cost moderate. 
In past applications, the full reference space was built without any truncation, as in the FCI method, and then a chosen p-h truncation was applied. 
Since the dimension of that space scales exponentially with the number of orbitals and particles, building the initial space in full quickly becomes unfeasible. 
To circumvent this issue, we apply truncations already during the construction of the reference space. 
Obviously, this truncation at construction has the drawback that the results depend on the order in which the orbitals are considered when building the reference space.

Below, we first describe how the reference space is built in practice, and then, using the well-bound $J^\pi = 0^+$ ground state of \isotope[4]{He} as an example, we discuss the convergence of a G-DMRG calculation with the size of the reference space, and how the new truncation scheme impacts results. 
Finally, to illustrate how the situation changes in OQSs, we consider the case of the broad $J^\pi = 2^-$ resonant ground state of \isotope[4]{H} and study the impact of the truncation scheme on its energy and width. 
In both cases, we do not attempt to reproduce experimental data.

\subsubsection{Construction of the reference space}

To build the FCI reference space, we first consider an orbital, denoted $s_0$, and generate all the possible SDs with 1 up to $A$ particles, denoted $\text{SD}(s_0)$, and calculate all the possible Hamiltonian matrix elements. 
At this point, we have $n$ SDs and a block of $n \times n$ matrix elements.

Then, we consider a second orbital $s_1$, generate all the SDs that can be built with this orbital ($\text{SD}(s_1)$), and then build all the SDs that can be built by coupling the new SDs with the previous ones ($\text{SD}(s_0,s_1)$). 
These two sets give us $k$ new SDs, and we update $n$ to be the total number of SDs. 
We calculate the Hamiltonian matrix elements between the $k$ new SDs ($\text{SD}(s_1)$ and $\text{SD}(s_0,s_1)$), with indices going from $n-k$ to $n-1$, and all the SDs with indices going from $0$ to $n-1$. 
This gives us a block of $k \times n$ matrix elements.

Finally, we calculate the matrix elements between the $(n-k)$ previous SDs ($\text{SD}(s_0)$) with indices between $0$ and $n-k$, and the $k$ new matrix elements ($\text{SD}(s_1)$ and $\text{SD}(s_0,s_1)$) with indices from $(n-k)$ and $n-1$. 
These constitute a block of size $(n-k) \times k$. 
Because the Hamiltonian matrix is complex symmetric, the elements in this block are identical to some of the elements calculated in the previous block.
The process followed for $s_1$ can be repeated for every following orbital until the reference space is fully built.

Once the FCI space is obtained, a particle-hole truncation can be applied by setting to zero all the matrix elements that involve an SD that does not satisfy it.

\subsubsection{Convergence with the size of the referene space}

In this section, we study the impact of the reference space $\mathcal{A}$ on the convergence of G-DMRG calculations. 
Specifically, we first look at the effect of the number of orbitals used to build $\mathcal{A}$ on the final energy. 
Then, we introduce a new truncation scheme applied during the construction of the reference space, and show that it is always more efficient to consider the largest space possible and apply the new truncation, than to reduce the number of orbitals entering $\mathcal{A}$.

The purpose of the reference space is to provide an approximation of the state of interest. 
For that reason, it must include at least all the orbitals expected to contribute in a mean field picture. 
Naively, in an ideal case such as the ground state of \isotope[4]{He}, where the variational principle applies and orbitals missing from the reference space are eventually included in the renormalized space, we expect that adding orbitals beyond the mean-field picture should have a minimal impact on the final result. 
However, using a minimal reference space is usually insufficient.

\begin{figure}[h!]
    \centering
    \includegraphics[width=\linewidth]{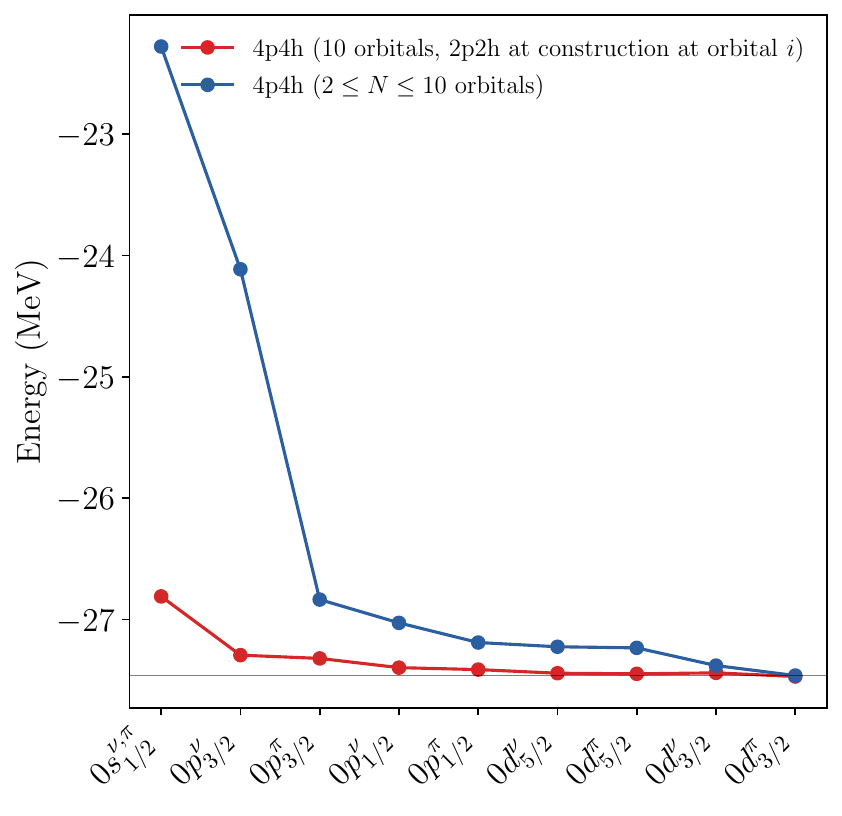}
    \caption{Blue: Binding energy of \isotope[4]{He} in an increasingly larger reference space ($2 \leq N \leq 10$ orbitals) and without any p-h truncation (4p4h). Red: Binding energy of \isotope[4]{He} in a reference space built upon $N=10$ orbitals, with a 2p2h truncation at construction applied at orbital $i$, and none otherwise. The exact result for $N=10$ and no p-h truncation is shown as a horizontal line.}
    \label{fig_4He_NH}
\end{figure}

To illustrate this point, the binding energy of the $J^\pi = 0^+$ state of \isotope[4]{He} is shown in Fig.~\ref{fig_4He_NH} (blue line) as a function of the orbitals included in the reference space. 
The results were obtained in the HO basis with $N_\text{max} = 12$, $l_\text{max} = 4$, $b_\text{HO} = 1.5 \, \text{fm}$, and with a DMRG truncation of $\varepsilon = 10^{-4}$. 
Orbitals in the environment were ordered using the $E$-ordering scheme presented in Sec.~\ref{sec_ordering}. 
The first value on the left side of the plot corresponds to the energy obtained starting from a minimal reference space composed of only the proton and neutron $0s_{1/2}$ orbitals. 
The subsequent values correspond to the energies obtained after adding the orbital indicated by the label to the previous reference space, \textit{i.e.} in a larger and larger reference space. 
The energy converges rapidly with respect to the size of the reference space and reaches a minimum when the $0d_{5/2}$ orbitals are included.

It is remarkable that, despite using the variational principle, the inclusion in the reference space of orbitals well above those expected to be occupied in a naive mean-field picture leads to a noticeable energy gain. 
This is due to the non-perturbative nature of the nuclear \textit{ab initio} problem. 
Even though n-\isotope[4]{He} and p-\isotope[4]{He} scattering phase shifts reveal $p_{3/2}$ resonances with widths of about 0.648 MeV and 1.23 MeV, respectively, forming the basis for the unbound systems \isotope[5]{He} and \isotope[5]{Li}, the fact remains that the wave function of \isotope[4]{He} itself has small but non-negligible occupations in the $p_{3/2}$ partial waves. 
When starting from the minimal reference space ($0s_{1/2}^{\nu}$, $0s_{1/2}^{\pi}$), during the first few iterations, the DMRG truncation removes small but important contributions from $p_{3/2}$ orbitals, which are then never recovered.

\subsubsection{Particle-hole truncation at construction}

Going back to the new truncation scheme, we note that a key point in the construction of the reference space is that most matrix elements in a given block are built using those in previously calculated blocks. 
For example, matrix elements in the block $\text{SD}(s_0,s_1) \otimes \text{SD}(s_0)$ rely on those in the block $\text{SD}(s_0) \otimes \text{SD}(s_0)$. 
Consequently, if some of the matrix elements in a given block are set to zero, most matrix elements in the subsequent blocks will be zero ``in cascade.'' 
Intuitively, if certain p-h excitations involving lower orbitals are suppressed, then the p-h excitations in higher orbitals, based on the suppressed p-h excitations in lower orbitals, will be suppressed too. 
The p-h truncation at construction leverages this cascade effect. 
Consequently, it must be used only in the highest orbitals considered to avoid truncating the reference space in an uncontrolled manner.

\begin{figure}[h!]
    \centering
    \includegraphics[width=\linewidth]{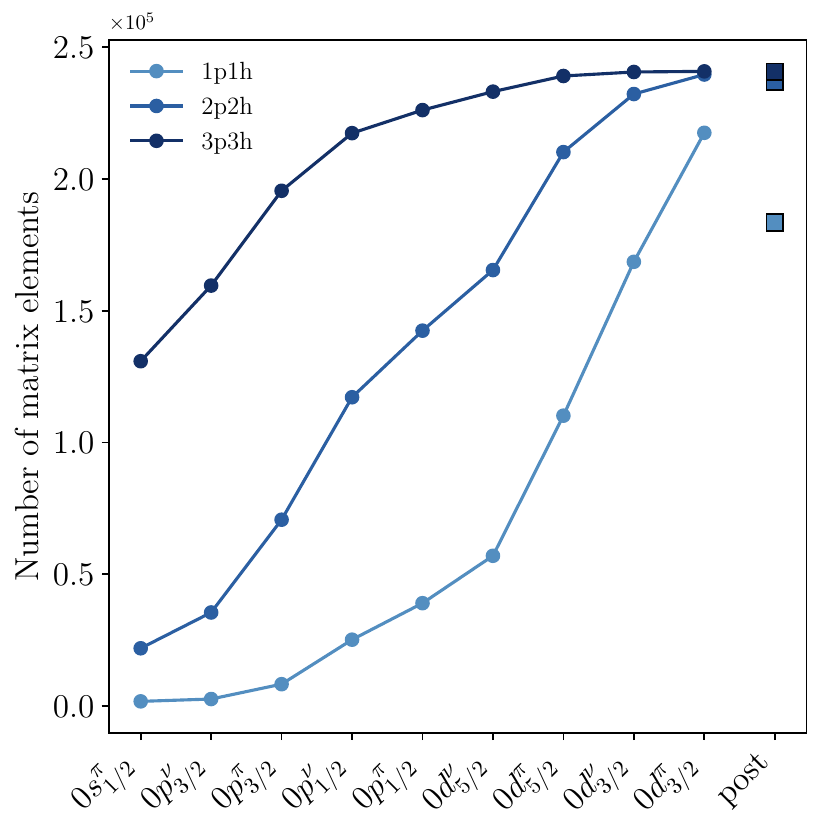}
    \caption{Number of retained Hamiltonian matrix elements in the reference space after applying the p-h truncation, as a function of the orbital at which the truncation is applied during the construction (left to right). The last column (squares) shows the number of matrix elements kept when the p-h truncation is applied after full construction.}
    \label{fig_me2b_H}
\end{figure}

To illustrate this cascading effect on the construction of the reference space for the isotope \isotope[4]{He}, in Fig.~\ref{fig_me2b_H}, we show the number of Hamiltonian matrix elements in the reference space after applying a given p-h truncation at construction, as a function of the orbital at which the truncation was applied. 
For reference, in the column denoted ``post,'' we also show the number of retained matrix elements when the p-h truncation is applied after the FCI reference space was built (no truncation at construction). 

As expected, we observed that in almost all cases the p-h truncation at construction reduces the number of matrix elements more than the p-h truncation applied post FCI construction. 
Given that the 1p1h truncation is always found to be excessively harsh in practice, we focus on the impact of the 2p2h truncation at construction on the energy.

In principle, the truncation applied in the full space should always be consistent with the one applied in the reference space. 
For example, if a 2p2h truncation at construction is applied starting at some orbital $i$ in the reference space, then the same p-h truncation should be applied post construction on the whole reference space (regular p-h truncation), and the same p-h truncation should also be applied in the rest of the space. 
However, for the sake of demonstration, in Fig.~\ref{fig_4He_NH} (red line), we show the binding energy of \isotope[4]{He} as a function of the orbital in the reference space at which a 2p2h truncation at construction was applied, but without any post-construction truncation applied in the reference space, and without any truncation applied in the rest of the space. 
By comparing this result to the exact result, shown as a horizontal line, we can quantify the loss of binding energy due to the effect of the 2p2h truncation at construction.

As the p-h truncation at construction is applied to higher and higher orbitals in the reference space, the energy converges toward the exact result. 
We also see that applying a 2p2h at construction in a reference space built upon $N$ orbitals always gives a lower energy than an exact result obtained from a reference space built upon $N' < N$ orbitals. 
This demonstrates the usefulness of the p-h truncation at construction. 
Indeed, we see that the truncation at construction has a smaller impact on the energy than a direct reduction of the dimension of the reference space, and it also provides a significant reduction in the number of non-zero matrix elements.

\subsubsection{Reference space for unbound states}

As explained in Sec.~\ref{sec_id}, in the absence of a variational principle, the method to identify the physical solution in OQSs requires a reference space that only contains orbitals associated with poles of the single-particle $S$-matrix. 
To be able to track the physical state during the renormalization, this method relies on the overlap between the solution obtained at the previous iteration and the candidate solutions. 
Consequently, if an important partial wave is omitted in the reference space, there is a risk that, at every iteration, candidate solutions including contributions from that partial wave are rejected due to their relatively low overlap with the suboptimal reference state.
In that case, the G-DMRG method can even appear to converge on a solution, like when an optimization process converges on a local minimum.
Thus, it is critical to check the convergence of G-DMRG resonance calculations with the size of the reference space whenever possible.

\begin{figure}[h!]
    \centering
    \includegraphics[width=.9\linewidth]{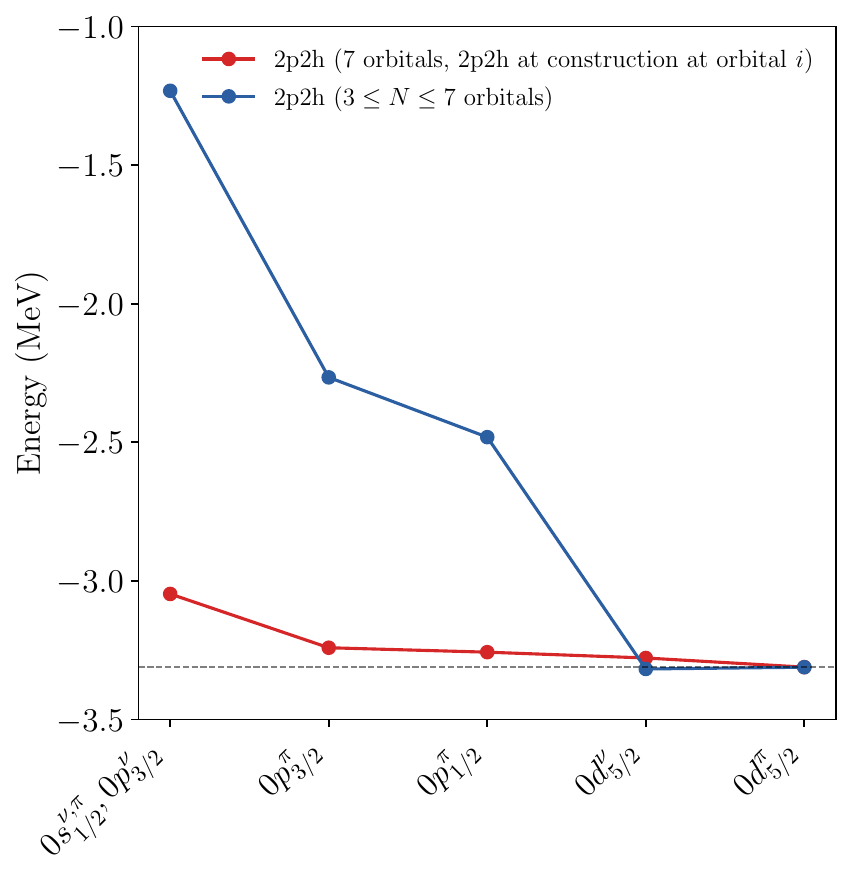}
    \caption{Binding energy of \isotope[4]{H} as a function of the orbitals included in the reference space.}
    \label{plot_4H_ref_E}
\end{figure}

\begin{figure}[h!]
    \centering
    \includegraphics[width=.9\linewidth]{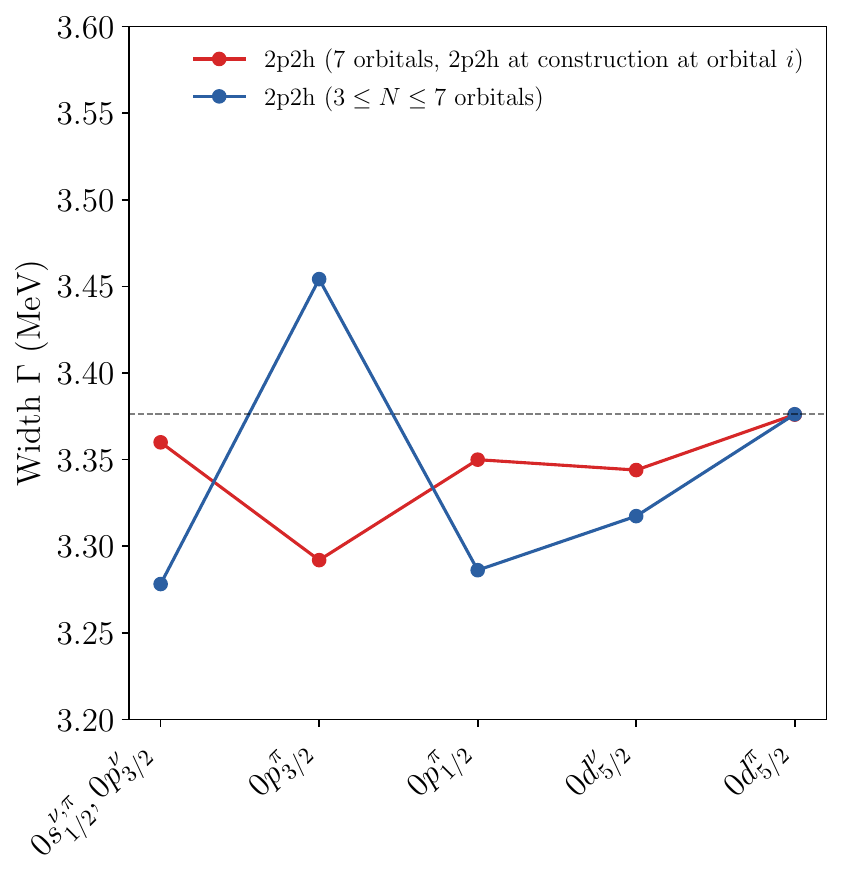}
    \caption{Width of \isotope[4]{H} as a function of the orbitals included in the reference space.}
    \label{plot_4H_ref_W}
\end{figure}

In Figs.~\ref{plot_4H_ref_E} and \ref{plot_4H_ref_W} (blue lines), we show the convergence of the energy and width, respectively, of the $J^\pi = 2^-$ state of \isotope[4]{H}, as a function of the orbitals included in the reference space. 
This state is a broad single-particle resonance. 

In both plots, the first result on the left is obtained in the minimal reference space built upon the orbitals $0s_{1/2}^{\pi}$, $0s_{1/2}^{\nu}$, and $0p_{3/2}^{\nu}$. 
Moving to the right, the subsequent results are obtained by adding the indicated orbital to the previous reference space (to the left). 
We note that there is no $0p_{1/2}^{\nu}$ discrete orbital in this case. 

Even though the energy decreases rapidly with the size of the reference space, we see that, in proportion, the effect of additional orbitals is considerably greater in \isotope[4]{H} than in the case of \isotope[4]{He}
(cf. Fig.~\ref{fig_4He_NH}).
From the $(0s_{1/2}^{\nu,\pi}, 0p_{3/2}^{\nu})$ minimal model space to the largest one considered, the binding energy of \isotope[4]{H} increases by about 160\%, compared to about 14\% in \isotope[4]{He}. 
This is somewhat expected as the inclusion of orbitals with higher $l$ helps build the \isotope[3]{H} cluster within \isotope[4]{H}, which has a spatially localized structure. 
The width, however, is not strongly affected by the size of the reference space, and oscillates around 3.3 MeV within a narrow range of only about 0.3 MeV, which is likely due to the fact that we use an optimized basis and that the width is purely governed by one-neutron decay.

Finally, we study the effect of applying a 2p2h truncation at construction in \isotope[4]{H}. 
The results for the energy are shown in Fig.~\ref{plot_4H_ref_E} (red line) and exhibit a weak dependence on the truncation, which once more demonstrates the usefulness of the truncation. 
The width, shown in Fig.~\ref{plot_4H_ref_W} (red line), is also barely affected by the truncation.

We can conclude from these results that the p-h truncation at construction can be used to approximate exact results reasonably well, provided that we consider a large enough reference space in the first place, and that it is applied in orbitals which are unlikely to be occupied in the Hartree-Fock picture. 
The main advantage of this truncation is that it allows us to add higher orbitals in the reference space, thus increasing the quality of the reference state at the price of introducing a small truncation error.

\subsection{Orbital ordering}
\label{sec_ordering}

Regardless of the reference space construction, by iteratively building the wave function one orbital (or a few) at a time, the DMRG method yields results that inherently depend on the orbital ordering. 
Even for bound states, where the variational principle applies, selecting the lowest state obtained at every DMRG iteration does not guarantee convergence to the exact solution. 
Fundamentally, the ordering reflects prior beliefs about the structure, or topology, of entanglement in the system~\cite{amico08_2705,horodecki09_3217,benatti20_2706}. 
The latter is elegantly reflected by the diagrammatic representation of tensor network states (TNS)~\cite{orus14_2015}. 
In the DMRG method, whether it is explicitly based on matrix product states (MPS) or not, orbitals are effectively mapped onto a one-dimensional chain. 
Once the orbital ordering is established, it is the DMRG truncation that imprints the order-dependence on the results. 
We note that here we should be careful to distinguish systematic uncertainties coming from the entanglement topology from the mapping of orbitals onto this topology.

In complicated cases, where local energy minima with respect to the ordering exist, one can attempt to optimize the ordering based on entanglement measures~\cite{legeza03_3199,rissler06_3191,barcza11_3247}. 
The idea is to first perform an approximate calculation to extract, for instance, the two-orbital mutual information of all the possible pairs of orbitals, reorder the orbitals according to their entanglement distance~\cite{legeza15_2055}, and then perform a second, more precise, calculation using the optimized ordering. 
This type of approach has been successfully applied in nuclear systems~\cite{legeza15_2055,tichai23_2940}. 
In quantum chemistry, entanglement measures have also been used to devise active-space selection methods to automate the DMRG ~\cite{stein16_3179,stein19_3181,khedkar19_3180}.

In that regard, quantum information concepts could also be relevant to accurately describe unbound states where the variational principle does not apply, and the number of orbitals is large. 
However, there are two related issues when describing resonances. 
The first is to converge toward the exact solution for both the energy and the decay width, and the second is to ensure the numerical stability of calculations. 
The optimization schemes mentioned above could certainly help with the former, but in this section, we will focus on the latter, which is specific to the problem of broad resonances, and also more pressing in practice.

When dealing with decaying resonances in the quasistationary formalism, we must build both the inner and outer (asymptotic) parts of the wave function, which are  mostly associated with the energy position and width, respectively. The asymptotic part of the wave function is considerably more difficult to build than the inner part, which can be understood intuitively: 
For example, for a single-particle decaying resonance with ${ \tilde{k} = \kappa - i\gamma }$, ${ \kappa, \gamma > 0 }$, the asymptotic of the wave function $e^{i\tilde{k}r = e^{i\kappa r} e^{\gamma r}}$ oscillates with a frequency $\kappa$ and grows exponentially at a rate $\gamma$. 
The localized inner part of the wave function can be efficiently represented by a few bound basis states, and the energy is expected to converge quadratically against the error on the wave function~\cite{wouters14_3248}. 
The asymptotic part, however, has an exponential character which magnifies errors and usually requires the superposition of many complex-energy scattering states that must be accurately represented over distances which are large compared to the range of the interaction. 
Due to the different sensitivities of the energy and width to details of calculations, we should always expect the orbital ordering to impact the width more strongly.

Consequently, for broad resonances, where the separation energy with respect to the relevant decay threshold and the width can be comparable in size, it is critical to optimize the ordering so that both are built as smoothly as possible during renormalization. 
We note that energy and width are correlated but, in the absence of the variational principle for complex energies, both quantities can either increase or decrease at each iteration. 
There is, however, a stationary principle~\cite{moiseyev98_92,rotureau09_140}, which can be used to show that near a solution the squared modulus of the energy is invariant under small perturbations of the wave function, namely $\delta_\Psi {|E|}^2 = 0$. 
This means that we should expect a stabilization of the complex energy when converging to a physical eigenstate.

In past G-DMRG applications, based on the Berggren basis, we associated to discrete states $(n,l,j)$ a number of quanta $N = 2n+l$ like in the HO basis, and then did the same for scattering states $(k,l,j)$ with $k > n$ along the discretized contours. 
Of course, for scattering states this number does not correspond to an actual number of quanta, but it allows us to define an approximate energy ordering for all orbitals regardless of the basis used. 
Below, we will refer to this ordering as ``energy-ordering'' or ``$E$-ordering.'' 
In this work, we found that, in practice, to be able to describe broad many-body resonances, we had to first use a different orbital ordering scheme to stabilize many-body calculations, allowing us to optimize the single-particle basis, and then return to the $E$-ordering to further stabilize and optimize our results. 
Below, we explain how this process works and how it was motivated.

The construction of the single-particle basis requires choosing a generating potential for each partial wave $(l,j)$ represented by the Berggren basis. 
In theory, it is possible to derive a Berggren basis from a Gamow Hartree-Fock calculation, but in light nuclei this is usually suboptimal. 
For that reason, in this work, we use Woods-Saxon potentials with an added spin-orbit term, and we initially adjust their parameters to reproduce basic experimental data whenever possible. 
Then, we refine the parameters until we obtain a basis that gives stable and meaningful results, namely a basis that ultimately leads to the same many-body results when we vary the generating potential within reason. 
The problem with this trial-and-error approach is that it relies on the many-body results being stable in a suboptimal basis, and this is not necessarily the case with broad resonances.

Ultimately, the stability issue depends on our ability to reliably track the physical state from one G-DMRG iteration to the next. 
In turn, this strongly depends on the balance between how much entanglement each new orbital adds, and how efficiently the renormalization reduces the entanglement (see Sec.~\ref{sec_id} for details). 
This is where orbital ordering is critical.

When trying to optimize the single-particle basis using $E$-ordering, we noticed that sometimes our calculations would suddenly lose track of the physical state, and that this almost always occurred when adding complex-energy scattering orbitals that contributed significantly to the asymptotic part of the wave function. 
Further investigation revealed that, in those instances, entanglement increased significantly, making the renormalization fail. 
As discussed in App.~\ref{app_cx_density}, continuum couplings do increase entanglement.

Following this realization, we experimented with different orbital orderings with the goal of stabilizing calculations first, and converging toward the exact result second. 
In practice, assuming that all the relevant discrete states are included in the reference space, we found that it is preferable to: 
\begin{enumerate}
  \item[i)] Group orbitals according to their partial wave $(l,j)$. 
  \item[ii)] Add all proton orbitals first in neutron-rich systems. 
  \item[iii)] Add orbitals relevant for decay last, and order them according to their position along the contour entering the Berggren basis. 
\end{enumerate}
These findings hold even if some partial waves are expressed in the Berggren basis and others in the harmonic oscillator basis, which suggests that their fundamental justification is probably based on entanglement considerations.

The first finding is not surprising given that all nuclei, even light ones, develop a self-consistent shell structure, which is at the basis of the nuclear shell model. 
By grouping orbitals by partial wave $(l,j)$, we essentially build the shell model orbitals, as was shown in Ref.~\cite{robin21_2471}. 
These orbitals can largely be identified with natural orbitals, as will be shown in Sec.~\ref{sec_nat}.

The second finding confirms the proton-neutron factorization of entanglement uncovered in Refs.~\cite{johnson23_3195,gorton24_3228} and observed in Refs.~\cite{perez23_3249,tichai24_3190}. 
In this work, where we consider neutron-rich systems, it is quite advantageous to converge the proton part of the wave function first to gain binding energy and stabilize the width. 
Similarly, adding neutron partial waves involved in the decay last further helps with stability and convergence.

The third point is critical to ensure the stability and convergence of the G-DMRG method, but somewhat at odds with results in bound systems. 
Indeed, it was shown in Ref.~\cite{legeza03_3199} that a near-optimal orbital ordering is obtained by moving the orbitals with the largest single-site entropies close to the center of the chain. 
This method was successfully used in nuclear systems~\cite{tichai24_3190}, where the most entangled orbitals lie near the Fermi surface. 
Here, these orbitals are precisely those important for decay. 
Perhaps, in broad resonances, the stabilization brought by point (ii) outweighs the benefits of placing highly entangled orbitals in the center of the chain.

Using the $(l,j)$-ordering, we were able to sufficiently stabilize our calculations to further refine the single-particle basis. 
However, we found that results obtained with this orbital ordering scheme were suboptimal when it came to generating natural orbitals (see Sec.~\ref{sec_nat}). 
These natural orbitals provided further convergence for the energy, but not for the width, suggesting that the $(l,j)$-ordering is suboptimal for the width as suspected.
Fortunately, the refinement of the single-particle basis made possible by the $(l,j)$-ordering allowed us to revert back to the $E$-ordering, which in turn gave optimal natural orbitals.

\begin{figure}[h!]
  \centering
  \includegraphics[width=\linewidth]{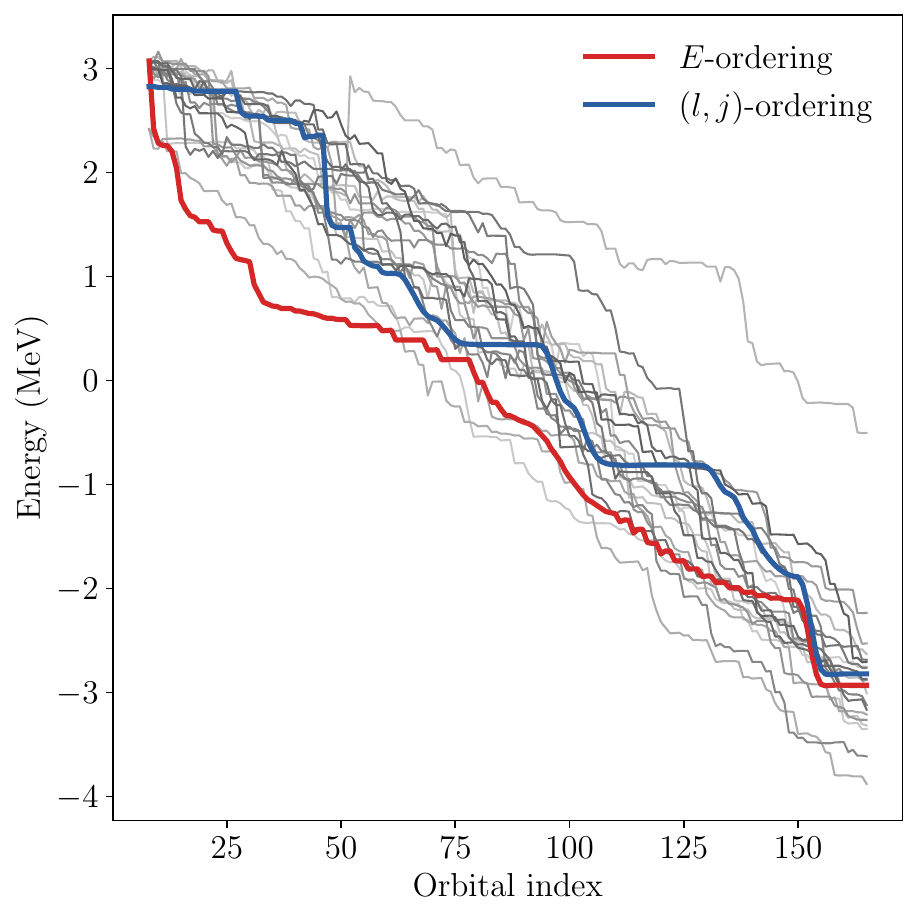}
  \caption{Energy of the ${J^\pi = 2^-}$ state of \isotope[4]{H} as a function of the index of the orbital included in the G-DMRG with no p-h truncation and $\varepsilon=10^{-4}$. Results for 20 random orderings of the orbitals are shown in shades of grey.}
  \label{fig_4H_ordering_E}
\end{figure}

\begin{figure}[h!]
  \centering
  \includegraphics[width=\linewidth]{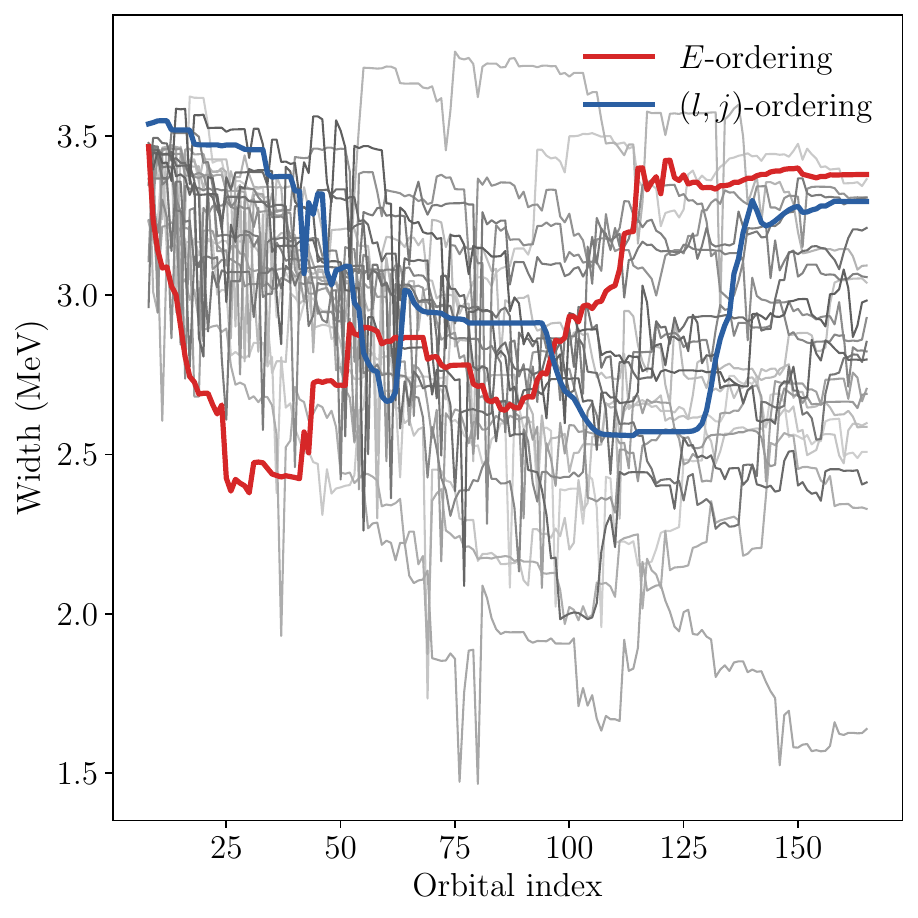}
  \caption{Same as in Fig.~\ref{fig_4H_ordering_E} for the width.}
  \label{fig_4H_ordering_W}
\end{figure}

To demonstrate that the $(l,j)$-ordering and post-optimization $E$-ordering are indeed good choices for G-DMRG calculations of broad resonances,  Figs.~\ref{fig_4H_ordering_E} and ~\ref{fig_4H_ordering_W} show the energy and width, respectively, of the ${J^\pi = 2^-}$ state of \isotope[4]{H} as a function of the index of the added orbital, for both ordering schemes as well as 20 random orderings. 

The model space is identical to the one used for the benchmark calculations in Sec.~\ref{sec_benchmark}.
The reference space included the proton $0s_{1/2}^{\pi}$, $0p_{3/2}^{\pi}$, $0p_{1/2}^{\pi}$, and $0d_{5/2}^{\pi}$ orbitals, and the neutron $0s_{1/2}^{\nu}$, $0p_{3/2}^{\nu}$, and $0d_{5/2}^{\nu}$ orbitals. 
For the $(l,j)$-ordering, all proton orbitals ($l=0$-4) were added in the usual order, followed by the $l=3,4$ neutron orbitals and the $0s_{1/2}$, $0p_{1/2}$ and $0p_{3/2}$ neutron partial waves. 

In Fig.~\ref{fig_4H_ordering_E}, we see that almost all energy trajectories steadily decrease with each added orbital while approximately remaining within a $2.0 \, \text{MeV}$ band. 
As expected, the energy obtained with $E$-ordering drops rapidly at first but, remarkably, ends in the middle of the band and very close to the energy obtained with the $(l,j)$-ordering.

The width trajectories are more interesting because they illustrate the problem of simultaneously optimizing the energy and the width. 
As mentioned earlier, building the asymptotic of the wave function, which gives the width, is considerably more difficult than building the inner part, which gives the energy. 
It follows that, as a heuristic, quasi-stationary calculations that minimize the energy and maximize the widths are usually the most converged. 
In Fig.~\ref{fig_4H_ordering_W}, we can see that the width trajectories for both ordering schemes considered in this work do maximize the width when compared against random orderings. 
Furthermore, while almost all random orderings show sharp variations with each orbital added, the $E$- and $(l,j)$-orderings remain relatively smooth in comparison.

\begin{figure}[h!]
  \centering
  \includegraphics[width=\linewidth]{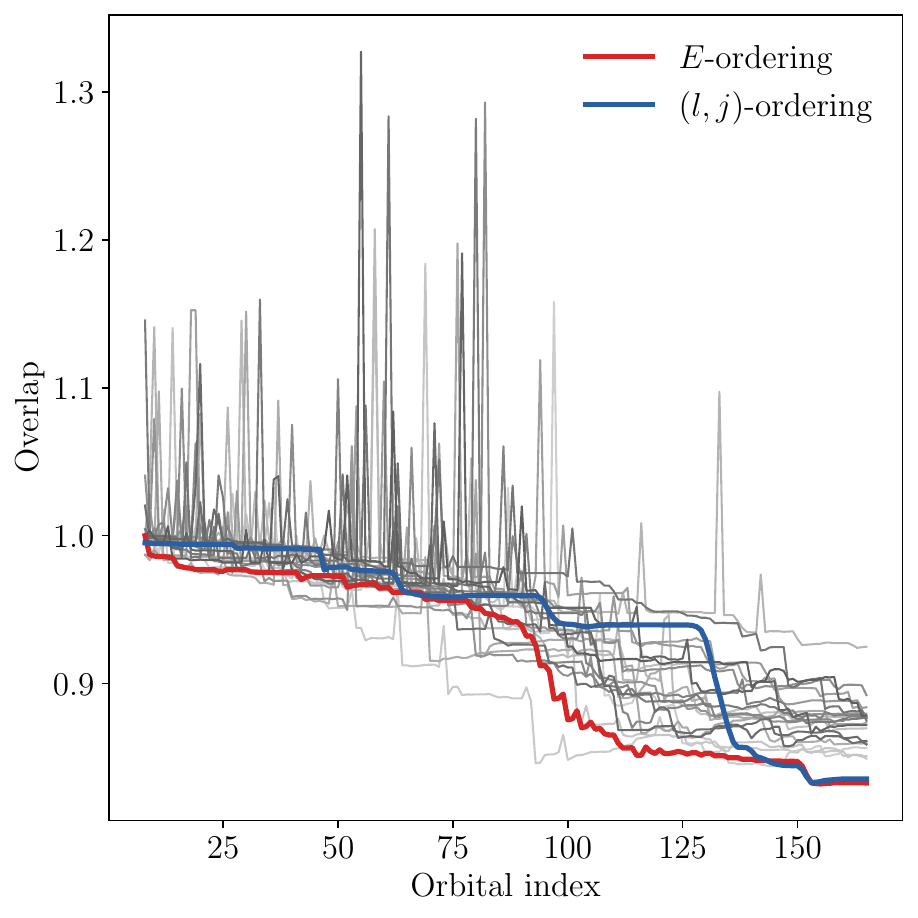}
  \caption{Same as in Fig.~\ref{fig_4H_ordering_E} for the overlap.}
  \label{fig_4H_ordering_O}
\end{figure}

Finally, in Fig.~\ref{fig_4H_ordering_O}, we show the absolute value of the overlap between the eigenstate $i$, selected after the addition of a specific orbital, and the eigenstate obtained in the reference space. 
We note that for Gamow states this value can be greater than 1.0. 
As will be discussed in the next section, while it is important for the final overlap to remain large, a value below 1.0 indicates that correlations between the reference space and the environment have been captured by the renormalization. 
It follows that, everything else considered, orderings that minimize the final overlap are the ones that maximize the amount of many-body correlations, and hence that approach the exact solution.
It is once again remarkable that both ordering schemes considered converge to the same value, and this value is lower than all the final overlaps obtained with random orderings.

\subsection{Identification of physical states}
\label{sec_id}

In quasistationary approaches, resonances are described by complex-energy Gamow states, which are discrete many-body eigenvectors of a non-Hermitian Hamiltonian matrix, as well as poles of the many-body $S$-matrix. 
The non-Hermitian nature of the matrix derives from expressing the Hamiltonian operator, which is Hermitian in the space of bound states, in a many-body basis that includes configurations built upon complex-energy orbitals with possibly incoming and outgoing asymptotics. 
Usually, a small number of these orbitals correspond to discrete states, but sometimes none of them do. 
Thus, most, if not all, configurations have one or more particles occupying complex-energy scattering orbitals. 
A critical point to understand is that, in the Berggren basis (or the uniform complex scaling method), single-particle continua are deformed (or rotated) in the complex momentum plane, and no unitary transformation can relate two different deformations (or rotations). 
This means that, contrary to the Hermitian case, eigenvectors and their eigenvalues are not necessarily invariant under basis changes.

A second key point is that, by definition, for an eigenstate to be a physical (discrete) Gamow state, its wave function must have a prominent localized inner part. 
In other words, the system must have a structure that lasts for a time before it decays. 
Therefore, we can expect the structure to be dominated by configurations built upon the discrete states included in the single-particle basis, if any. 
It follows that, in practice, properties of Gamow states should be far less sensitive to basis changes compared to states dominated by scattering orbitals, notably changes that involve deforming scattering continua.
This means that when representing a Gamow state in two complete but different Berggren bases, even though we obtain eigenvectors that are not related by a unitary transformation, we should still obtain identical energies and widths. 
Formally, the stationary principle~\cite{moiseyev98_92,rotureau09_140} mentioned in Sec.~\ref{sec_ordering} states that around an eigenstate of a non-Hermitian Hamiltonian, the complex energy must be stationary under small variations of the wave function.

The identification problem, common to all quasistationary approaches, thus consists in telling apart a small number of discrete physical solutions, with eigenvalues that are invariant under basis changes, from a large basis-dependent background of complex-energy eigenvectors that are dominated by scattering orbitals. 
At this point, the solution to the identification problem seems simple. 
We could solve the problem for two different but equivalent deformations of the continuum, and find those eigenvectors whose eigenvalues remain invariant. 
However, for large-scale problems, it is not feasible to fully diagonalize dense complex-symmetric matrices repeatedly.

In the context of the GSM, an elegant solution to the identification problem leverages resonant states in the single-particle basis~\cite{michel02_8,michel03_10}. 
This solution, known as the \emph{overlap method,} consists in building a truncated many-body basis using only resonant single-particle states, \textit{i.e.} states associated with poles of the single-particle $S$-matrix, and diagonalizing the Hamiltonian in this basis. 
This initial space is usually called the \emph{pole space}.
By construction, all the obtained eigenvectors are approximations of eigenstates associated with the poles of the many-body $S$-matrix. 
Then, the approximation of the eigenstate of interest $\ket{\Psi_0}$ obtained in the pole space is selected and used as pivot (starting vector) for diagonalizing the Hamiltonian in the full space, using, for example, the Lanczos or Davidson method so that only one extremal eigenvector can be extracted. 
The overlap method thus ensures that the eigenstate $\ket{\Psi}$ obtained is the eigenvector that has the largest overlap $\braket{\Psi | \Psi_0}$ with the pole space solution.

The overlap method was adapted to the G-DMRG by identifying the reference space $\mathcal{A}$ with the pole space, and by updating in each G-DMRG iteration the pivot used to diagonalize the renormalized Hamiltonian with the solution retained at the previous iteration. 
The update of the pivot is not only essential to reduce the risk of selecting an unphysical state in any given iteration, but also to accelerate the convergence of each diagonalization and hence the entire calculation.

The overlap method works well for bound states and low-lying narrow resonances, where the wave function is largely dominated by its inner part. 
In contrast, the description of broad resonances requires a large continuum space as the outer and inner parts of their wave functions can be equally important. 
Consequently, the pole space approximation can be rather poor, especially when the single-particle basis is not optimized. 
In the G-DMRG method, if the overlap between the pole space solution and the targeted state decreases to 50\% or below at any given iteration, the selection of the physical state may fail. 
Moreover, as explained below, when the single-particle basis itself includes broad resonances, calculations can become numerically unstable.

To overcome the limitations of the overlap method and stabilize G-DMRG calculations, in Sec.~\ref{ssec_kappa} we introduce a simple truncation scheme based on the eigenvalues of the reduced density matrix, which is motivated by entanglement considerations, and helps us to stabilize and accelerate calculations. 
Before introducing these developments, we first show how the identification method operates in relation to the eigenvalues of the reduced density matrix in the Hermitian case, using the well-bound ground state of $\isotope[4]{He}$ as an example. 
These results will serve as a reference when it comes to understanding how continuum couplings affect entanglement in the non-Hermitian many-body problem in practice, and the resulting impact on the identification problem.

\begin{figure}[h!]
  \centering
  \includegraphics[width=\linewidth]{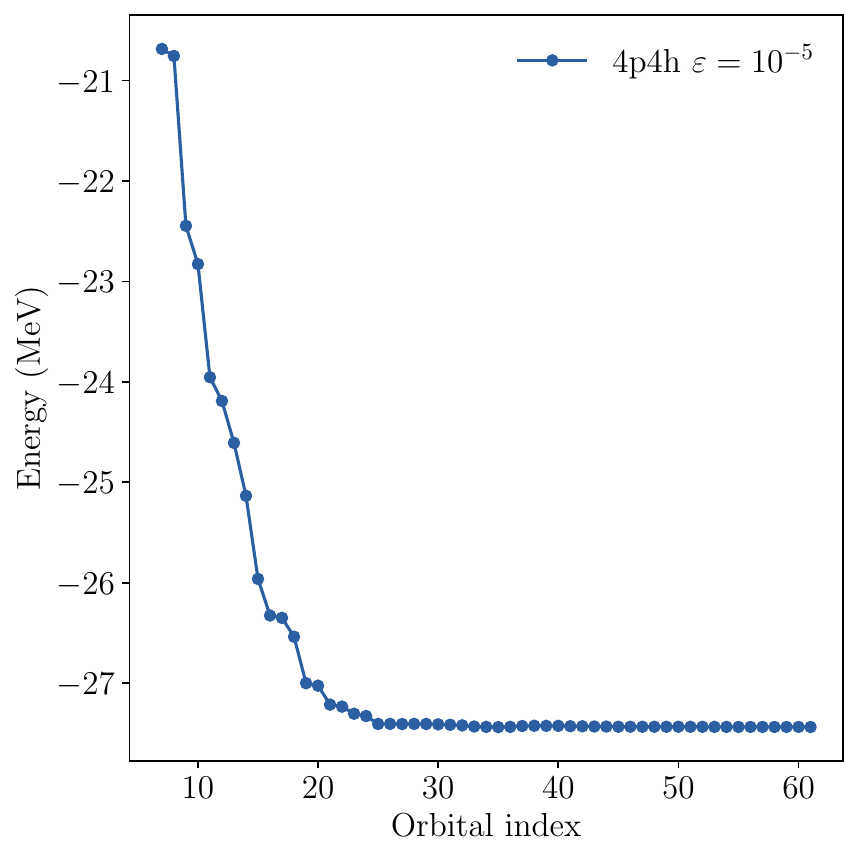}
  \caption{Ground-state energy of $\isotope[4]{He}$ as a function of the orbitals included in the G-DMRG calculation.}
  \label{fig_4He_E}
\end{figure}

In the bound-state case, the identification method is unnecessary since the variational principle is available, but we can still use it for the sake of demonstration. 
In Fig.~\ref{fig_4He_E}, we show the ground-state energy of $\isotope[4]{He}$ as a function of the added orbitals in the G-DMRG calculation. 
The results were obtained in the HO basis with $N_\text{max} = 12$, $l_\text{max} = 4$, and $b_\text{HO} = 1.5 \, \text{fm}$. 
We used the $E$-ordering and a G-DMRG truncation of $\varepsilon = 10^{-4}$. 
As expected, the energy decreases rapidly with each additional orbital and eventually stabilizes.

\begin{figure}[h!]
  \centering
  \includegraphics[width=\linewidth]{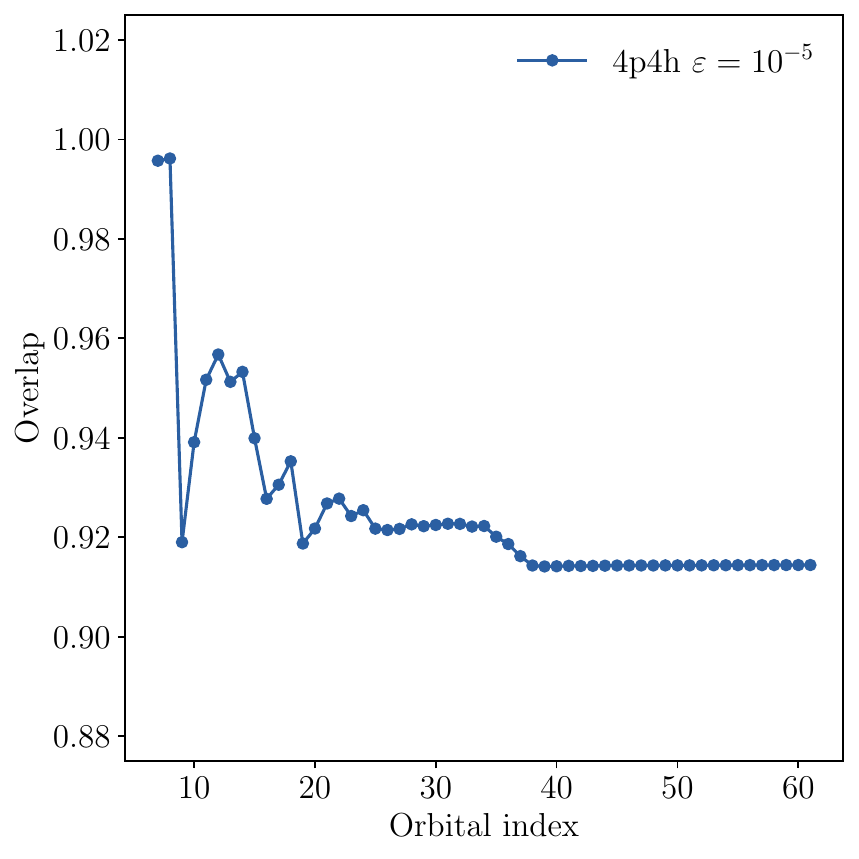}
  \caption{Overlap between the identified ground state of $\isotope[4]{He}$ and the reference state as a function of the orbitals included in the G-DMRG calculation.}
  \label{fig_4He_O}
\end{figure}

In this ideal case, we can also expect the overlap between the state identified at each iteration and the reference state to remain high. 
This is indeed what we observe, as shown in Fig.~\ref{fig_4He_O}. 
Interestingly, the overlap varies up and down before converging slightly below 0.92.

\begin{figure}[h!]
  \centering
  \includegraphics[width=\linewidth]{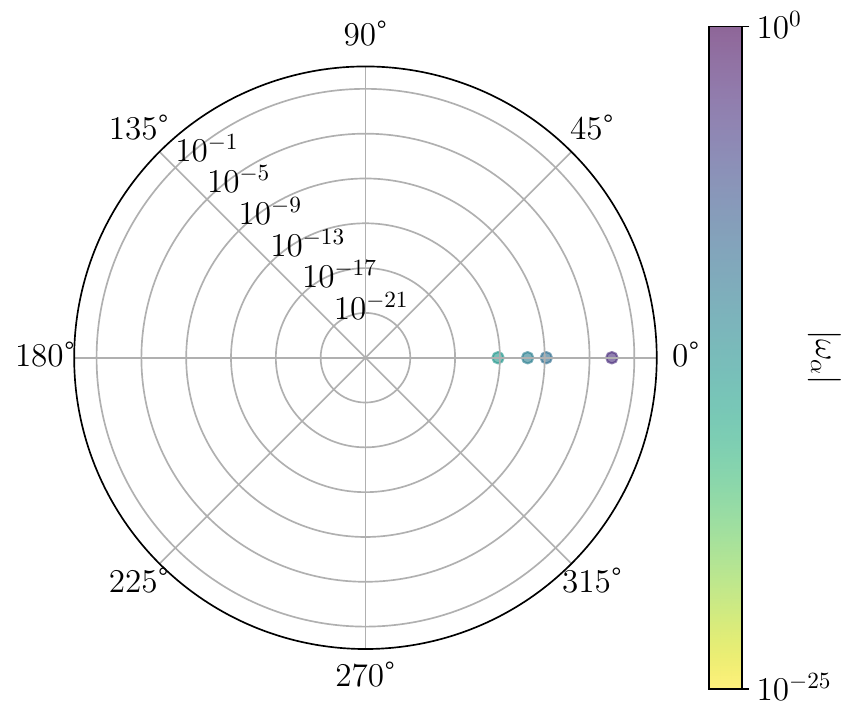}
  \caption{Eigenvalues of the reduced density matrix for the ground state of $\isotope[4]{He}$, plotted in the complex plane in polar representation and with the radial component in log-scale.}
  \label{fig_4He_rho_eigvals}
\end{figure}

In anticipation of the next case, in Fig.~\ref{fig_4He_rho_eigvals}, we show the eigenvalues $\omega_\alpha = |\omega_\alpha| e^{i\theta}$ of the reduced density matrix obtained at the end of the G-DMRG calculation, \textit{i.e.} once the calculation is converged and all orbitals have been taken into account. 
These eigenvalues correspond to the weights of configurations in the wave function. 
The eigenvalues are shown in the complex plane using the polar representation with the radial component $|\omega_\alpha|$ in log-scale. 
We can see that all eigenvalues are real ($\theta = 0 \degree$), and that the wave function is dominated by one optimized SD state with eigenvalue $|\omega_\alpha| \approx 0.8$.

The next case we consider is the unbound ${J^\pi = {3/2}^-}$ ground state of \isotope[5]{He}, which has a neutron single-particle resonance with a width of about $\Gamma = 0.648 \, \text{MeV}$, and a binding energy of about $E = -27.56 \, \text{MeV}$. 
Its one-neutron separation energy is $S_{n} = -0.735 \, \text{MeV}$~\cite{papadimitriou13_441} and thus it has a ratio $|\Gamma/(2 S_{n})| \approx 0.44$. 
Of course, here the goal is not to reproduce experiment but solely to illustrate how the identification of the physical state works with a typical decaying resonance.

\begin{figure}[h!]
  \centering
  \includegraphics[width=\linewidth]{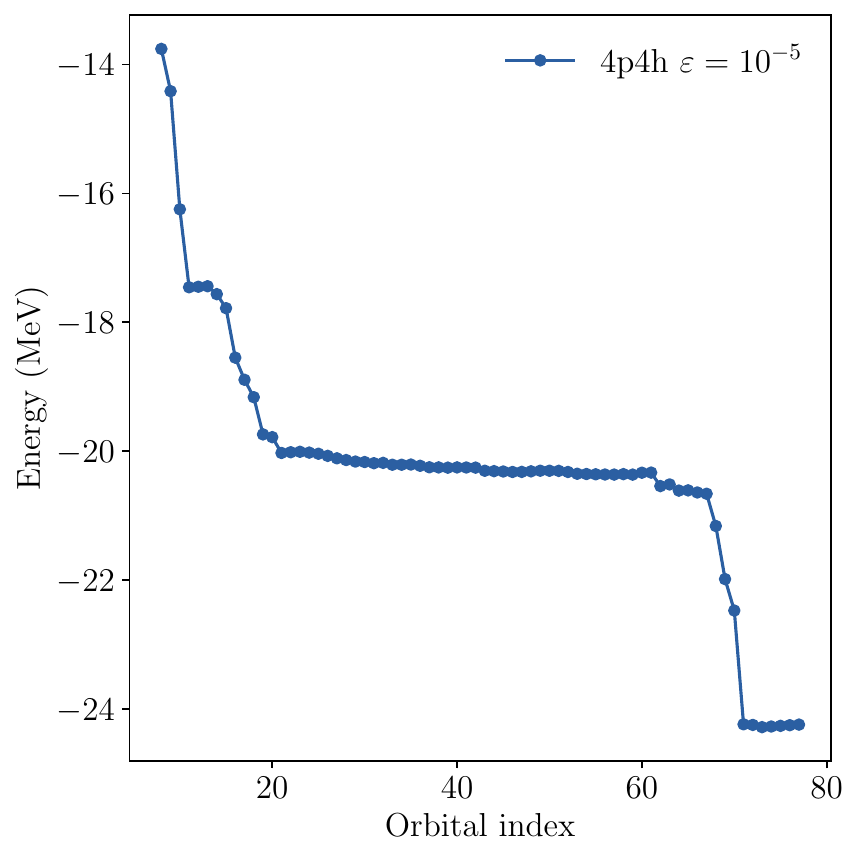}
  \caption{Binding energy of the ${J^\pi = {3/2}^-}$ ground state of \isotope[5]{He} as a function of the orbitals included in the G-DMRG calculation.}
  \label{fig_5He_E}
\end{figure}

\begin{figure}[h!]
  \centering
  \includegraphics[width=\linewidth]{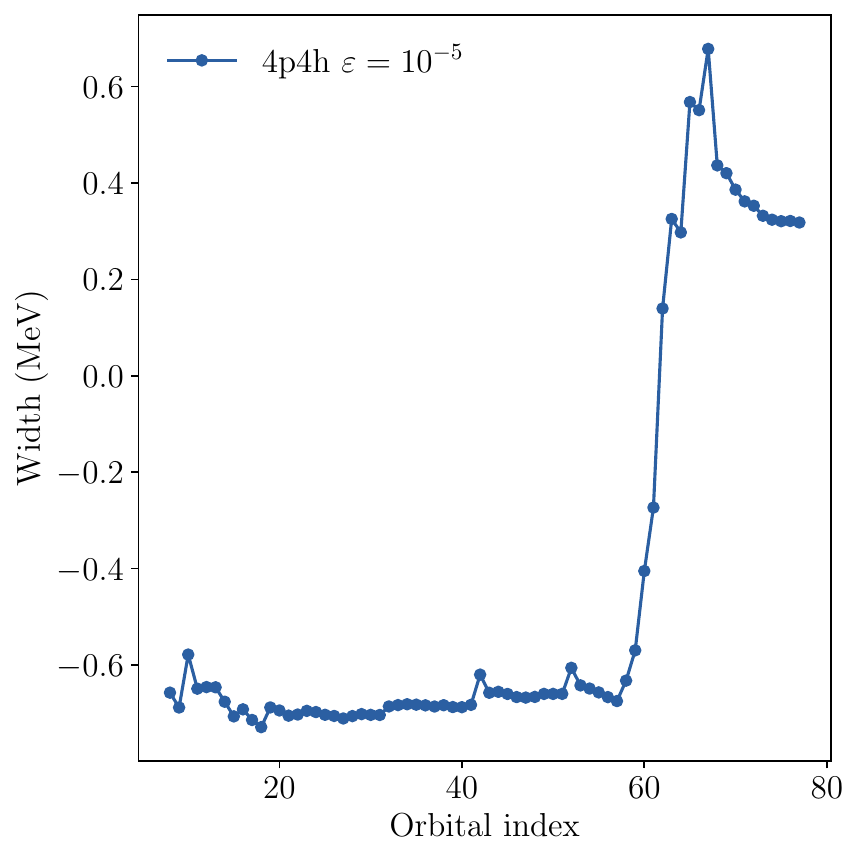}
  \caption{Width of the ${J^\pi = {3/2}^-}$ ground state of \isotope[5]{He} as a function of the number of orbitals included in the G-DMRG calculation.}
  \label{fig_5He_G}
\end{figure}

The binding energy and width of \isotope[5]{He} are shown in Figs.~\ref{fig_5He_E} and \ref{fig_5He_G}, respectively, as a function of the orbitals included in the G-DMRG calculation. 
This calculation was performed in a model space defined by $N_\text{max} = 12$ and $l_\text{max} = 2$, except for the neutron $l=1$ partial waves that were represented in the Berggren basis. 
Both the $p_{3/2}^{\nu}$ and $p_{1/2}^{\nu}$ partial waves included 15 scattering states along a contour extending into the \nth{4} quadrant of the complex-momentum plane with a cutoff at $ 4.0 \, \text{fm}^{-1}$, but only the $p_{3/2}^{\nu}$ partial wave included a resonant pole. 
The reference space included the proton and neutron $0s_{1/2}$, $0p_{3/2}$, and $0d_{5/2}$ orbitals, as well as the $0p_{1/2}^{\pi}$ orbital. 
All subsequent orbitals were $E$-ordered. 
The results were obtained with a p-h truncation of 4p4h and a G-DMRG truncation of $\varepsilon = 10^{-4}$.
At first, the ground-state energy of \isotope[5]{He} follows a similar pattern as that of \isotope[4]{He}, which is expected given that the structure of the \isotope[4]{He} cluster must emerge at some point, but then the energy drops around orbital index $i=60$, when $p_{3/2}^{\nu}$ scattering orbitals that are important for the asymptotic of the wave function are added. 
We want to emphasize that the energy and width are not expected to exhibit specific convergence patterns as more orbitals are added in the calculation. Obviously, if an important orbital is added last, which is not optimal, the energy will drop. The convergence pattern is with the DMRG truncation $\varepsilon$ as shown in Fig.~\ref{fig_bench_3He}.

\begin{figure}[h!]
  \centering
  \includegraphics[width=\linewidth]{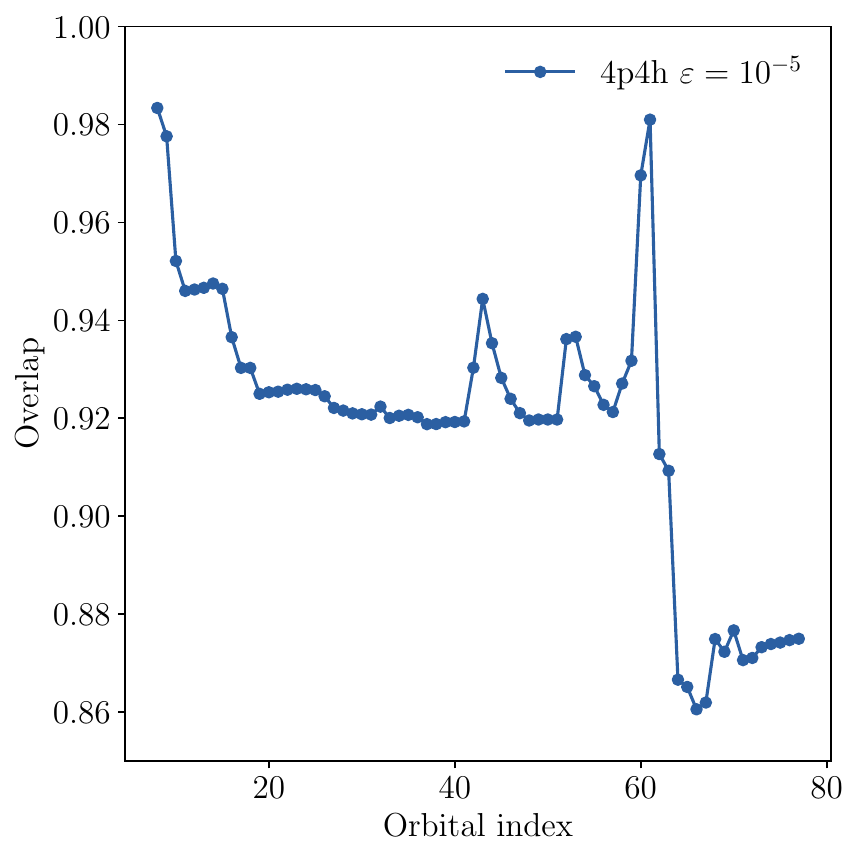}
  \caption{Overlap between the selected ground state of \isotope[5]{He} and the reference state as a function of the number of orbitals included in the G-DMRG calculation.}
  \label{fig_5He_O}
\end{figure}

The drop in energy is mirrored by a sudden increase in the width, and eventually both quantities stabilize. 
It is clear that the inclusion of important continuum couplings leads to many-body correlations being integrated into the wave function. 
This change in the composition of the wave function is apparent when looking at the overlap shown in Fig.~\ref{fig_5He_O}. 
Around orbital index $i=60$, we see a sharp increase in the overlap followed by a drop toward about 0.87. 

\begin{figure}[h!]
  \centering
  \includegraphics[width=\linewidth]{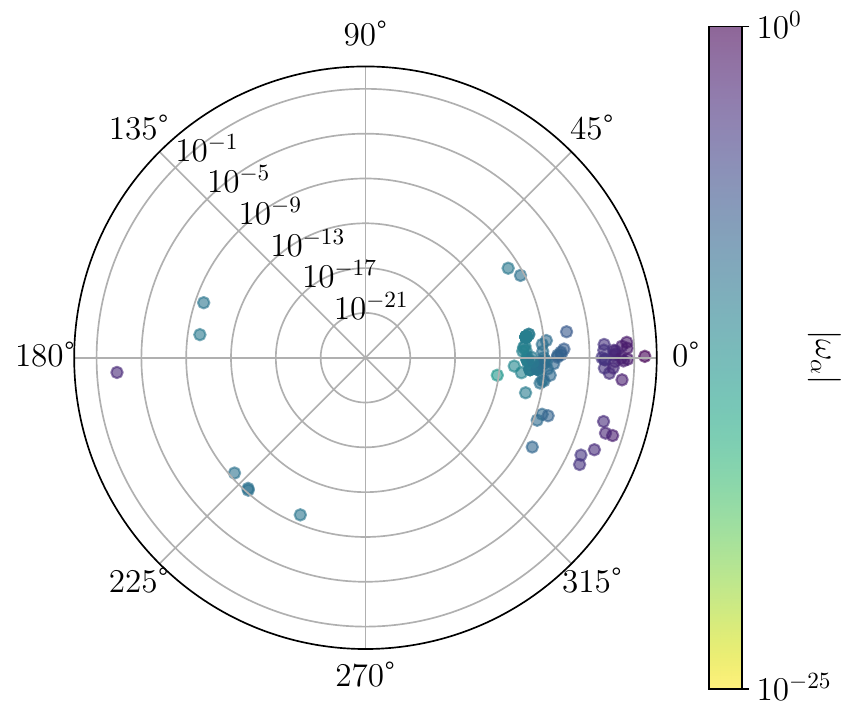}
  \caption{Eigenvalues of the reduced density matrix for the ground state of $\isotope[5]{He}$ at orbital index $i=50$, plotted in the complex plane in polar representation and with the radial component in log-scale.}
  \label{fig_5He_rho_eigvals_i50}
\end{figure}

\begin{figure}[h!]
  \centering
  \includegraphics[width=\linewidth]{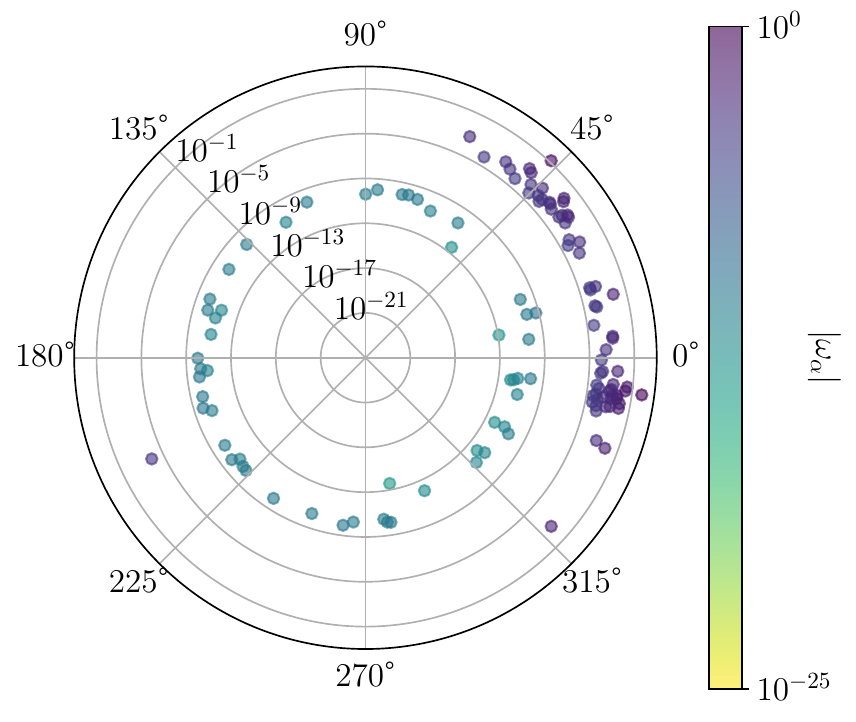}
  \caption{Eigenvalues of the reduced density matrix for the ground state of $\isotope[5]{He}$, plotted in the complex plane in polar representation and with the radial component in log-scale.}
  \label{fig_5He_rho_eigvals}
\end{figure}

To better visualize the reorganization of the wave function, in Figs.~\ref{fig_5He_rho_eigvals_i50} and \ref{fig_5He_rho_eigvals}, we show the eigenvalues of the reduced density matrix at $i=50$ and at the last iteration, respectively, \textit{i.e.} before and after $p_{3/2}^{\nu}$ scattering states are included. 
As in the case of \isotope[4]{He}, we used the polar representation with the radial component in log-scale. 

Before important continuum couplings are added, most eigenvalues of the reduced density matrix are positive with a small imaginary part ($\theta \approx 0$). 
There are two clusters with $|\omega_\alpha| \approx 10^{-10}$ and $\approx 10^{-3}$, respectively, and one large eigenvalue stands out at about 0.9. 
Notably, there is one large negative eigenvalue with a small imaginary part (more on this later). 
Except for the negative eigenvalue, at this state of the calculation, the profile of the wave function of \isotope[5]{He} as seen from the eigenvalues of the reduced density matrix looks quite similar to that of the ground-state of \isotope[4]{He}. 
This is not surprising given that no significant continuum couplings have been added yet. 
We remark that $\trace(\rho) = 1.0$ at any point during the G-DMRG renormalization.

After the inclusion of $p_{3/2}^{\nu}$ scattering states, the cluster of small eigenvalues spreads around a circle defined by $|\omega_\alpha| \approx 10^{-10}$, while the one with larger eigenvalues splits into two groups with $\theta \approx -10$ and $\theta \approx 40$, and the dominant eigenvalue goes down to about 0.8. 
The first group of large eigenvalues at $\theta \approx -10$ is most likely responsible for the localized inner part of the wave function, while the other one, with values where $|\text{Re}(\omega_\alpha)| \approx |\text{Im}(\omega_\alpha)|$, contributes to the asymptotic part.

\begin{figure}[h!]
  \centering
  \includegraphics[width=\linewidth]{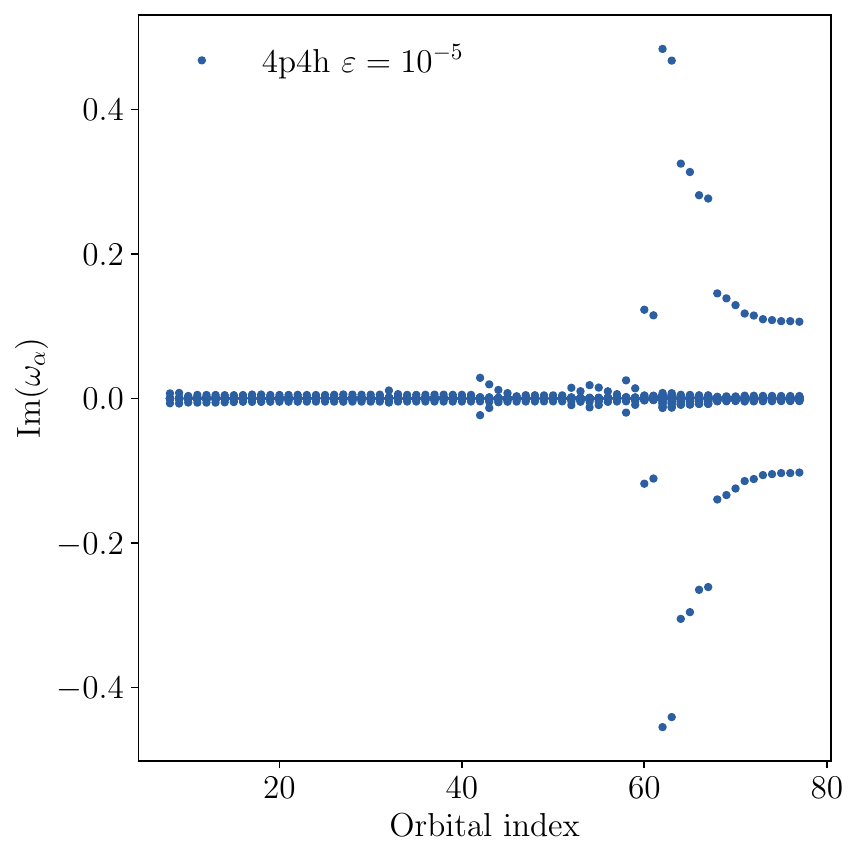}
  \caption{Imaginary part of the eigenvalues of the reduced density matrix for the ground state of $\isotope[5]{He}$ as a function of the orbital added in the G-DMRG calculation.}
  \label{fig_5He_rho_eigvals_im}
\end{figure}

This is more apparent in Fig.~\ref{fig_5He_rho_eigvals_im} where the distribution of the imaginary part of the eigenvalues is shown at each G-DMRG iteration, \textit{i.e.} everytime a new orbital is added. 
We can see one group of eigenvalues with near-zero imaginary parts, and another group with imaginary parts symmetric with respect to the imaginary axis. 
This also shows that $\trace(\rho) = 1.0$ does result in all imaginary parts cancelling. 
Note that eigenvalues with imaginary parts that mirror each other generally do not have the same real part. 
The sudden change associated with the inclusion of important continuum couplings around orbital index $i=60$ is unmistakable.

Before we move to the problematic case of a broad resonance, we provide the reader with some intuition about what is happening here. 
As we have seen with the ground state of \isotope[5]{He}, in the quasi-stationary formalism, density matrices become complex-symmetric and acquire complex eigenvalues. 
Following Berggren's prescription~\cite{berggren68_32,gyarmati72_775,berggren96_23}, the real part of the eigenvalue is interpreted as a probability amplitude, while the magnitude of the imaginary part is interpreted as an uncertainty on the amplitude due to the inherently time-dependent nature of the state. 
In App.~\ref{app_cx_density}, we use a toy model to demonstrate that the inclusion of continuum couplings, modeled as increasingly complex off-diagonal matrix elements in the reduced density matrix, leads to more entanglement, and specifically to a saturation of entanglement at an exceptional point of the reduced density.

As noted in App.~\ref{app_cx_density}, this is easily understood if we picture, for example, a system that can decay \textit{via} two resonances, and then, little by little, we broaden these resonances until they overlap significantly. 
In this situation, the probability to decay \textit{via} either resonance will approach 50\%. 
Furthermore, if these resonances are broad and decay on a timescale comparable to the interaction timescale, there will be considerable uncertainty on the probabilities themselves, meaning many measurements will be needed to determine the branching ratios accurately. 
In short, the two overlapping resonances will effectively behave as one broad state, hence the large entanglement.

An important point to understand is that the process described above is driven by increasing the imaginary part of the off-diagonal matrix elements of the reduced density matrix, but in practice scattering orbitals also contribute to the real part. 
Increasing the real part has a different effect and leads to repulsion between the eigenvalues of the reduced density matrix. 
This can cause issues in quasi-stationary calculations when the real and imaginary parts of the off-diagonal matrix elements have comparable magnitudes, because the repulsion becomes strong enough to generate eigenvalues with negative real parts, namely, negative probabilities. 
As mentioned in Sec.~\ref{sec_formalism}, the problem of the positivity of the reduced density matrix is a known~\cite{cuevas13_3237,kshetrimayum17_3211,weimer21_3210} and difficult problem~\cite{kliesch14_3236,werner16_3235}. 
Below, we show how this problem affects the identification of the physical state for broad resonances, and how we solve it in the G-DMRG approach.

We finally move to the case of a broad single-particle resonance, given here by the $J^\pi = 2^-$ state of \isotope[4]{H}. 
According to state-of-the-art calculations~\cite{lazauskas19_2363}, this state has a ratio $|\Gamma/(2 S_{n})| \approx 1.82$, placing it well into the broad resonance region. 
While we have optimized results for this state, presented in Sec.~\ref{sec_results}, for illustrative purposes, here the Berggren basis representing the $p_{3/2}^{\nu}$ partial wave is chosen so that the contour, starting at the origin in the complex-momentum plane, goes down into the \nth{4} quadrant along a line forming an excessively large angle with respect to the real axis ($\approx -32.5 \degree$). 
As a result, towards the end of the calculation, the G-DMRG method is forced to integrate continuum couplings into the wave function too rapidly.

Specifically, this calculation was performed in a model space defined by $N_\text{max} = 12$ and $l_\text{max} = 4$, except for the neutron $l=0$ and 1 partial waves that were represented in the Berggren basis. 
We also limited the $l=3$ and 4 partial waves to $n=0$ and 1. 
The $s_{1/2}^{\nu}$, $p_{3/2}^{\nu}$, and $p_{1/2}^{\nu}$ partial waves included 20, 30, and 20 scattering states, respectively. 
The $s_{1/2}^{\nu}$ and $p_{1/2}^{\nu}$ scattering states were taken along the real axis in the complex-momentum plane, while the $p_{3/2}^{\nu}$ sattering states were taken along a contour extending into the \nth{4} quadrant. 
We applied a cutoff at $ 4.0 \, \text{fm}^{-1}$. 
Only the $p_{3/2}^{\nu}$ partial wave included a resonant pole. 
The reference space included the proton and neutron $0s_{1/2}$, $0p_{3/2}$, and $0d_{5/2}$ orbitals, as well as the $0p_{1/2}^{\pi}$ orbital, and all subsequent orbitals were ordered using the $(l,j)$-ordering. 
The results were obtained with by a 4p4h truncation and a G-DMRG truncation of $\varepsilon = 10^{-4}$.
We note in passing that our optimized results still rely on the truncation scheme presented below, but to a lesser degree.

Given the previous discussion for \isotope[5]{He}, we can anticipate issues with the identification of the physical state due to increased entanglement and possibly the emergence of negative probabilities. 
In addition, we saw that the introduction of strong continuum couplings generated many small complex eigenvalues, which could also affect the effectiveness of the DMRG truncation itself. 
Finally, when the magnitude of the real and imaginary parts of the elements of the reduced density matrix are comparable in size, the diagonalization can become sensitive to numerical errors, so that small eigenvalues are less precise.

As a matter of fact, we were unable to calculate the energy and width of the $J^\pi = 2^-$ state of \isotope[4]{H} directly, \textit{i.e.} without using the truncation scheme introduced below. 
In short, once eigenvalues of the reduced density matrix with a negative real part appear, the DMRG truncation in Eq.~\eqref{eq_DMRG_epsilon} can fail because, to maintain $\trace(\rho) = 1$, negative real parts are compensated by positive real parts that add to a number $w > 1.0$. 
It is thus possible that, when adding those real part from the largest to the smallest, we reach a point where $|w-1.0| > \varepsilon$, in which case nearly all subsequent negative eigenvalues will be added. 
As a consequence, not only unimportant couplings are kept, but irrelevant states associated with numerical noise are included. 
This further destabilizes the next G-DMRG iteration and ultimately leads to more spurious mixing in the wave function.

\begin{figure}[h!]
  \centering
  \includegraphics[width=\linewidth]{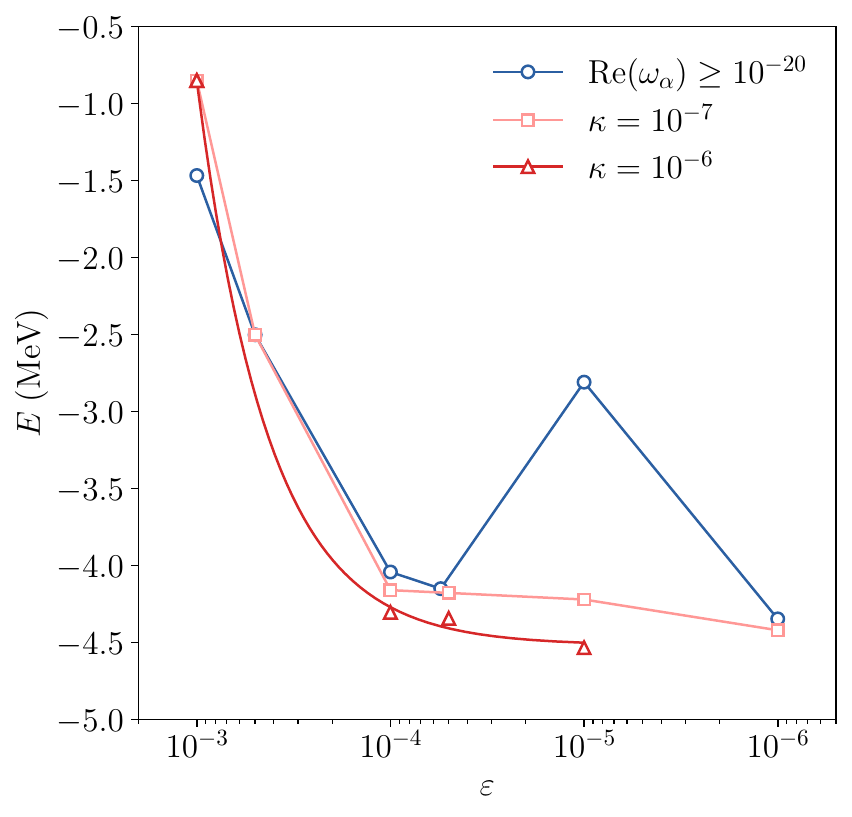}
  \caption{Binding energy of the ${J^\pi = 2^-}$ state of \isotope[4]{H} as a function of the DMRG truncation for various values of the truncation parameter $\kappa$.}
  \label{fig_E_eps_kappa}
\end{figure}

For the sake of illustration, here we truncate all eigenvalues that satisfy $\text{Re}(\omega_\alpha) < 10^{-20}$, \textit{i.e.} eigenvalues with a negative real part or with a positive real part that is well below numerical precision. 
Doing so does not remove all instabilities, but at least it makes calculations feasible. 
To demonstrate this point, in Fig.~\ref{fig_E_eps_kappa} (blue line) we show the energy of the $J^\pi = 2^-$ state of \isotope[4]{H} as a function of the DMRG truncation. 
We immediately see that the energy does not converge smoothly with $\varepsilon$.

\begin{figure}[h!]
  \centering
  \includegraphics[width=\linewidth]{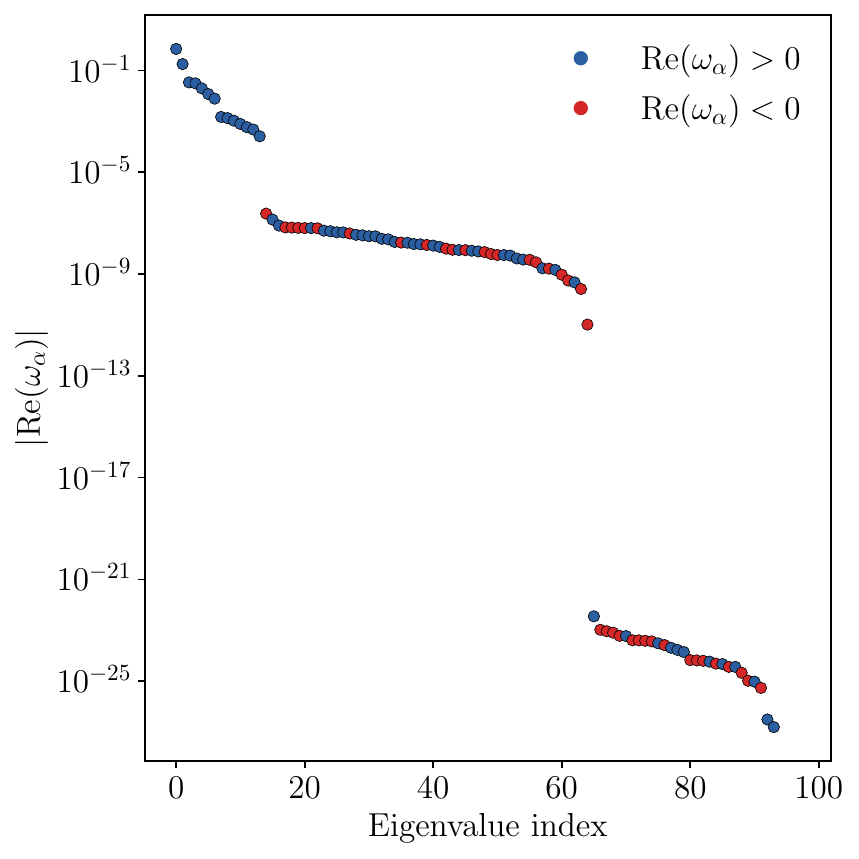}
  \caption{Absolute values of the real part of the eigenvalues of the reduced density matrix in the ${J^\pi = 2^-}$ state of \isotope[4]{H} ordered from largest to smallest.}
  \label{fig_Re_eigvals_abs}
\end{figure}

In fact, as shown in Fig.~\ref{fig_Re_eigvals_abs}, if we inspect the real part of the eigenvalues of the reduced density matrix at the end of a G-DMRG calculation with truncation $\text{Re}(\omega_\alpha) < 10^{-20}$, we see that new eigenvalues with negative real parts are generated after adding the very last orbital.
There is one small group of eigenvalues with $\text{Re}(\omega_\alpha) > 10^{-5}$ that dominate the wave function, then a large group that includes eigenvalues with either positive or negative real parts satisfying $10^{-6} > |\text{Re}(\omega_\alpha)| > 10^{-12}$, and finally another large group with real parts within numerical noise. 
A similar profile is observed in almost all late iterations, where continuum couplings are strong.

We note that in the standard DMRG approach, a weak entanglement between the reference space and the environment is usually reflected in the real and positive eigenvalues $\omega_\alpha$ having an exponentially decreasing distribution. 
Instead, here we observe that only the first small group of eigenvalues has rapidly decreasing $\text{Re}(\omega_\alpha)$, while the second group characterized by $10^{-6} > |\text{Re}(\omega_\alpha)| > 10^{-12}$ has a weakly decreasing distribution. 
The latter is a hallmark of strong entanglement, which we know is due to continuum couplings.

\begin{figure}[h!]
  \centering
  \includegraphics[width=\linewidth]{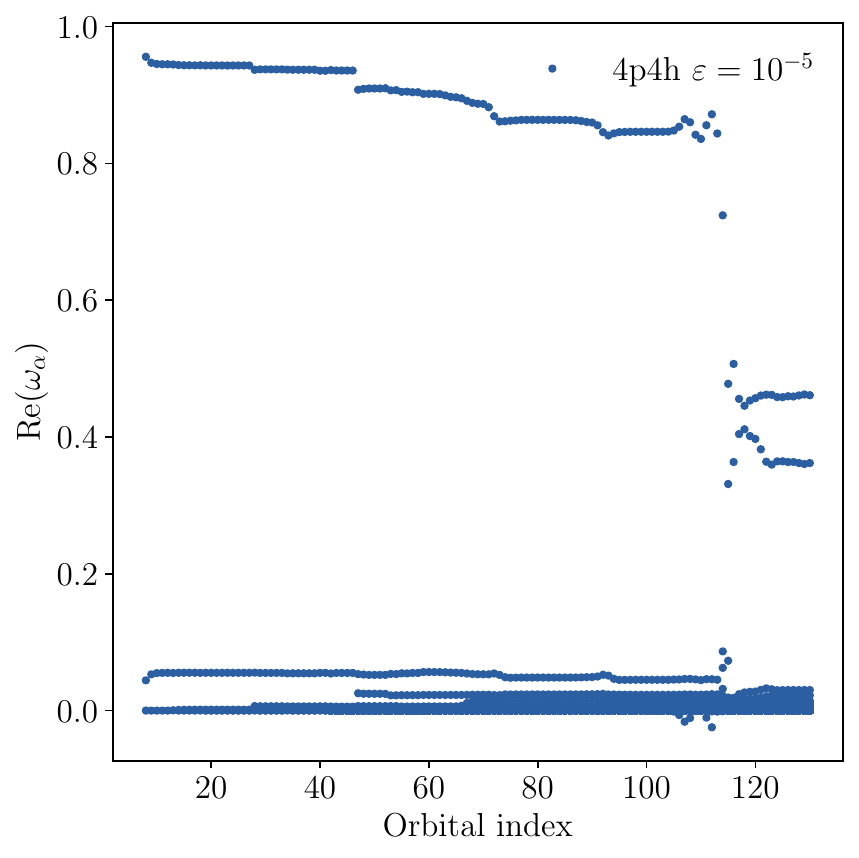}
  \caption{Real part of the eigenvalues of the reduced density matrix for the ground state of $\isotope[4]{H}$ as a function of the orbital added in the G-DMRG calculation}
  \label{fig_4H_rho_eigvals_im}
\end{figure}

Besides for the accumulation of small eigenvalues, the excessive growth of entanglement with increasing continuum couplings actually leads to a bigger issue, namely, the collapse of the DMRG ansatz. 
Indeed, as shown in Fig.~\ref{fig_4H_rho_eigvals_im}, there can be a point during the calculation where the renormalization fails to find a better representation for the wave function. 
When this happens, the wave function is not dominated by one SD anymore, as it should, but by two with nearly equal weights. 
This competition between two configurations is shown in Fig.~\ref{fig_4H_rho_eigvals_im} around orbital index $i=115$. 
When that happens, the overlap method is likely to fail because, in practice, the diagonalization of the renormalized Hamiltonian gives two states with similar energies and overlaps with the reference state. 
This is easy to understand given that these eigenstates can be approximately written as $a_0 \ket{\text{SD}_0} + a_1 \ket{\text{SD}_1}$ and $a_0' \ket{\text{SD}_0} - a_1' \ket{\text{SD}_1}$, where $\ket{\text{SD}_{0,1}}$ are the dominant SDs.

\begin{figure}[h!]
  \centering
  \includegraphics[width=\linewidth]{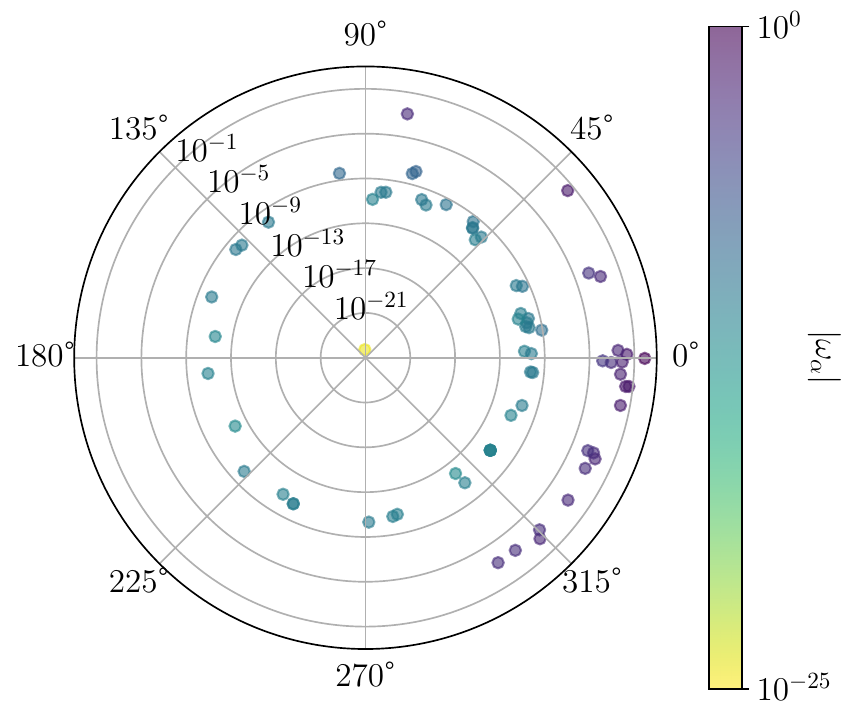}
  \caption{Eigenvalues of the reduced density matrix for the ground state of $\isotope[4]{H}$ at orbital index $i=100$, plotted in the complex plane in polar representation and with the radial component in log-scale.}
  \label{fig_4H_rho_eigvals_i100}
\end{figure}

\begin{figure}[h!]
  \centering
  \includegraphics[width=\linewidth]{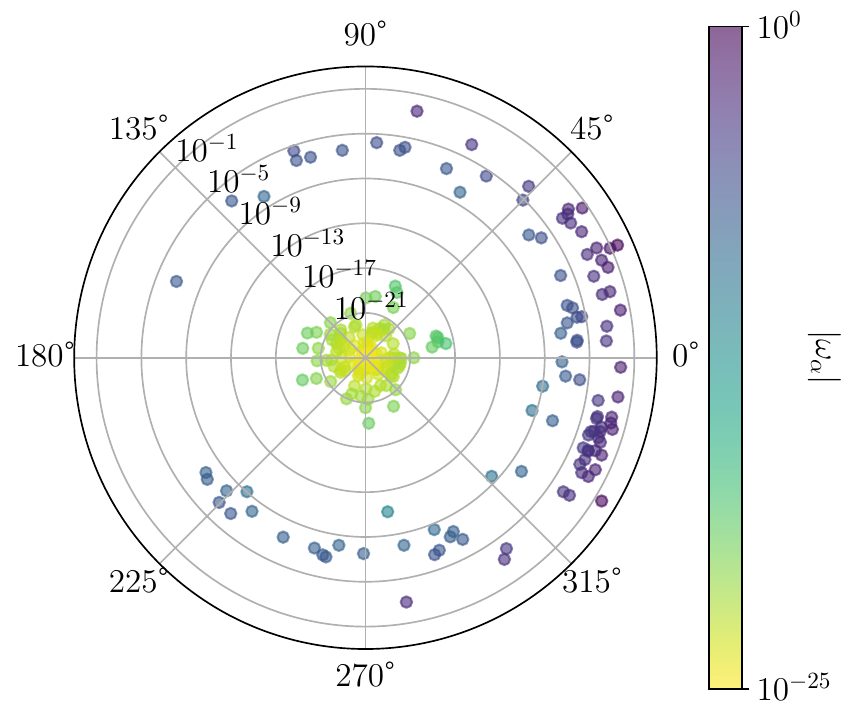}
  \caption{Eigenvalues of the reduced density matrix for the ground state of $\isotope[4]{H}$ at orbital index $i=120$, plotted in the complex plane in polar representation and with the radial component in log-scale.}
  \label{fig_4H_rho_eigvals_i120}
\end{figure}

A better way to visualize what happens to the wave function and how this is related to the second group of eigenvalues characterized by $10^{-6} > |\text{Re}(\omega_\alpha)| > 10^{-12}$ is to use the polar representation. 
In Figs.~\ref{fig_4H_rho_eigvals_i100} and \ref{fig_4H_rho_eigvals_i120} we show the eigenvalues of the reduced density matrix before and after the collapse of the DMRG ansatz, respectively.
Before the collapse, as in the case of \isotope[5]{He}, there is a cluster of large eigenvalues in the $[-\pi/4,\pi/4]$ range with an accumulation near the real axis ($\theta = 0$), and a second group of small eigenvalues with $|\omega_\alpha| \approx 10^{-10}$ forming a circle. 
The latter corresponds to the second group mentioned earlier that satisfied $10^{-6} > |\text{Re}(\omega_\alpha)| > 10^{-12}$, but we see here that, in fact, it is better characterized by the condition $10^{-6} > |\omega_\alpha| > 10^{-12}$.

After the collapse, the group of large eigenvalues spreads further within the $[-\pi/4,\pi/4]$ range and the accumulation near the real axis disappears. 
Basically, the localized inner part of the wave function is disappearing. 
The magnitude of the eigenvalues in the second group grows to $|\omega_\alpha| \approx 10^{-6}$, increasing numerical instabilities, which is confirmed by the increase in the number of eigenvalues with magnitudes within numerical noise.

\subsubsection{Stabilization of the renormalization of complex-symmetric matrices}
\label{ssec_kappa}

The observations made so far suggest a simple and elegant truncation scheme that consists in keeping only eigenvalues that satisfy
\begin{equation}
  |\omega_\alpha| \geq \kappa
  \label{eq_kappa}
\end{equation}
where $\kappa \geq 0$ is large enough that it removes the group of eigenvalues identified earlier that were distributed along a circle defined by $10^{-6} > |\omega_\alpha| > 10^{-12}$ in polar coordinates. 
To avoid interference between this $\kappa$ and the regular DMRG truncation $\varepsilon$, we must always have $\kappa \ll \varepsilon$. 
In well-behaved G-DMRG calculations, where the distribution of $\text{Re}(\omega_\alpha)$ is rapidly decreasing, the truncation $\kappa$ will never be used in practice. 
It is only in those cases where a group of eigenvalues has a weakly decreasing distribution of $\text{Re}(\omega_\alpha)$ that this truncation scheme will apply.

The main drawback of this new truncation is that it might cut correlations that are ``under construction'' but, as shown in Fig.~\ref{fig_E_eps_kappa}, it does not seem to be the case. 
The energy decreases exponentially with $\varepsilon$ again, and converges at a lower value than before. 
Furthermore, calculations are significantly faster as more unnecessary contributions are removed.

\begin{figure}[h!]
  \centering
  \includegraphics[width=\linewidth]{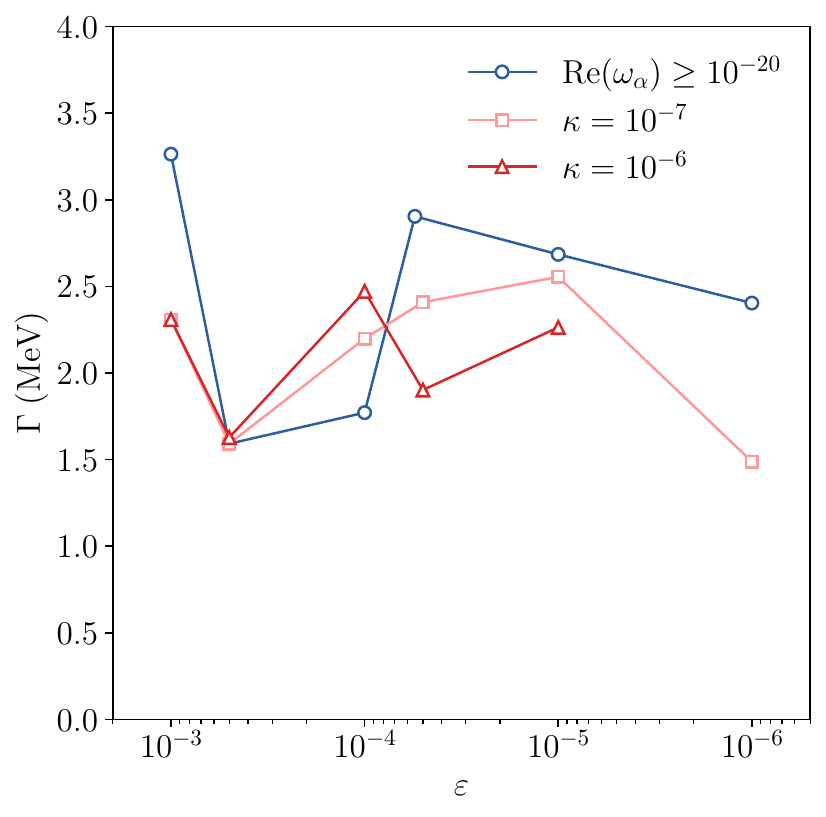}
  \caption{Width of the ${J^\pi = 2^-}$ state of \isotope[4]{H} as a function of the DMRG truncation for various values of the truncation parameter $\kappa$.}
  \label{fig_G_eps_kappa}
\end{figure}

We can also see in Fig.~\ref{fig_G_eps_kappa} that the new truncation does not negatively impact the width. 
We obviously do not expect to obtain a smooth behavior for the width using a suboptimal basis, but we see that the width does not collapse when using the new truncation. 
These results demonstrate that the new scheme can be used effectively to stabilize G-DMRG calculation for broad resonances.

Interestingly, we found that the truncation $\kappa$ can also be used to dramatically accelerate calculations for bound states and narrow resonances without introducing a significant error on the final energy. 
This is again due to the distribution of the eigenvalues of the reduced density matrix which, in many cases, looks like a decreasing exponential followed by a long tail of small contributions. 
The latter is probably caused by the fact that entanglement in nuclear systems likely obeys a volume law~\cite{pazy23_3230,gu23_3201}.

\begin{figure}[h!]
  \centering
  \includegraphics[width=\linewidth]{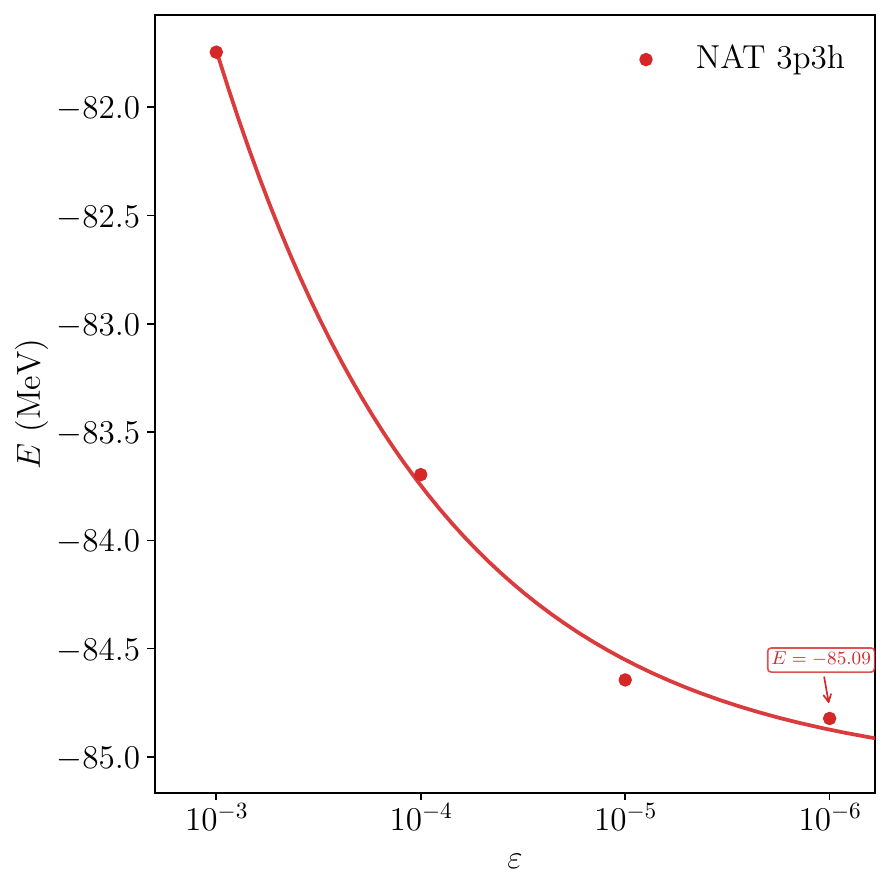}
  \caption{Binding energy of the ${J^\pi = 0^+}$ ground state of \isotope[12]{C} as a function of the DMRG truncation for various values of the truncation parameter $\kappa$.}
  \label{fig_E_12C}
\end{figure}

To illustrate this finding, we first calculated the ground-state energy of \isotope[12]{C} in an HO space defined by $N_\text{max} = 12$ and $l_\text{max} = 4$, starting from a reference space including the neutron and proton $n=0$ orbitals with $l=0$, 1, and 2. 
All orbitals outside the reference space were $E$-ordered. 
We applied a 6p6h truncation-at-construction starting in the $l=0$ orbitals, and a 2p2h in the medium. 
To make this calculation possible, we combined a modest truncation $\kappa = 10^{-10}$ together with a DMRG truncation of $\varepsilon = 10^{-4}$. 
We then generated a natural orbital basis (see Sec.~\ref{sec_nat}) and recalculated the ground-state energy with a 3p3h truncation in the renormalized medium, $\kappa = 10^{-10}$, and different values of $\varepsilon$. 
The results are shown in Fig.~\ref{fig_E_12C}.

We would like to point out that, before this work, performing such large \textit{ab initio} calculations using the G-DMRG method was unthinkable, but it is now possible on just 128 processor cores. 
The results exhibit a nearly exponential convergence with the DMRG truncation, with an extrapolated value at $E = -85.09 \, \text{MeV}$. 
For context, this is a relative error of about 7.7\% with experiment. 
We are still far from the 1-2\% relative error of state-of-the-art calculations, but this is a promising development given the absence of three-body forces and the interaction dependence.

\subsection{Natural orbital basis}
\label{sec_nat}

In previous sections, we highlighted the importance of having a good starting reference state and an orbital ordering that helps build both the energy and the width smoothly. 
We also discussed the identification problem for physical states, and how it relates to the way the renormalization handles the increase in entanglement due to continuum couplings. 
In this section, we show that once a reasonable low-fidelity result has been obtained, generating natural orbitals~\cite{brillouin33,lowdin55_2499,lowdin56_2498} can considerably simplify all the problems encountered earlier, and help converge calculations more efficiently and at a lower computational cost.

The natural orbital (NAT) basis can be obtained by diagonalizing the one-body density matrix, whose matrix elements in the quasi-stationary formalism are defined as:
\begin{equation}
    \rho_{\alpha,\beta}^{(1)} = \smel{\tilde{\Psi}}{\hat{a}_{\beta}^{\dagger} \hat{a}_{\alpha}}{\Psi},
    \label{eq_rho1_mel}
\end{equation}
where $\ket{\Psi}$ is an (approximate) eigenstate, and $\alpha$, $\beta$ denote orbitals. 
The resulting single-particle eigenvectors are, by definition, similar to the orbitals that we would obtain in a mean-field calculation. 
For that reason, solving the many-body problem in this basis can yield substantial benefits. 
When further convergence is needed, it also possible to generate a new set of NAT orbitals from the new many-body eigenvector obtained in the first NAT basis. 
We denote this second set $\text{NAT}^2$.

The NAT basis was originally introduced in the G-DMRG method in Ref.~\cite{shin16_1860}, and used in several subsequent applications~\cite{fossez17_1916,fossez18_2171,fossez22_2540}, but it was never applied to the calculation of broad resonances. 
In all cases, the NAT orbitals within each partial wave $(l,j)$ were ordered according to the magnitude of their associated eigenvalue $|\omega|$, and the same reference space and orbital ordering were used in both the original calculation and the one in the NAT orbitals. 
The energy would converge exponentially while the width would remain within a narrow band.

In \textit{ab initio} calculations of bound states based on the HO basis, the NAT basis helps converge the energy earlier with respect to the number of orbitals added in the calculation, but it does not significantly lower the energy. 
The same observation was made in quantum chemistry using optimized sets of orbitals~\cite{rissler06_3191}. 
For example, if we calculate the ground state of \isotope[4]{He} in the same HO basis as used in Sec.~\ref{sec_id}, and with a 2p2h truncation and $\varepsilon = 10^{-4}$, we obtain $E = -27.93 \, \text{MeV}$. 
Then, if we generate the NAT basis from this result and recalculate the energy without changing anything else, we obtain $E = -28.09 \, \text{MeV}$, which is barely lower. 
Note that when generating the NAT basis, within each partial wave $(l,j)$ we assigned the principal quantum number $n$ in descending order of occupation. 
Interestingly, if we generate a new NAT basis ($\text{NAT}^2$) and recalculate the energy, due to the loss of information after each renormalization, we actually obtain a slightly higher result of $E = -28.02 \, \text{MeV}$.

However, when considering narrow resonances, the NAT basis helps converge faster and improves the energy. 
This is not surprising given that, in this case, it is often difficult to find a quasi-optimal single-particle basis through trial-and-error, and thus there is always room for improvement. 
Another important aspect noticed in the past is that the NAT basis makes the renormalization more stable, with the energy converging exponentially with each orbital added, and the width remaining in a narrow band. 
Below we show that most of these features can be preserved when calculating broad resonances, but for that we must revisit the problems of the reference space and of the orbital ordering.

In the quasi-stationary formalism, the one-body density matrix is complex-symmetric, and the NAT basis is not related to the original basis by a unitary transformation, but a similarity transformation. 
Nevertheless, the NAT orbitals are linear combinations of the original orbitals. 
Consequently, the association between poles of the single-particle $S$-matrix and certain orbitals in the original basis is, in principle, lost in the NAT basis. 
The NAT orbitals are expected to approximate the discrete eigenstates of the mean-field, but there is no guarantee that they are eigenstates of a single-particle potential. 
This means that in the NAT basis the reference space is not the pole space.

This problem is insignificant for narrow resonances because, in that case, continuum couplings act as a small correction on a structure otherwise dominated by discrete orbitals. 
It follows that as long as NAT orbitals are assigned a principal quantum number $n$ within each partial wave $(l,j)$, based on either the magnitude or the real part of their occupations in descending order, we can keep the same reference space and the $E$-ordering, and we expect the G-DMRG approach to converge to the same state as in the original basis. 
Basically, the overall structure of entanglement in the final wave function should be preserved.

However, for broad resonances, it can happen that a discrete NAT orbital with a high occupation emerges from the continuum as a result of many-body correlations. This suggests that the original basis was suboptimal, and it may be advantageous to include this state in the reference space. 
In fact, reordering all orbitals based on their occupation number may be the best strategy\footnote{Here, we continue to rely on Berggren's interpretation and treat the real part of a complex eigenvalue of the one-body density matrix as an occupation, and the magnitude of the imaginary part as an uncertainty on that occupation.}.

To demonstrate that this is indeed the case, we first calculate the energy and width of the $J^\pi = 2^-$ state of \isotope[4]{H} in a model space similar to the one used in Sec.~\ref{sec_id}, except that we increase the number of scattering states in the $s_{1/2}^{\nu}$, $p_{3/2}^{\nu}$, and $p_{1/2}^{\nu}$ partial waves to 30, 45, and 30, respectively, and adjusted the $p_{3/2}^{\nu}$ contour entering the Berggren basis to make an angle of $\approx -24.4 \degree$ with the real axis. 
The reference space included the proton and neutron $0s_{1/2}$, $0p_{3/2}$, and $0d_{5/2}$ orbitals, as well as the $0p_{1/2}^{\pi}$ orbital, and all subsequent orbitals were ordered using the $E$-ordering. 
The results were obtained with a 2p2h truncation and a G-DMRG truncation of $\varepsilon = 10^{-4}$.

\begin{figure}[h!]
  \centering
  \includegraphics[width=.9\linewidth]{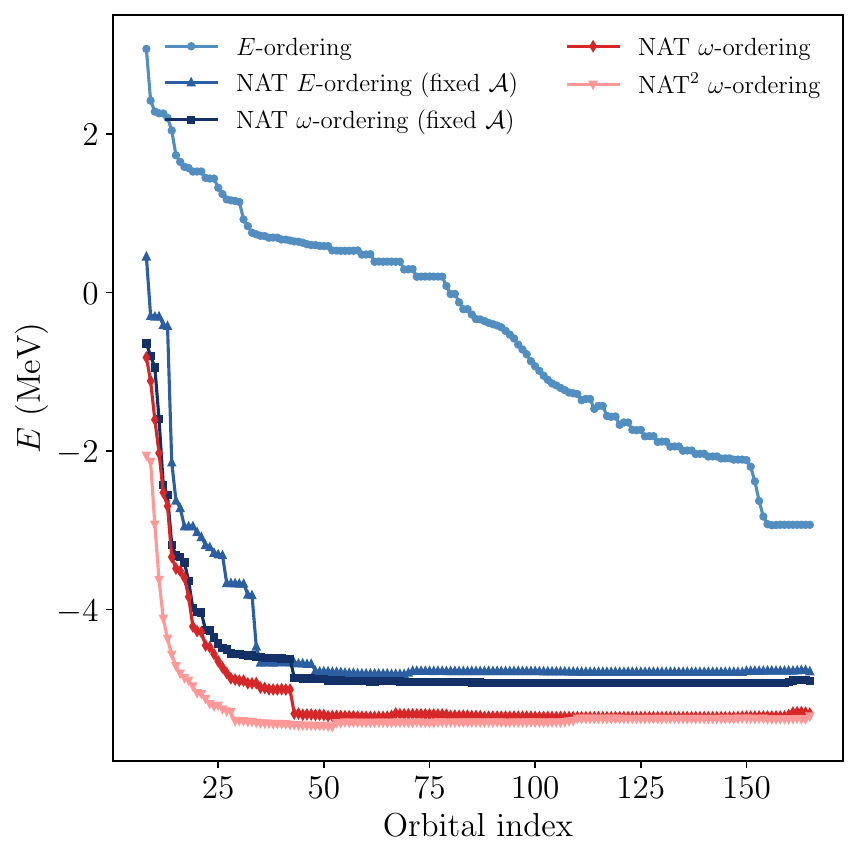}
  \caption{Energy of the $J^\pi = 2^-$ state of $\isotope[4]{H}$ as a function of the number of orbitals included in the G-DMRG method for different bases, reference space, and ordering choices.}
  \label{fig_4H_E_NAT}
\end{figure}

\begin{figure}[h!]
  \centering
  \includegraphics[width=.9\linewidth]{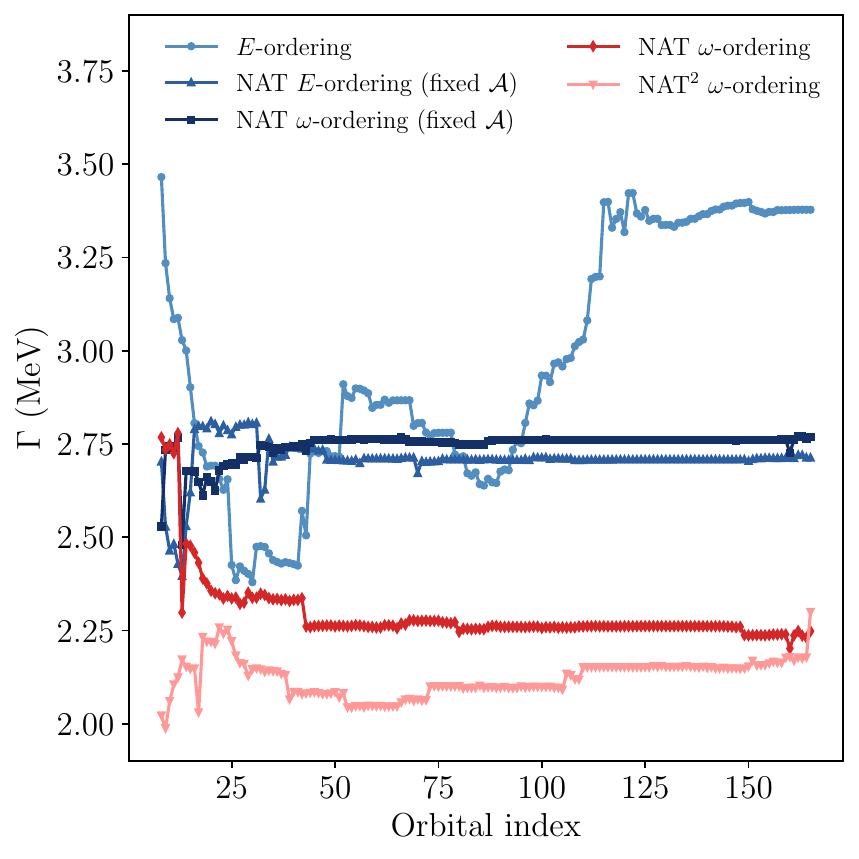}
  \caption{Width of the $J^\pi = 2^-$ state of $\isotope[4]{H}$ as a function of the number of orbitals included in the G-DMRG method for different bases, reference space, and ordering choices.}
  \label{fig_4H_G_NAT}
\end{figure}

From this calculation, we generated a NAT basis, and the NAT occupations revealed the emergence of a discrete $0p_{1/2}^{\nu}$ orbital which was not included in the original basis. 
Then, we compared the original result with calculations in the NAT basis where we keep the reference space $\mathcal{A}$ fixed, \textit{i.e.} we do not add the $0p_{1/2}^{\nu}$ in it, and where the rest of the orbitals are either $E$-ordered or occupation-ordered ($\omega$-ordered). 
Finally, we compare these results with a calculation in which we keep the same number of orbitals in the reference space, but let the $\omega$-ordering determine which orbitals are included in it. 
The results for the energy and the width as a function of the orbitals added are shown in Figs.~\ref{fig_4H_E_NAT} and \ref{fig_4H_G_NAT}, respectively.

Clearly, the use of the NAT basis yields several benefits. 
First, even without changing the orbitals entering the reference space, the NAT basis improves the quality of the reference state and lowers the initial energy. 
Second, the energy approaches its minimum after adding only about a quarter of the total number of orbitals available, and then it remains nearly constant. 
Third, letting the $\omega$-ordering determine which orbitals should enter the reference space first provides a significantly lower energy than all other schemes.

A similar stabilization of the renormalization is observed with the width. 
Results in the NAT basis all exhibit some variations when adding the first 40 orbitals, and then they remain nearly constant. 
As expected, the width in the NAT basis is lower than in the original one because the energy is also lower.

For completeness, in Figs.~\ref{fig_4H_E_NAT} and \ref{fig_4H_G_NAT} we also show results in the $\text{NAT}^2$ basis, generated from the optimal NAT result with $\omega$-ordering. 
While there is no significant gain for the energy, the width shows some instabilities. 
This shows that, for broad resonances, the $\text{NAT}^2$ basis actually degrades the quality of the results when generated from truncated calculations. 
It is not surprising that only the width degrades significantly given how sensitive it is to details of the wave function.

\begin{figure}[h!]
  \centering
  \includegraphics[width=.9\linewidth]{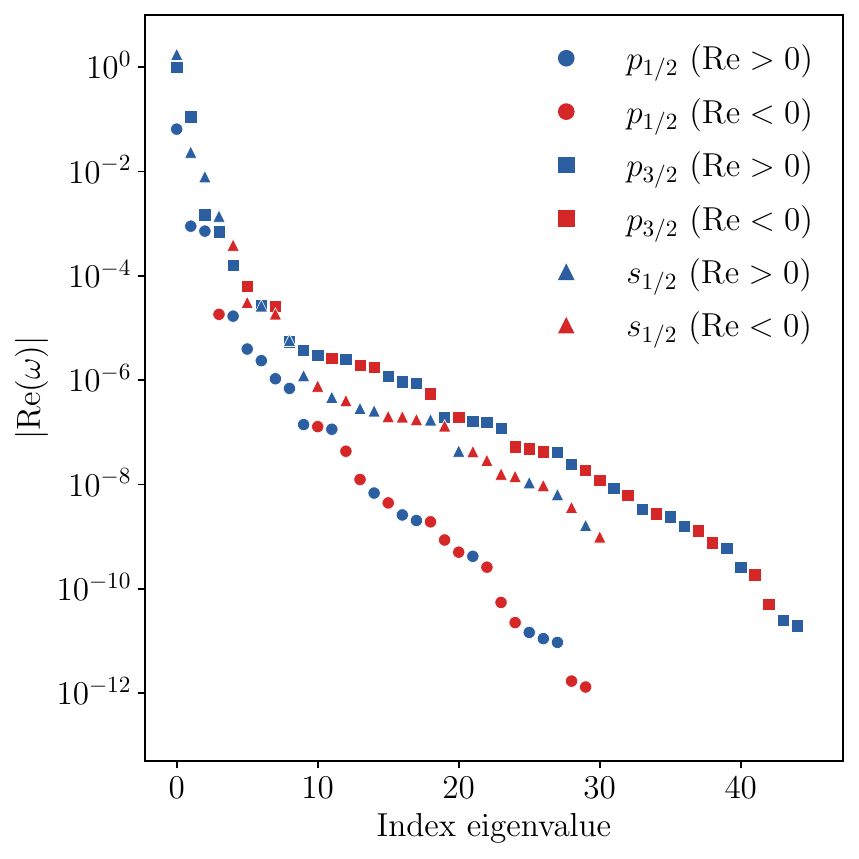}
  \caption{Distribution of the real part of the eigenvalues of the neutron one-body density matrix in the $J^\pi = 2^-$ state of $\isotope[4]{H}$. The sign of the real part is indicated in blue and red for positive and negative values, respectively.}
  \label{fig_4H_Rew_NAT}
\end{figure}

Having shown that using the $\omega$-ordering in the NAT basis is optimal for the description of broad resonances, we now highlight the fact that the profiles of the energy and width in Figs.~\ref{fig_4H_E_NAT} and \ref{fig_4H_G_NAT} suggest a simple and efficient truncation scheme. 
Indeed, these profiles reveal that, even in light unbound nuclei, NAT orbitals can play the role that mean-field orbitals do in larger systems. 
This means that even if many-body entanglement obeys a volume law in nuclei, the eigenvalues of NAT orbitals, essentially associated with occupations of the mean-field orbitals, should roughly follow a decreasing exponential distribution. 
In all past and present applications considered, this was the case regardless of whether eigenvalues were ordered by their real part or magnitude. 
As an example, in Fig.~\ref{fig_4H_Rew_NAT}, we show the distribution of the real part of the eigenvalues of the neutron one-body density matrix for the $J^\pi = 2^-$ state of \isotope[4]{H}, ordered by real part. 

Given the distribution of the NAT occupations, it is natural to truncate the NAT basis by removing all orbitals that satisfy:
\begin{equation}
  \text{Re}(\omega) < \eta
  \label{eq_eta_NAT}
\end{equation}
where $\eta > 0$ is a small number controlling the error on the occupations. 
When using the $\omega$-ordering, this is equivalent to stopping the calculation at a given number of orbitals $N_\text{orb}$. 
We have checked that this truncation on the magnitude of the occupation $\omega$ did not change the results significantly.

\begin{figure}[h!]
  \centering
  \includegraphics[width=\linewidth]{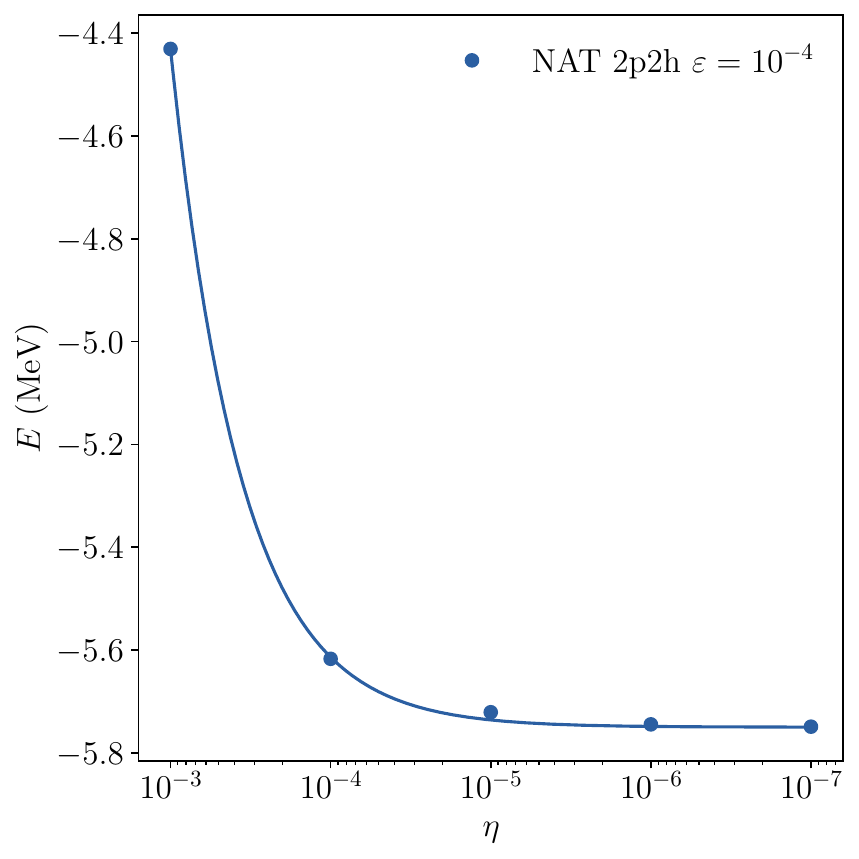}
  \caption{Energy of the $J^\pi = 2^-$ state of $\isotope[4]{H}$ as a function of the NAT truncation $\eta$.}
  \label{fig_4H_E_eta}
\end{figure}

\begin{figure}[h!]
  \centering
  \includegraphics[width=\linewidth]{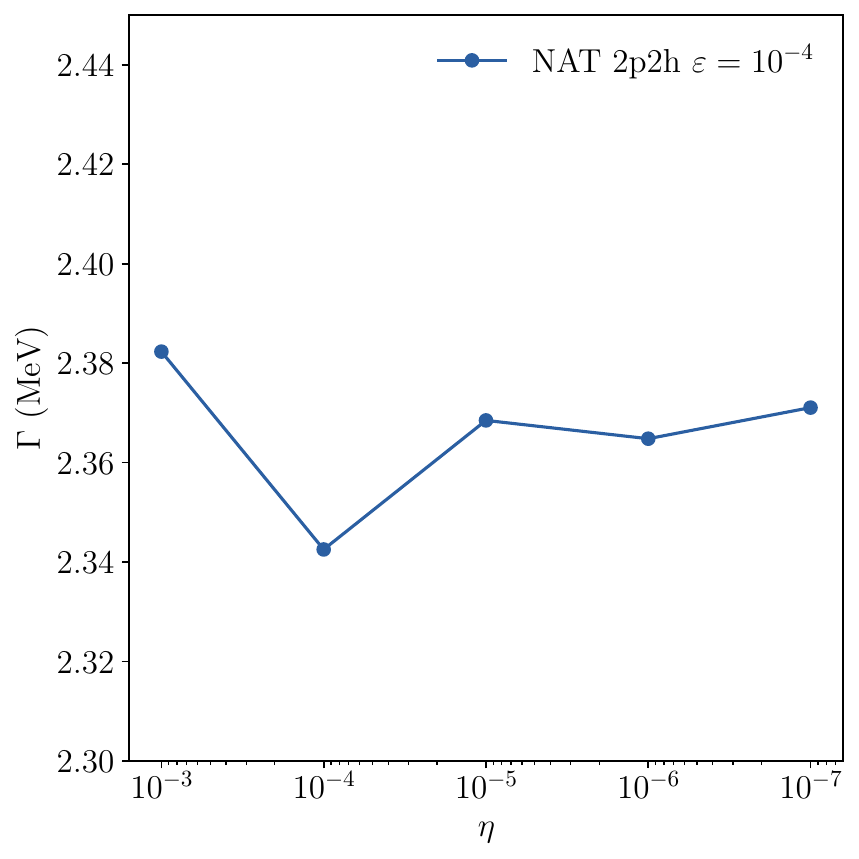}
  \caption{Width of the $J^\pi = 2^-$ state of $\isotope[4]{H}$ as a function of the NAT truncation $\eta$.}
  \label{fig_4H_G_eta}
\end{figure}

In Figs.~\ref{fig_4H_E_eta} and \ref{fig_4H_G_eta} we show the energy and width, respectively, of the $J^\pi = 2^-$ state of \isotope[4]{H} as a function of the NAT truncation $\eta$.
As expected, the near-exponential decrease of the energy with respect to the number of NAT orbitals observed in Fig.~\ref{fig_4H_E_NAT} translates into a near-exponential convergence with $\eta$, providing it is sufficiently small. 
In practice, we found that convergence is reached for $\eta \approx 10^{-6}$ regardless of the targeted state.

While the width shows some variations with $\eta$, compared to the relatively large value of the width in the present case, these are negligible. 
The NAT truncation happens to be so efficient at reducing the computational cost without losing important information that it allows us to turn a decent low-fidelity calculation with particle-hole truncations, into a high-fidelity one in which few or no truncations are applied. 
We will show in the next section how this truncation, combined with the other truncation schemes presented earlier, allows us to extend the reach of the G-DMRG method well beyond what was possible in the past.

\section{Demonstration in few- and many-body systems}
\label{sec_results}

The tests presented in the previous sections showed that, for resonances, the most important convergence factor is the G-DMRG truncation $\varepsilon$, followed by the size of the reference space, and finally the p-h truncation. 
To avoid stability issues and lower the cost of calculations, we also found that a truncation $\kappa = \varepsilon/10$ can safely be applied. 
In addition, we noticed that, in practice, it is preferable to limit the p-h truncation in the original basis to 2p2h to preserve stability, use the result obtained to generate the NAT basis, truncate it, and then remove the p-h truncation to the extent that it is computationally feasible.

Assuming that a stable single-particle basis has been found using the two orbital orderings presented in Sec.~\ref{sec_ordering}, the quasi-optimal convergence path for the G-DMRG method can thus be summarized as follows: 
\begin{enumerate}
  \item Set a minimal initial reference space.
  \item Extrapolate 2p2h results in the limit $\varepsilon \to 0$, with $\kappa = \varepsilon/10$ at most.
  \item Enlarge the reference space and repeat point 2. If the result is not converged, repeat point 3.
  \item Generate the NAT basis.
  \item Converge results with respect to the p-h truncation.
\end{enumerate}

In this section, we apply this recipe on progressively more challenging states and show that, despite using restricted model spaces and modest computing resources, we are able to obtain a convergence pattern in each considered case. 
In the future, further numerical optimizations will be necessary to fully leverage the developments presented in this work.

We note that in all calculations presented below, we applied a stabilizing truncation $\kappa = \varepsilon/10$ and the NAT basis was generated with a truncation $\eta = 10^{-6}$. 
For reference, we provided the NCGSM results with a 2p2h truncation in systems with $A < 6$ where calculations were feasible. 
The difference observed with our direct calculations with a 2p2h truncation provide an estimate of the error due to the DMRG truncation, the size of the reference space, and the orbital ordering.

\subsection{$J^\pi = {3/2}^-$ one-neutron resonance in $\isotope[5]{He}$}

We first consider the $J^\pi = {3/2}^-$ ground state of $\isotope[5]{He}$. 
As mentioned in Sec.~\ref{sec_id}, this is a single-particle $p$-wave resonance with a ratio $|\Gamma/(2 S_{n})| \approx 0.44$, meaning it is a typical resonance. 
Many high-precision \textit{ab initio} studies already exist for this state and it is not our intent to compare our results with these works.

It was first shown that a renormalized two-body interaction cannot reproduce $n$-\isotope[4]{He} phase shifts~\cite{quaglioni08_755}, and later that the inclusion of three-body (3N) forces considerably reduced the discrepancy~\cite{nollett07_944}. 
In Ref.~\cite{papadimitriou13_441}, it was observed that the width was indeed sensitive to the renormalization of nuclear forces and hence how $3N$ were effectively included or not. 
This sensitivity to $3N$ force was then used in Ref.~\cite{lynn16_2018} to find forces that could reproduce light nuclei and nuclear matter properties simultaneously.
The importance of many-body correlations in \isotope[5]{He} and their impact on phase shifts was studied in Ref.~\cite{lazauskas18_2420}, and more recently the problem of the reproduction of $p$-wave phase shifts in high-precision calculation was revisited using effective field theory~\cite{bagnarol23_3026}.

\begin{figure}[h!]
  \centering
  \includegraphics[width=\linewidth]{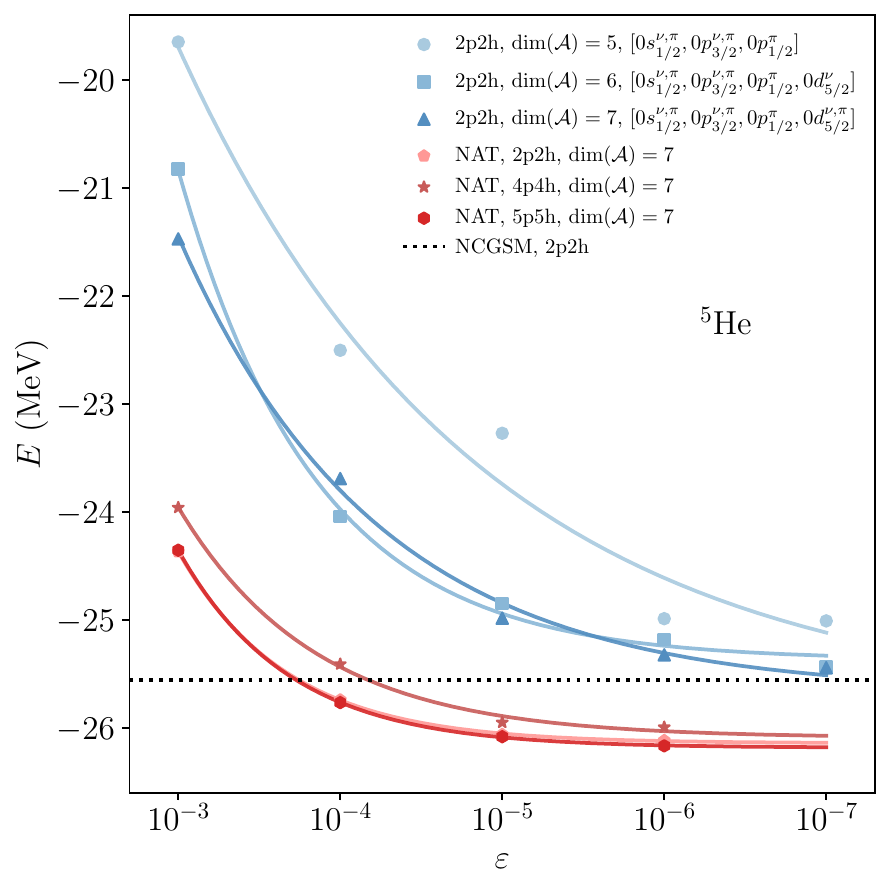}
  \caption{Energy of the $J^\pi = {3/2}^-$ state of $\isotope[5]{He}$ as a function of the G-DMRG truncation $\varepsilon$.}
  \label{fig_5He_E_final}
\end{figure}

\begin{figure}[h!]
  \centering
  \includegraphics[width=\linewidth]{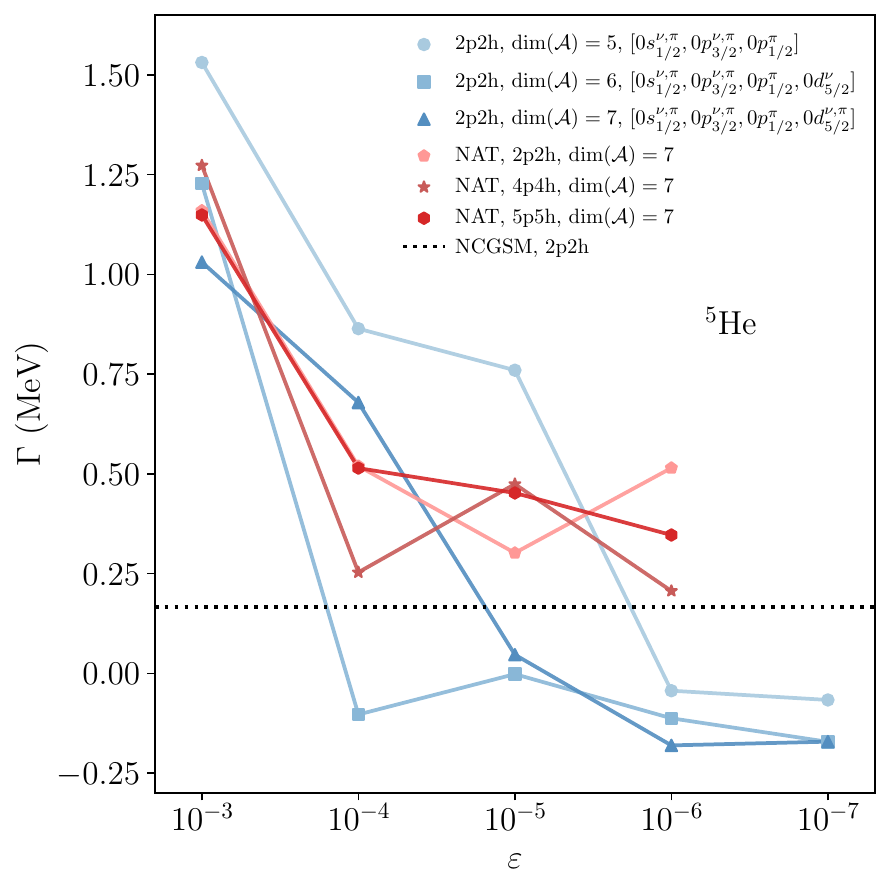}
  \caption{Width of the $J^\pi = {3/2}^-$ state of $\isotope[5]{He}$ as a function of the G-DMRG truncation $\varepsilon$.}
  \label{fig_5He_G_final}
\end{figure}

Given that we use renormalized two-body forces in this work, we do not expect the present results to closely match experiment. 
Results for the energy and width of the $J^\pi = {3/2}^-$ of $\isotope[5]{He}$ are shown in Figs.~\ref{fig_5He_E_final} and \ref{fig_5He_G_final}, respectively, as a function of the DMRG truncation.
We first considered a minimal reference space including the $0s_{1/2}^{\nu,\pi}$, $0p_{3/2}^{\nu,\pi}$, and $0p_{1/2}^{\pi}$ orbitals, which we eventually extended by adding the neutron and proton $0d_{5/2}$ orbitals. 
We applied a particle-hole truncation of 2p2h and used the $E$-ordering. 
Using the most converged result, we generated a NAT basis and switched to $\omega$-ordering, kept the same number of states in the reference space, and progressively removed the particle-hole truncation. 

These calculations were performed in a model space including partial waves up to $l_\text{max} = 4$, with the neutron $l=0$ and 1 partial waves expressed in the Berggren basis, and all others in the HO basis with $N_\text{max} = 10$ and $b_\text{HO} = 1.6 \, \text{fm}$. 
Both the $s_{1/2}^{\nu}$ and $p_{1/2}^\nu$ contours were taken along the real axis in the complex-momentum plane with a cutoff of $4.0 \, \text{fm}^{-1}$, while the $p_{3/2}^\nu$ contour was formed by three segments defined by the points $k_0=0.0$, $k_1 = 0.2-i0.07$, $k_2 = 0.5$, and $k_4 = 4.0$ (all in $\text{fm}^{-1}$). 
All three contours were discretized into 30 scattering states. 
In addition, we included the $0s_{1/2}^{\nu}$ and $0p_{3/2}^\nu$ poles as bound and resonance states, respectively, for a total of 160 single-particle states.

In all calculations we obtain a near-exponential convergence pattern. 
There are some unstabilities in the smallest model space which are likely due to the reference state being a poor approximation of the physical state. 
As expected, we observe a significant gain in energy when using the NAT basis, but there is almost no gain from going beyond the 2p2h truncation. 
To further converge this calculation, we would need to consider larger reference and model spaces.

The width also exhibits convergence, albeit not near-exponential nor smooth against the DMRG truncation. 
This behavior is likely due to the limited size of our model space, which limits our ability to accurately reproduce the width. 
Nevertheless, we see that calculations in the original basis, even though they are higher in energy, lead to small and even negative widths due to the lack of convergence, while those in the NAT basis remain positive and within a reasonable range.

\subsection{$J^\pi = 0^+$ and $2^+$ two-neutron states in $\isotope[6]{He}$}

As a next step, we consider the first $J^\pi = 0^+$ and $2^+$ states of $\isotope[6]{He}$. 
The $J^\pi = 0^+$ ground state is weakly bound and exhibits a two-neutron halo structure. 
This is an ideal situation to test how our approach handles the emergence of few-body degrees of freedom from the many-body dynamics in the absence of decay. 
The first $2^+$ state is, however, a resonance with a ratio of about $|\Gamma/(2 S_{n})| = |\frac{0.113}{(2 \times 0.087)}| \approx 0.65$, making its asymptotic behavior more challenging to describe than that of the $J^\pi = {3/2}^-$ ground state of $\isotope[5]{He}$.

As for \isotope[5]{He}, several high-precision \textit{ab initio} studies of the low-lying states of $\isotope[6]{He}$ are available in the literature. 
We mention here quantum Monte Carlo~\cite{pudliner95_1636,pieper04_2711} and no-core shell model~\cite{navratil01_2493,caurier06_1924} (NCSM) results. 
The NCSM was also used with the Lorentz integral transform to calculate the response function~\cite{bacca02_2497,bacca04_2496}, and extended using the resonating group method to handle the continua between \isotope[4]{He} and the neutrons~\cite{romero14_1583,romero16_3287}. 
More recently, natural orbitals techniques were used to improve the convergence of the NCSM in \isotope[6]{He}~\cite{fasano22_3285}.

All our G-DMRG calculations were performed in the same model space as for \isotope[5]{He}, except that the $p_{3/2}^\nu$ contour was taken along the real axis in the momentum plane for the $0^+$ state, and along a contour defined by the points $k_0=0.0$, $k_1 = 0.21-i0.1$, $k_2 = 0.5$, and $k_4 = 4.0$ (all in $\text{fm}^{-1}$) for the $2^+$ state. 
In addition, we note that for results with $\text{dim}(\mathcal{A})$ we applied a 3p3h truncation-at-construction starting at the $0p_{1/2}^\pi$ orbital.

\begin{figure}[h!]
  \centering
  \includegraphics[width=\linewidth]{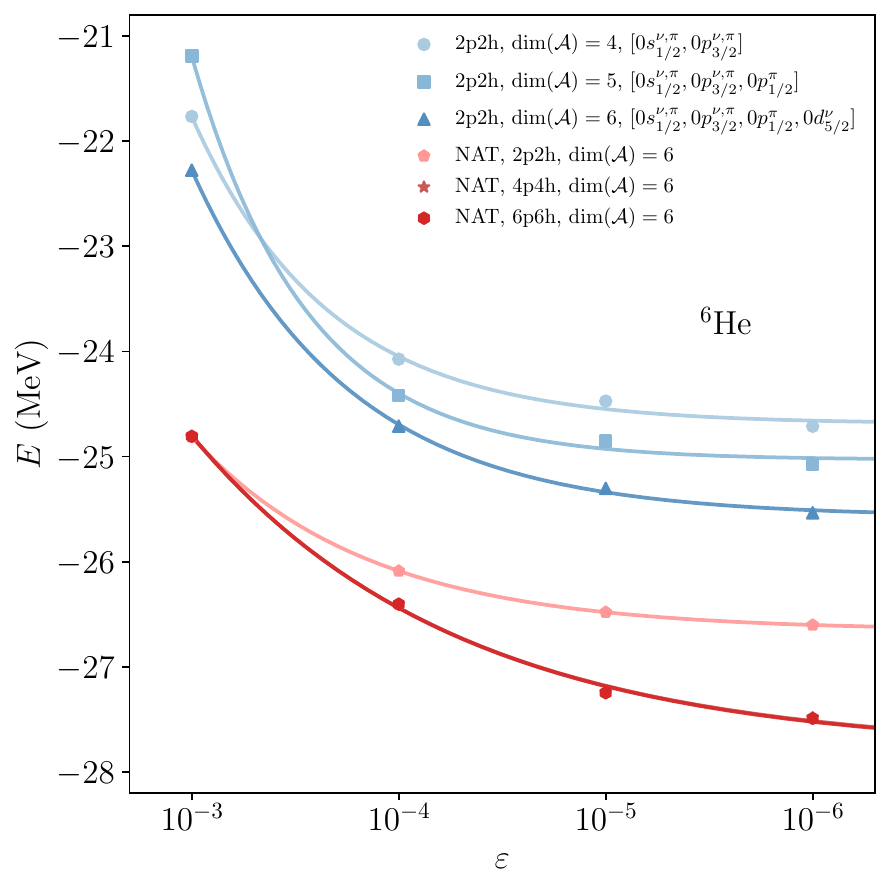}
  \caption{Energy of the $J^\pi = {0}^+$ state of $\isotope[6]{He}$ as a function of the G-DMRG truncation $\varepsilon$.}
  \label{fig_6He_E_final_0+}
\end{figure}

As for \isotope[5]{He}, we first considered a minimal space which we progressively extended to include the $0s_{1/2}^{\nu,\pi}$, $0p_{3/2}^{\nu,\pi}$, $0p_{1/2}^{\pi}$, and $0d_{5/2}^{\nu}$ orbitals, and solve with a 2p2h truncation. 
We were unable to enlarge the reference space further at this stage. 
After generating the NAT basis, we kept the same number of reference orbitals and progressively removed the particle-hole truncation. 
As shown in Fig.~\ref{fig_6He_E_final_0+}, all curves show near-exponential convergence as a function of the DMRG truncation parameter $\varepsilon$. 
We note that the result in the NAT basis with a 4p4h truncation is indistingushable from the one without any particle-hole truncation (6p6h). 

\begin{figure}[h!]
  \centering
  \includegraphics[width=\linewidth]{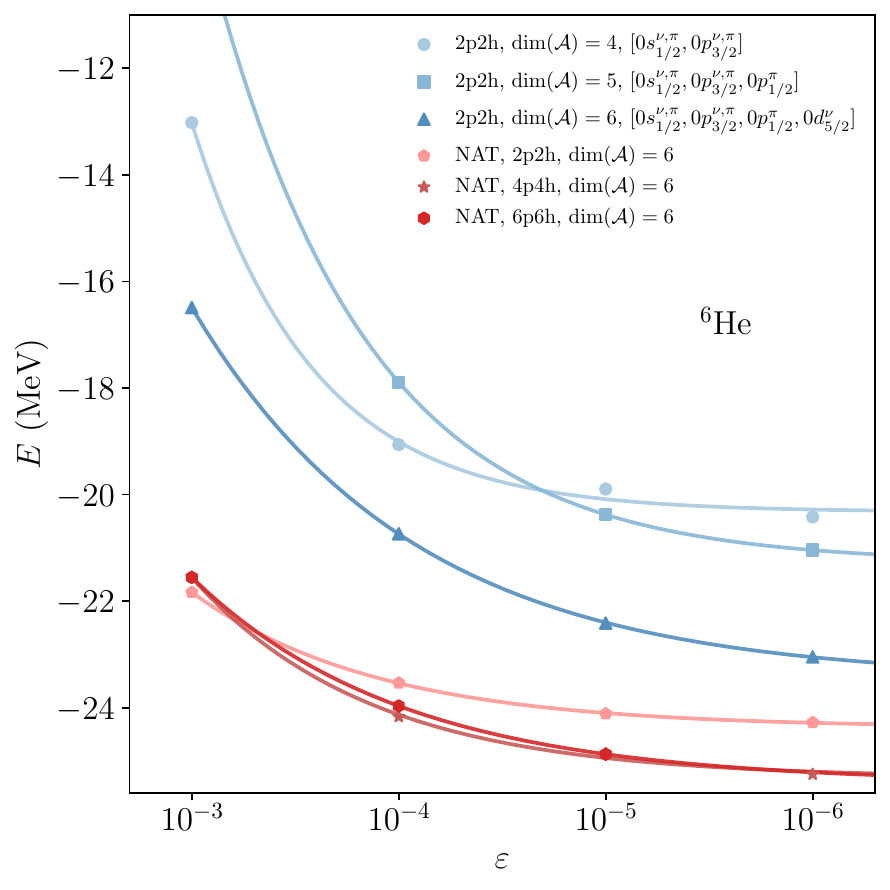}
  \caption{Energy of the $J^\pi = {2}^+$ state of $\isotope[6]{He}$ as a function of the G-DMRG truncation $\varepsilon$.}
  \label{fig_6He_E_final_2+}
\end{figure}

The energy of the $2^+$ state, shown in Fig.~\ref{fig_6He_E_final_2+}, exhibits similar convergence, with a saturation of the energy obtained in the NAT basis with a 4p4h truncation. 
Using our most converged results, we obtain an excitation energy of $E_x(2^+) = 2.39 \, \text{MeV}$ instead of the experimental value $1.80 \, \text{MeV}$.

\begin{figure}[h!]
  \centering
  \includegraphics[width=\linewidth]{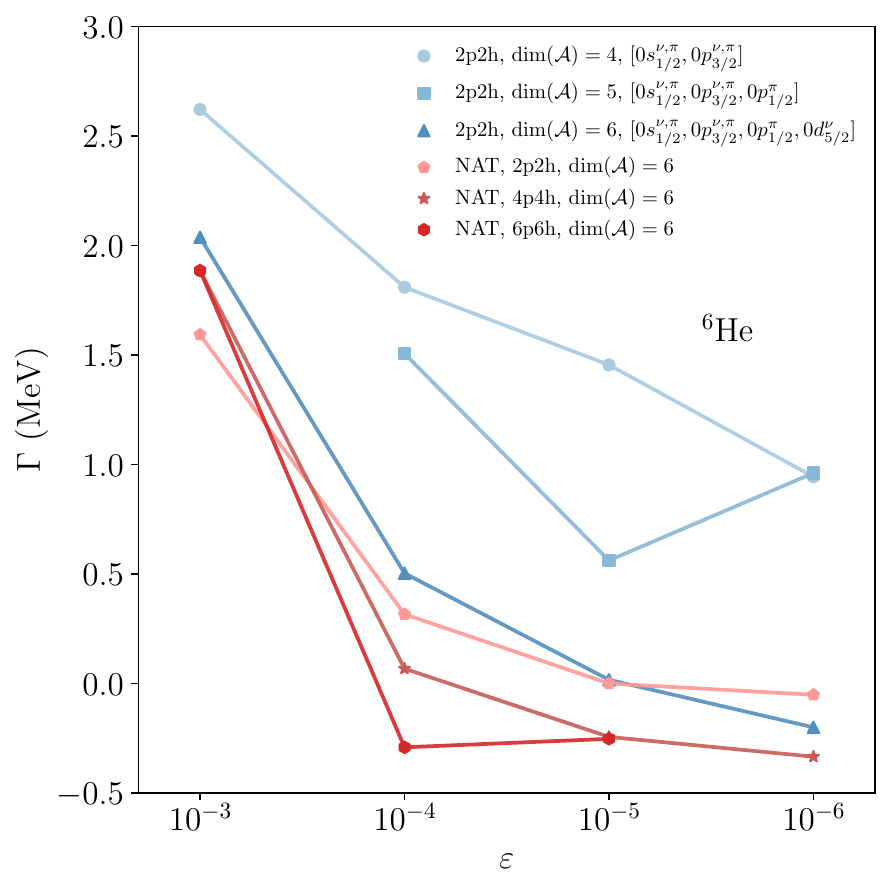}
  \caption{Width of the $J^\pi = {2}^+$ state of $\isotope[6]{He}$ as a function of the G-DMRG truncation $\varepsilon$.}
  \label{fig_6He_G_final_2+}
\end{figure}

The width, shown in Fig.~\ref{fig_6He_G_final_2+}, almost follows a near-exponential convergence profile once the $0d_{5/2}^{\nu}$ orbital is included. 
This improvement compared to \isotope[5]{He} is likely due to the relatively small width of this state ($\Gamma = 0.11 \, \text{MeV}$). 
In our calculations, we actually obtain a small but negative width, which is probably due to the limited model space used.

\subsection{$J^\pi = 2^-$ one-neutron resonance in $\isotope[4]{H}$}

To test our ability to handle broad resonances, we now apply our recipe to the $J^\pi = 2^-$ one-neutron resonance of $\isotope[4]{H}$. 
We recall that the isotope \isotope[4]{H} supports two nearly degenerate broad resonances with $J^\pi = 1^-,2^-$. 
According to state-of-the-art complex-scaled Faddeev-Yakubovsky calculations~\cite{lazauskas19_2363}, they lie about $1.1 \, \text{MeV}$ above the one-neutron particle emission threshold, with energies around $E \approx -7.3 \, \text{MeV}$ and widths of about $\Gamma \approx 4.0 \, \text{MeV}$. 
Considerably smaller widths were obtained in Ref.~\cite{li21_2395}.

Our G-DMRG calculations were performed in a model space including partial waves up to $l_\text{max} = 4$, with the neutron $l=0$ and 1 partial waves expressed in the Berggren basis, and all others in a HO basis with $N_\text{max} = 12$ and $b_\text{HO} = 1.8 \, \text{fm}$. 
We further limited the $l \geq 3$ orbitals to $n_\text{max} = 1$. 
Both the $s_{1/2}^{\nu}$ and $p_{1/2}^\nu$ contours were taken along the real axis in the complex-momentum plane with a cutoff of $4.0 \, \text{fm}^{-1}$ and 30 scattering states for discretization. 
The $p_{3/2}^\nu$ contour was formed by three segments defined by the points $k_0=0.0$, $k_1 = 0.22-i0.1$, $k_2 = 0.5$, and $k_4 = 4.0$ (all in $\text{fm}^{-1}$) and discretized into 45 scattering states. 
In addition, we included the $0s_{1/2}^{\nu}$ and $0p_{3/2}^\nu$ poles as bound and resonance states, respectively, for a total of 166 single-particle states.

\begin{figure}[h!]
  \centering
  \includegraphics[width=\linewidth]{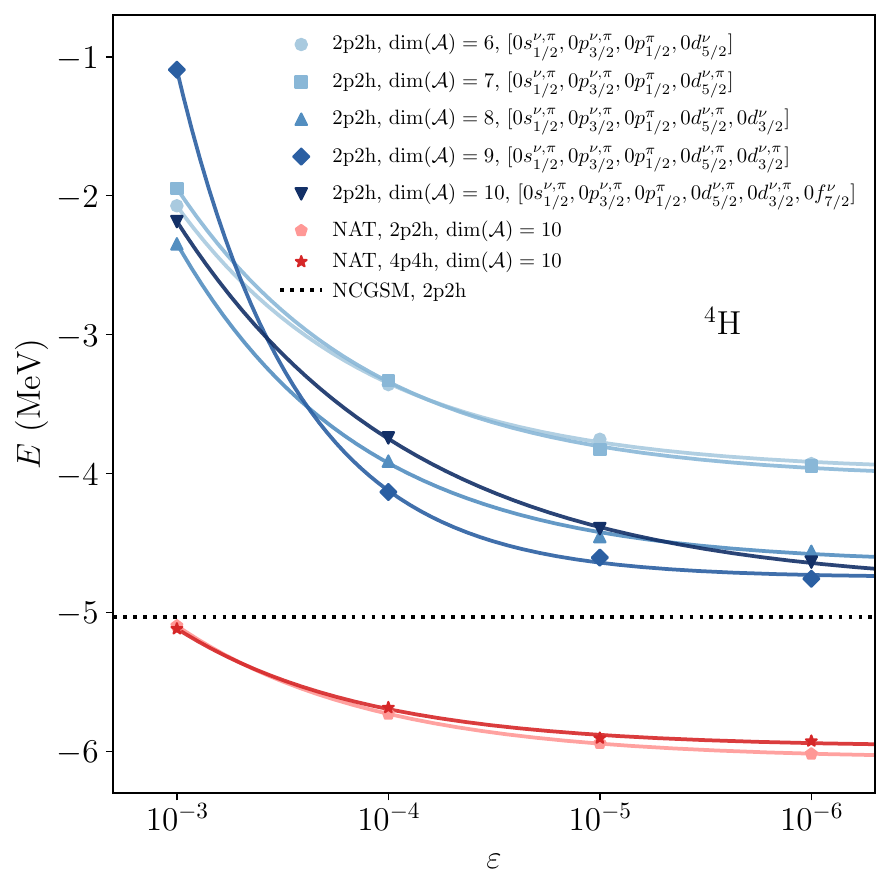}
  \caption{Energy of the $J^\pi = 2^-$ state of $\isotope[4]{H}$ as a function of the G-DMRG truncation $\varepsilon$.}
  \label{fig_4H_E_final}
\end{figure}

In Fig.~\ref{fig_4H_E_final}, we show the energy of the $J^\pi = 2^-$ state as a function of the G-DMRG truncation. 
The smallest reference space in which we could obtain stable results is composed of the $0s_{1/2}^{\nu,\pi}$, $0p_{3/2}^{\nu,\pi}$, $0p_{1/2}^{\pi}$, and $0d_{5/2}^{\nu}$ orbitals. 
We then increased the size of the reference space by adding one-by-one the $0d_{5/2}^{\pi}$, $0d_{3/2}^{\nu,\pi}$, and $0f_{7/2}^{\nu}$ orbitals. Only the addition of the $0d_{3/2}^{\nu}$ had a significant effect.

The NAT basis was generated from the most converged result and, using the $\omega$-ordering, we included in the new reference space the 10 orbitals with the largest occupations. 
We note that this translated into the inclusion of emergent neutron $0p_{1/2}$ and $1s_{1/2}$ NAT orbitals. 
This is quite interesting in light of the known parity inversion in \isotope[9]{He}, which is precisely driven by these two orbitals. 
In the NAT basis, the energy dropped by more than $1.0 \, \text{MeV}$ and was almost converged with a 2p2h truncation. 
This large relative drop reflects the suboptimal nature of the original basis. 
Our most converged value is $E = -5.97 \, \text{MeV}$, about $1.3 \, \text{MeV}$ above the result obtained in Ref.~\cite{lazauskas19_2363}, which can once again be attributed to the limited model space used in this work. 

\begin{figure}[h!]
  \centering
  \includegraphics[width=\linewidth]{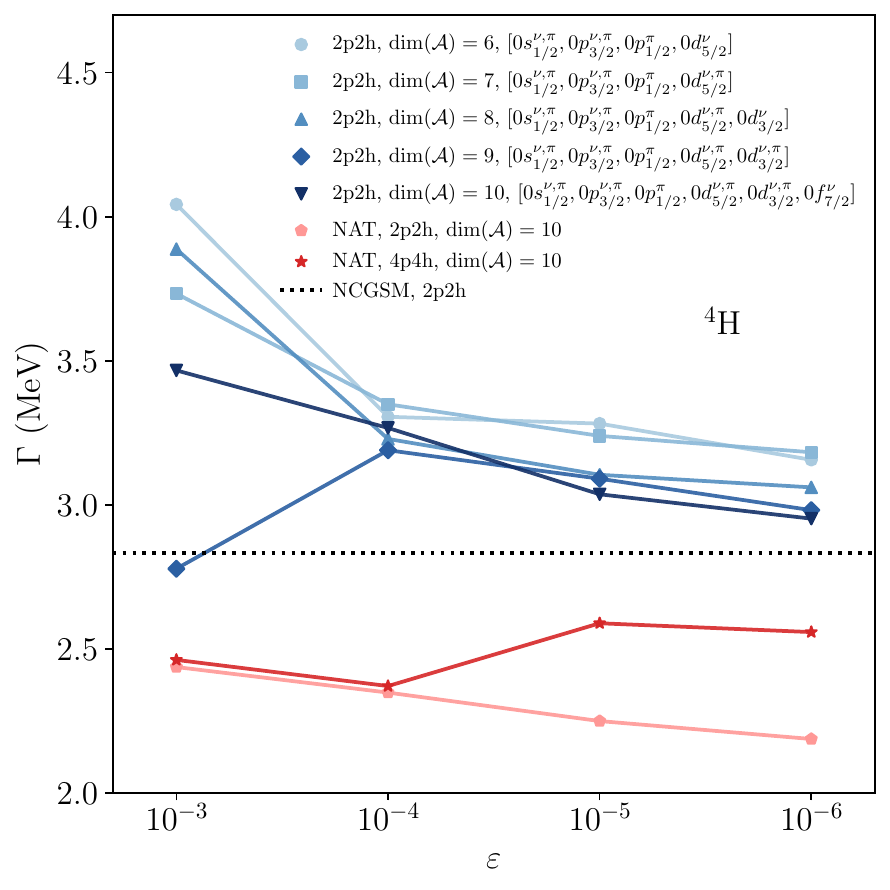}
  \caption{Width of the $J^\pi = 2^-$ state of $\isotope[4]{H}$ as a function of the G-DMRG truncation $\varepsilon$.}
  \label{fig_4H_W_final}
\end{figure}

The width, shown in Fig.~\ref{fig_4H_W_final}, behaves in a somewhat different way than in the previous cases. 
First, we note that regardless of the reference space used, we obtain a rather large width of about $3.25 \, \text{MeV}$ in the original basis, which drops to about $2.5 \, \text{MeV}$ in the NAT basis. 
This result is significantly larger than the one obtain in Ref.~\cite{li21_2395}, around $1.0 \, \text{MeV}$, and we believe that in a larger continuum space our value would grow closer to the $4.0 \, \text{MeV}$ width reported in Ref.~\cite{lazauskas19_2363}.

\subsection{$J^\pi = {1/2}^+$ two-neutron resonance in $\isotope[5]{H}$}

We finally test our \textit{ab initio} approach on the $J^\pi = {1/2}^+$ ground state of $\isotope[5]{H}$, a state that has never been calculated directly. 
To our knowledge, the only \textit{ab initio} results on this system are the complex-scaled Faddeev-Yakubovsky calculations in Ref.~\cite{lazauskas19_2363}. 
In that study, the system was confined using a fictitious five-body hyperradial potential, and its energy and width were calculated for different depths of the potential, and extrapolated in the physical limit using the analytic continuation
in a coupling constant method. 
These high-precision calculations estimated that the ground state of \isotope[5]{H} lies about $1.8 \, \text{MeV}$ above the $\isotope[3]{H}+n+n$ threshold, with absolute energy and width of about $E \approx -6.7 \, \text{MeV}$ and $\Gamma \approx 2.4 \, \text{MeV}$, respectively.

Here, we considered a limited model space including partial waves up to $l_\text{max} = 4$, with the neutron $l=0$ and 1 partial waves expressed in the Berggren basis, and all others in the HO basis with $N_\text{max} = 10$ and $b_\text{HO} = 1.5 \, \text{fm}$. 
Both the $s_{1/2}^{\nu}$ and $p_{1/2}^\nu$ contours were taken along the real axis in the complex-momentum plane with a cutoff at $4.0 \, \text{fm}^{-1}$ and discretized by 30 scattering states. 
The $p_{3/2}^\nu$ contour was formed by three segments defined by the points $k_0=0.0$, $k_1 = 0.20-i0.08$, $k_2 = 0.5$, and $k_4 = 4.0$ (all in $\text{fm}^{-1}$) and was discretized by 45 scattering states. 
In addition, we included the $0s_{1/2}^{\nu}$ and $0p_{3/2}^\nu$ poles as bound and resonance states, respectively.

\begin{figure}[h!]
  \centering
  \includegraphics[width=\linewidth]{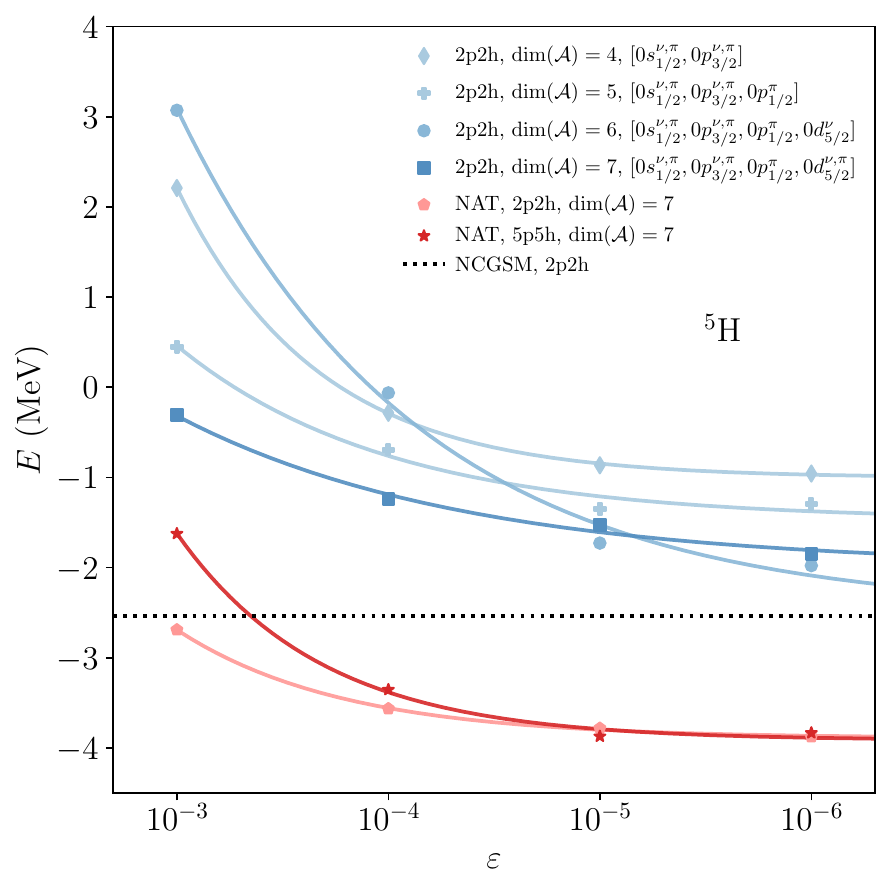}
  \caption{Energy of the $J^\pi = {1/2}^+$ state of $\isotope[5]{H}$ as a function of the G-DMRG truncation $\varepsilon$.}
  \label{fig_5H_E_final}
\end{figure}

The energy of the $J^\pi = {1/2}^+$ state is shown in Fig.~\ref{fig_5H_E_final} as a function of the DMRG truncation. 
We first considered a minimal reference space including the $0s_{1/2}^{\nu,\pi}$ and $0p_{3/2}^{\nu,\pi}$ orbitals, and then increased its size by progressively including the $0p_{1/2}^{\pi}$, $0d_{5/2}^{\nu}$, and $0d_{5/2}^{\pi}$ orbitals. 
The inclusion of the $0d_{5/2}^{\nu}$ orbital led to a different and somewhat less stable convergence pattern than for the other reference spaces. 
Interestingly, including the $0d_{5/2}^{\pi}$ orbital reestablished the convergence trend but did not change the final value significantly, suggesting convergence with the reference space size. 
We suspect that this unusual behavior is due to strong dineutron correlations, which require contributions from many higher partial waves to be properly described. 
Unfortunately, we were unable to increase the dimension further to estimate the impact of higher orbitals. 

From our results in the largest reference space considered, we generated the NAT basis and, using the $\omega$-ordering, changed the reference space to include the $0s_{1/2}^{\nu,\pi}$, $0p_{3/2}^{\nu,\pi}$, $0p_{1/2}^{\pi}$ orbitals, which were included before, and the emergent $0p_{1/2}^{\nu}$ and $1p_{3/2}^{\pi}$ orbitals. 
We recall that for \isotope[4]{H}, we had an emergent $1s_{1/2}^{\nu}$ orbital instead of a $1p_{3/2}^{\pi}$ one. 
It is unclear at present why we observe this difference. 
We then calculated the ground state with a 2p2h truncation, and finally removed the p-h truncation (5p5h), observing no significant difference in the final result, which indicates that we reached convergence in the NAT basis.

\begin{figure}[h!]
  \centering
  \includegraphics[width=\linewidth]{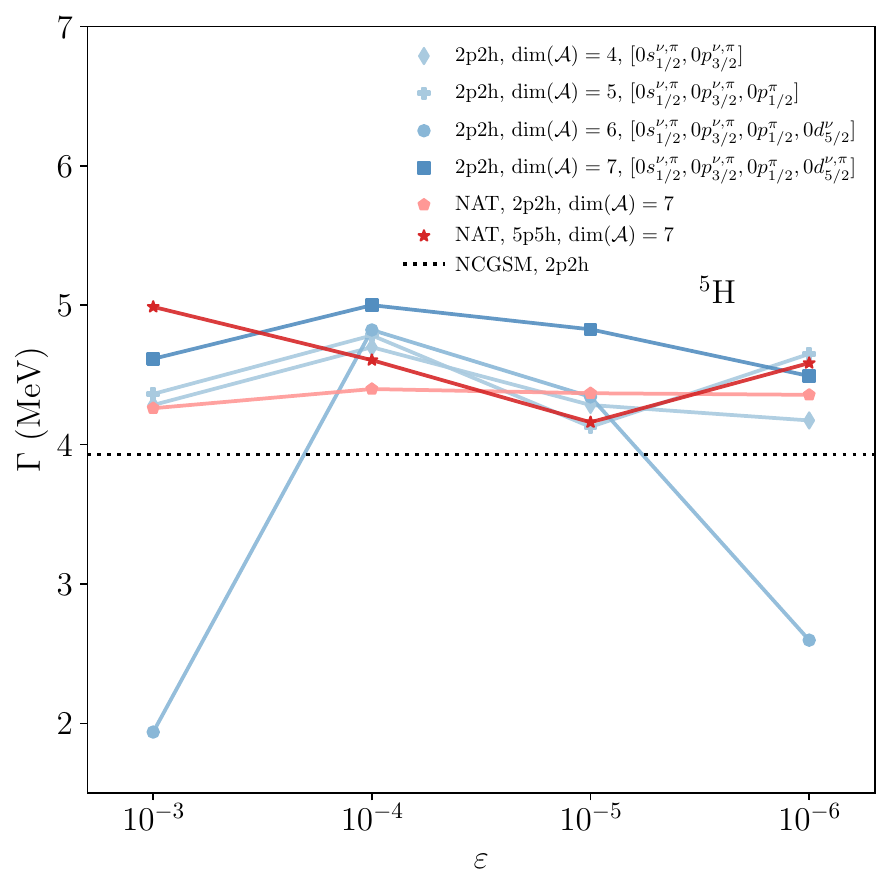}
  \caption{Width of the $J^\pi = {1/2}^+$ state of $\isotope[5]{H}$ as a function of the G-DMRG truncation $\varepsilon$.}
  \label{fig_5H_W_final}
\end{figure}

The width, shown in Fig.~\ref{fig_5H_W_final}, follows a different pattern. 
Except for the result using the reference space with $\text{dim}(\mathcal{A}) = 6$, where the $0d_{5/2}^{\nu}$ orbital is the last one included in the reference space, all other results are essentially constant with the DMRG truncation, giving a width of about $4.5 \, \text{MeV}$. 
In the future, we will attempt to describe the $J^\pi = {1/2}^+$ state in a larger model space to check whether our energy and width decrease and approach those reported in Ref.~\cite{lazauskas19_2363}.

\section{Conclusions}
\label{sec_conclusion}

In this work, we successfully extended the \textit{ab initio} G-DMRG approach to the description of broad many-body resonances. 
First, we developed a new truncation scheme that is applied during the construction of the reference space, allowing us to better approximate target states such as many-body resonances, for which a minimal reference space based on mean-field occupation numbers is insufficient.
Then, we studied the problem of orbital ordering when both the energy and width must be optimized, and calculations must be kept stable at every DMRG iteration. 
Using results from studies of entanglement in nuclei, we designed an ordering scheme that leverages the proton-neutron factorization in the nuclear many-body problem and the emergent shell structure in nuclei. 
This new ordering scheme stabilized our calculations of broad resonances, allowing us to further improve the single-particle basis and switch back to our original ordering scheme based on energy considerations to then obtain better natural orbitals. 
We also demonstrated that continuum couplings increase entanglement in the many-body problem, and introduced a new truncation scheme at the level of the reduced density matrix that eliminates negligible contributions not captured by the regular DMRG truncation. This scheme further stabilizes the renormalization, and greatly accelerates calculations. 
To illustrate this finding, we showed that we are now able to approximate the ground-state energy of \isotope[12]{C} using modest computational resources.
We then showed how natural orbitals can be used efficiently to describe broad resonances, by introducing a new ordering scheme and redefining the reference space based on occupation considerations, which then allows us to truncate low-occupation natural orbitals so that typical DMRG many-body truncations can be lifted. 
Finally, we proposed a recipe combining all of the previous techniques to safely converge \textit{ab initio} G-DMRG calculations of broad many-body resonances.

Using limited computational resources, we demonstrated control of the renormalization and the emergence of a convergence pattern for the $J^\pi = {3/2}^-$ ground state of \isotope[5]{He}, the first $J^\pi = {2}^+$ state of \isotope[6]{He}, and the $J^\pi = {2}^-$ ground state of \isotope[4]{H}. 
We also provided the first direct \textit{ab initio} calculation of the $J^\pi = {1/2}^+$ ground state of \isotope[5]{H}. 

In the future, we will address our most pressing computational bottlenecks to provide systematic results in light exotic nuclei, and test our method in proton-rich systems where the Coulomb interaction makes calculations more challenging. 
In addition, we will consider calculating additional observables such as radii, and start probing the impact of three-body body forces in extreme $N/Z$ condition using normal ordering techniques.

\section*{Data availability statement}
The data that support the findings of this article will be made available upon request.

\section*{acknowledgments} 

This material is based upon work supported by the U.S. Department of Energy, Office of Science, Office of Nuclear Physics under awards No. DE-SC0013617 (FRIB Theory Alliance), No. DE-SC0023516, and No. DE-SC0023175 (SciDAC-5 NUCLEI Collaboration), and by the National Science Foundation under award No. PHY-2238752 (CAREER).

\appendix

\section{Evolution of the single-particle energy and width with the position of its associated $S$ matrix pole}
\label{app_cx_energy}

To help the reader picture the relation between complex energies the poles of the $S$ matrix in the quasi-stationary formalism, 
this appendix discusses how the energy and width of a single-particle Gamow state evolve as the position of its momentum changes in the momentum plane.

Here, we only consider poles of the single-particle $S$-matrix along the positive imaginary axis (bound states) and in the \nth{4} quadrant (decaying resonances). 
These have momentum ${ k = \kappa - i\gamma }$ with ${ \kappa, \gamma > 0 }$. 
In polar coordinates, the momentum can be written $k = |k| e^{i\theta}$, where $|k| = \sqrt{\kappa^2 + \gamma^2}$ and $\tan(\theta) = -\gamma/\kappa$.

\begin{figure}[h!]
  \centering
  \includegraphics[width=\linewidth]{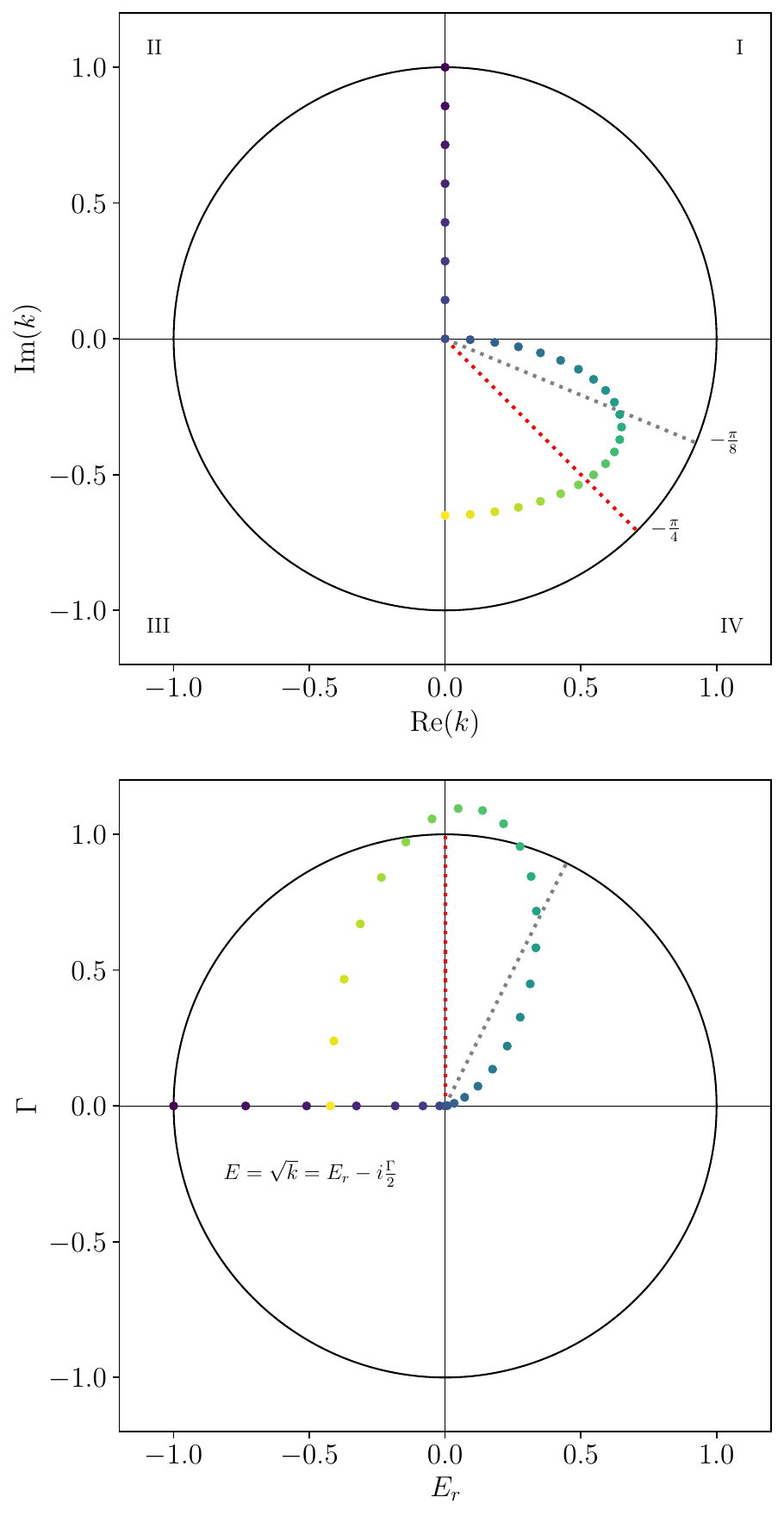}
  \caption{Evolution of the momentum (panel (a)) and corresponding energy (panel (b)) in the complex momentum plane and $(E_r, \Gamma)$ plane, respectively.}
  \label{fig_cx_k_E}
\end{figure}

The energy can be directly obtained from the momentum using ${ E = (\hbar^2 \tilde{k}^2)/(2m) }$, but for simplicity we will assume that $E = k^2 = E_r - i\Gamma/2$. 
To gain intuition about how energies and widths relate to the poles of the $S$-matrix, we consider the typical trajectory that the momentum $k$ of a particle would have in the momentum plane if the particle was initially bound in some attractive potential with a barrier, the potential was made shallower until the particle can decay by leaking through the barrier, and finally the barrier was removed altogether.
In this scenario, the state of the particle evolves from a bound state to a narrow resonance, a broad resonance, and ends up as an antibound state. 
The trajectory of the momentum for this scenario is shown in panel (a) of Fig.~\ref{fig_cx_k_E}, and the corresponding energy-width trajectory is shown in panel (b) in the $(E_r,\Gamma)$ plane instead of the complex $E$-plane. 

The line marking the limit between narrow and broad resonances is given by the condition $E_r = \Gamma/2$. 
Given that $E = E_r - i\Gamma/2 = |k|^2 e^{2i\theta}$, we immediately see that this condition is satisfied when $\tan(2\theta) = -1$, namely when $\theta = -\pi/8$. 
In the complex $E$-plane, $E = |E| e^{i\phi}$ where $\tan(\phi) = \text{Im}(E)/\text{Re}(E) = -\Gamma/(2E_r)$ and $\phi = 2\theta$.

To find the corresponding line in the $(E_r, \Gamma)$ plane, we note that $\tan(\varphi) = \Gamma/E_r = -2\tan(\phi)$, where $\varphi$ is the rescaled angle. 
It follows that the narrow-to-broad resonance line is at $ \varphi = \tan^{-1}(-2\tan(2\theta)) \approx 63.43\degree $ in the $(E_r, \Gamma)$ plane.

Finally, we can consider the limit between broad and subthreshold resonances given by the condition $E_r = 0$. 
Following the same reasoning, we find $\theta = -\pi/4$ and, of course, $\varphi = \pi/2$.

\section{Complex-symmetric density matrices}
\label{app_cx_density}

In non-Hermitian quantum mechanics, Gamow states are normalized according to the rigged Hilbert space metric. 
In practice, this is done using the so-called c-product~\cite{moiseyev11_b72}. 
For example, if we consider a pure state $\ket{\psi}$ with components $(a,b)$ in some representation, the norm is given by:

\begin{equation}
  \braket{\tilde{\psi}|\psi} = 
  \begin{pmatrix}
    (a^*)^*  \\
    (b^*)^*
  \end{pmatrix}
  \begin{pmatrix}
    a & b
  \end{pmatrix} 
  = a^2 + b^2 = 1
  \label{eq_norm_RHS}
\end{equation}
where the tilde represents the time-reversal operation. 
The density matrix is obtained using the outer product instead:
\begin{equation}
  \rho = \ket{\tilde{\psi}}\bra{\psi}
  \begin{pmatrix}
    a \\
    b
  \end{pmatrix}
  \begin{pmatrix}
    (a^*)^*  & (b^*)^*
  \end{pmatrix} = 
  \begin{pmatrix}
    a^2  & ab \\
    ba  & b^2
  \end{pmatrix}
  \label{eq_rho_2x2_cprod}
\end{equation}
For comparison, in the Hermitian case, where the norm is given by $|a|^2 + |b|^2 = 1$, we would have obtained a Hermitian density matrix $\rho = \rho^\dagger$ where:

\begin{equation}
  \rho = 
  \begin{pmatrix}
    |a|^2  & ab^* \\
    ba^*  & |b|^2
  \end{pmatrix}
  \label{eq_rho_2x2_cprod_Hermitian}
\end{equation}

In the case of a mixed state $\psi = p_1 \ket{\psi_1} + p_2 \ket{\psi_2}$, where $p_1 + p_2 = 1$ are probability amplitudes, the density matrix is obtained as the weighted sum of the density matrices $\rho_1 = \ket{\tilde{\psi_1}}\bra{\psi_1}$ and $\rho_2 = \ket{\tilde{\psi_2}}\bra{\psi_2}$ associated with the states $\ket{\psi_1}$ and $\ket{\psi_2}$, respectively:

\begin{align}
  \rho 
  &= p_1 + \rho_1 + p_2 + \rho_2 \\
  &= 
  \begin{pmatrix}
    p_1 a_1^2 + p_2 a_2^2  & p_1 a_1 b_1 + p_2 a_2 b_2 \\
    p_1 a_1 b_1 + p_2 a_2 b_2  & p_1 b_1^2 + p_2 b_2^2
  \end{pmatrix}
  \label{eq_rho_2x2_mixed}
\end{align}
We note that if we had expressed the mixed-state density matrix in the (non-orthogonal) basis of the Gamow states $\ket{\psi_1}$ and $\ket{\psi_2}$, the density would be:

\begin{equation}
  \rho =
  \begin{pmatrix}
    p_1 + p_2 S_{1,2} S_{2,1}  & S_{1,2} \\
    S_{2,1}  & p_2 + p_1 S_{2,1} S_{1,2}
  \end{pmatrix}
  \label{eq_rho_2x2_mixed_bis}
\end{equation}
where $S_{i,j} = \braket{\tilde{\psi_i}|\psi_j} \neq 0$ are the transition amplitudes, \textit{i.e.} the probability amplitudes for the system to transition from one state to the other through the continuum. 
Obviously, the trace cannot be one in this non-orthogonal basis, but if these amplitudes are small, $\rho_{i,i} \approx p_i$, and $\trace(\rho) \approx 1$. 
It is then clear that the off-diagonal elements $S_{i,j}$ control the strength of continuum couplings and are responsible for decay. 
In practice, this still holds to a good extent in the general case.

Moving to a general complex-symmetric density matrix of dimension 2, we can always write:

\begin{equation}
  \rho = 
  \begin{pmatrix}
    \rho_{1,1}  & \rho_{1,2} \\
    \rho_{2,1}  & \rho_{2,2}
  \end{pmatrix}
  \label{eq_rho_2x2}
\end{equation}
where $\trace(\rho) = 1$, $\rho_{2,2} = 1 - \rho_{1,1}$, and $\rho_{2,1} = \rho_{1,2}$ instead of $\rho_{2,1} = \rho_{1,2}^*$ in the Hermitian case. 
All matrix elements can be complex numbers.

The eigenvalues $\lambda_{1,2}$ of the density matrix in Eq.~\eqref{eq_rho_2x2} are solutions of the characteristic equation $\det(\rho - \lambda I) = 0$, or equivalently:
\begin{equation}
  \lambda^2 - \trace(\rho) \lambda + \det(\rho) = 0
  \label{eq_rho_2x2_eig_eq}
\end{equation}
Solving this equation gives:
\begin{align}
  \lambda_{1,2} 
  &= \frac{1}{2} \trace(\rho) \pm \frac{1}{2} \sqrt{ {\trace(\rho)}^2 - 4 \det(\rho)} \\
  &= \frac{1}{2} \pm \frac{1}{2} \sqrt{ 1 - 4 \det(\rho)}
  \label{eq_rho_2x2_eig}
\end{align}
where we used the fact that $\trace(\rho) = 1$. 
The determinant is given by
\begin{equation}
  \det(\rho) = \rho_{1,1}(1-\rho_{1,1}) - \rho_{1,2}^2
  \label{eq_rho_det}
\end{equation}
Alternatively, the eigenvalues can be expressed in terms of the components:
\begin{equation}
  \lambda_{1,2} = \frac{1}{2} \pm \frac{1}{2} \sqrt{ {(2\rho_{1,1} - 1)}^2 + 4 \rho_{1,2}^2}
  \label{eq_rho_2x2_eig_bis}
\end{equation}
To understand the structure of the spectrum, we fix the off-diagonal matrix element $\rho_{1,2}$, which is roughly equivalent to fixing the strength of continuum couplings, and look for a critical point where the derivative of the determinant vanishes as a function of $\rho_{1,1}$:
\begin{equation}
  \frac{d \det(\rho)}{d \rho_{1,1}} = \frac{d}{d \rho_{1,1}} (\rho_{1,1}(1-\rho_{1,1}) - \rho_{1,2}^2) = -2\rho_{1,1} + 1
  \label{eq_rho_det_max}
\end{equation} 
We find a critical point for $\rho_{1,1} = 1/2$, which is the point of maximum entanglement. 
At this point, the eigenvalues are:

\begin{equation}
  \lambda_{1,2}^\text{crit} = \frac{1}{2} \pm \rho_{1,2}
  \label{eq_rho_2x2_eig_critical}
\end{equation}
In the Hermitian case, the eigenvalues would be $\lambda_{1,2} = 1/2 \pm |\rho_{1,2}|$, but to preserve the trace, we must have $\rho_{1,2} = 0$. 
This is not the case in non-Hermitian quantum mechanics, where only $\Re{\rho_{1,2}} = 0$ is required. 
This means that, at the point of maximum entanglement, the eigenvalues of the density matrix can be complex numbers, and the imaginary part is entirely due to continuum couplings. 
Following Berggren's prescription, we interpret the real part of the eigenvalues of the density as occupation probabilities, and the magnitude of the imaginary part as uncertainties due to treating an inherently time-dependent phenomenon as quasi-stationary. 
We argue that this interpretation remains valid even outside of the critical point.

To understand the effect of continuum couplings on the density, we fixed the diagonal matrix elements by setting $\rho_{1,1} = 1$, and varied the value of $\rho_{1,2} = |\rho_{1,2}| e^{i\theta}$ numerically. 
The choice for $\rho_{1,2}$ is a matter of convenience as it does not affect the conclusion.

We considered three scenarios defined by $\theta = \pi/2$, $\pi/2-\pi/32$, and $\pi/4$, with $|\rho_{1,2}|$ increasing from 0 to 1. 
In the first scenario, $\rho_{1,2}$ is purely imaginary, with $\Im{\rho_{1,2}} \geq 0$, and thus satisfies the condition that preserves the trace at the critical point. 
In the second scenario, the situation is similar, except that $\rho_{1,2}$ acquires a small, positive real part that grows as $|\rho_{1,2}| \to 1$. 
This could, for example, simulate small numerical innacuracies or truncations in realistic calculations. 
Finally, in the third scenario, the real and imaginary parts of $\rho_{1,2}$ grow at the same rate. 
This could be a situation in which continuum couplings are extremely strong and the basis is becoming strongly non-orthogonal.

\begin{figure}[h!]
  \centering
  \includegraphics[width=\linewidth]{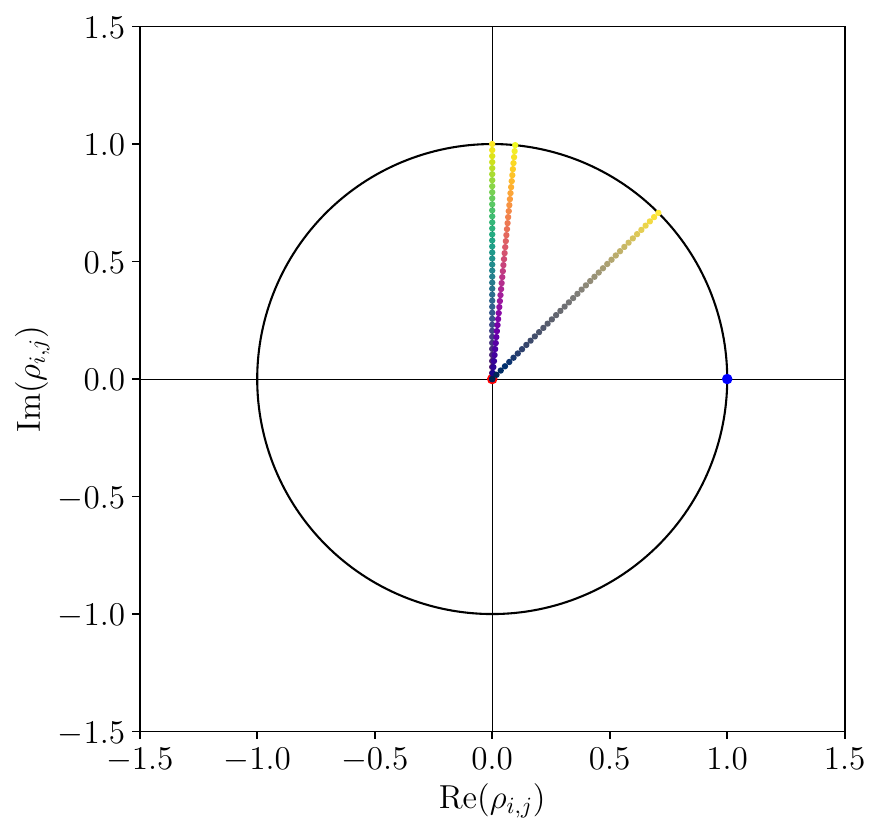}
  \caption{Evolution of the off-diagonal matrix elements $\rho_{1,2}$ in the complex plane for three different scenarios. The values of $\rho_{1,1} = 1$ and $\rho_{2,2} = 0$ are indicated by blue and red dots, respectively.}
  \label{fig_cx_symm_density_0}
\end{figure}

These three scenarios are shown in Fig.~\ref{fig_cx_symm_density_0}, where each trajectory in the complex plane is highlighted using a different colormap for convenience. 
The values of $\rho_{1,1} = 1$ and $\rho_{2,2} = 0$ are indicated by blue and red dots, respectively.

\begin{figure}[h!]
  \centering
  \includegraphics[width=\linewidth]{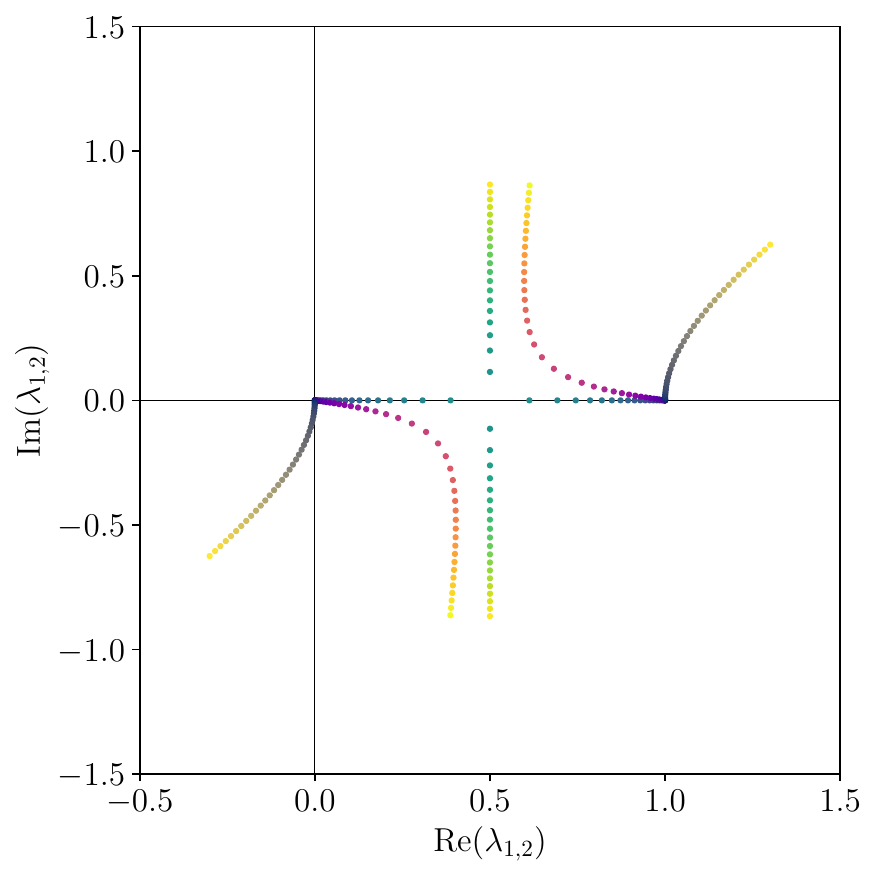}
  \caption{Evolution of the eigenvalues $\lambda_{1,2}$ in the complex plane for the three different scenarios shown in Fig.~\ref{fig_cx_symm_density_0}.}
  \label{fig_cx_symm_density_1}
\end{figure}

The associated eigenvalue trajectories are shown in Fig.~\ref{fig_cx_symm_density_1} using the same colormaps. 
In the first scenario, defined $\theta = \pi/2$, the two eigenvalues start at 0 and 1, respectively, and move along the real axis toward the critical point as $|\rho_{1,2}| \to 0.5$, \textit{i.e.} $\rho_{1,2} \to +i0.5$. 
At the critical point, they coalesce, forming a so-called exceptional point~\cite{bergholtz21_3302,ahida21_3301}. 
Finally, when $\rho_{1,2} \to +i$, the real part of both eigenvalues remains fixed at 0.5, but the first eigenvalue acquires a negative imaginary part while the second one acquires a positive imaginary part equal in magnitude to preserve the trace. 

In short, in this scenario, \emph{increasing the continuum couplings first enhances entanglement until it saturates, and then it increases the uncertainty on occupation probabilities.} 
This makes perfect sense, because increasing the overlap between two resonances would make them act more and more like one effective state. 

In the second scenario, a similar picture emerges, but an avoided crossing centered on the critical point appears. 
The eigenvalues never coalesce and, in fact, acquire an imaginary part as soon as $|\rho_{1,2}| \neq 0$. 
This is expected as the real part of $\rho_{1,2}$ must lead to coherent couplings (constructive interference) between the states $\ket{\psi_1}$ and $\ket{\psi_2}$, resulting in repulsion between the eigenvalues.

At the extreme, considered in the third scenario with $\theta = \pi/4$, we observe that $\lambda_1 > 1$ and $\lambda_2 < 0$ when $|\rho_{1,2}| > 0$. 
This result is obviously incompatible with any statistical interpretation. 
Physically, this could be a situation beyond the quasi-stationary formalism, like a system decaying through two broad resonances, where the decay time of each resonance is comparable to the timescale of the interaction.


%

\end{document}